\documentclass[draft]{agujournal2019}
\usepackage{url} 
\usepackage{lineno}
\usepackage[inline]{trackchanges} 
\usepackage{soul}
\usepackage{physics}

\newcommand\apjl{ApJL}     
\newcommand\apjs{ApJS}
\newcommand\aap{A\&A}

\draftfalse

\journalname{JGR: Planets}

\begin{document}

%
%


\title{Hazy blue worlds: A holistic aerosol model for Uranus and Neptune, including Dark Spots}

%
%




\authors{P.G.J. Irwin\affil{1}, N.A. Teanby\affil{2}, L.N. Fletcher\affil{3}, D. Toledo\affil{4}, G.S. Orton\affil{5}, M.H. Wong\affil{6}, M.T. Roman\affil{3}, S. P\'{e}rez-Hoyos\affil{7}, A. James\affil{1}, J. Dobinson\affil{1} }


\affiliation{1}{Department of Physics, University of Oxford, Parks Rd, Oxford, OX1 3PU, UK}
\affiliation{2}{School of Earth Sciences, University of Bristol, Wills Memorial Building, Queens Road, Bristol, BS8~1RJ, UK}
\affiliation{3}{School of Physics \& Astronomy, University of Leicester, University Road, Leicester, LE1 7RH, UK}
\affiliation{4}{Instituto Nacional de T\'ecnica Aeroespacial (INTA), 28850, Torrej\'on de Ardoz (Madrid), Spain.}
\affiliation{5}{Jet Propulsion Laboratory, California Institute of Technology, 4800 Oak Grove Drive, Pasadena, CA~91109, USA}
\affiliation{6}{Center for Integrative Planetary Science, University of California, Berkeley, CA 94720, USA}
\affiliation{7}{University of the Basque Country UPV/EHU, 48013 Bilbao, Spain}




\correspondingauthor{Patrick Irwin}{patrick.irwin@physics.ox.ac.uk}




\begin{keypoints}
\item Ice Giant reflectivity spectra from 0.3 to 2.5 $\mu$m well approximated by a single aerosol model comprised of three to four distinct layers.
\item Static stability region at 1--2 bar, caused by methane condensation, seems to lead to build-up of haze and seeds CH$_4$ snow at its base.
\item Darkening of deepest H$_2$S/haze layer, based at $p>$ 5--7 bar, found to account well for spectral properties of dark spots.
\end{keypoints}

%
%

%
%


\begin{abstract}
We present a reanalysis (using the Minnaert limb-darkening approximation) of visible/near-infrared (0.3 -- 2.5 $\mu$m) observations of Uranus and Neptune made by several instruments. We find a common model of the vertical aerosol distribution that is consistent with the observed reflectivity spectra of \textbf {both}  planets,  consisting of:  1) a deep aerosol layer with a base pressure $>$5--7 bar, assumed to be composed of a mixture of H$_2$S ice and photochemical haze; 2) a layer of photochemical haze/ice, coincident with a layer of high static stability at the methane condensation level at 1--2 bar; and 3) an extended layer of photochemical haze, likely mostly of the same composition as the 1--2-bar layer, extending from this level up through to the stratosphere, where the photochemical haze particles are thought to be produced. For Neptune, we find that we also need to add a thin layer of micron-sized methane ice particles at $\sim$0.2 bar to explain the enhanced reflection at longer methane-absorbing wavelengths. We suggest that methane condensing onto the haze particles at the base of the 1--2-bar aerosol layer forms ice/haze particles that grow very quickly to large size and immediately `snow out' \cite<as predicted by>{carlson88}, re-evaporating at deeper levels to release their core haze particles to act as condensation nuclei for H$_2$S ice formation. In addition, we find that the spectral characteristics of `dark spots', such as the Voyager-2/ISS Great Dark Spot and the HST/WFC3 NDS-2018, are well modelled by a darkening or possibly clearing of the deep aerosol layer only. 
\end{abstract}

\section*{Plain Language Summary}
Previous studies of the reflectance spectra of Uranus and Neptune have concentrated on individual, narrow wavelength regions and the conclusions have been difficult to compare with each other. Here, we analyse a combined set of observations from three different instruments covering the wavelength range 0.3 -- 2.5 $\mu$m to arrive at a \textbf{single} aerosol model that matches the observations at all wavelengths simultaneously for \textbf{both} planets. We conclude that photochemical haze produced in the upper atmospheres of both planets is steadily mixed down to lower layers, where it forms part of a vertically-thin layer in a statically stable region above the methane condensation level at 1--2 bar. We suggest that methane condenses so rapidly upon these haze particles that it efficiently `snows' out at the base of this layer, falling to lower, warmer levels, where the methane evaporates, releasing the core haze particles to `seed' H$_2$S condensation. For Neptune we need to add an additional layer of moderately large methane ice particles at $\sim$0.2 bar. Intriguingly, we find that a darkening (or perhaps clearing) of the lowest H$_2$S/haze layer matches very well the observed properties of the dark spots seen occasionally in Neptune's atmosphere and very occasionally in Uranus's atmosphere.

%
%

%


%
%
%
%


\section{Introduction} \label{introduction}

The visible and near-infrared spectra of the Solar System's `Ice Giants', Uranus and Neptune, have fascinated planetary astronomers for many years. The atmospheres of the Ice Giants are observed to have similar atmospheres with similar tropospheric temperature profiles and He/H$_2$ ratios, determined from Voyager-2 observations and post analyses \cite{lindal87,lindal92,sromovsky11}, and similar, high mole fractions of methane of $\sim$ 4\% \cite{kark09,kark11,sromovsky14,sromovsky19,irwin19methane}. Both planets appear blue or bluish-green to the naked eye, in contrast to the more yellowish appearance of Jupiter and Saturn. We now know this blueness comes from a combination of this higher abundance of gaseous methane, which has strong absorption bands in the infrared and red portion of the visible spectrum, and enhanced Rayleigh-scattering from  atmospheres that have low abundances of aerosols. 
In addition to CH$_4$, \citeA{irwin18} detected the presence of gaseous H$_2$S in Uranus's atmosphere, which is also probably present in Neptune's atmosphere \cite{irwin19h2s}. Hence, it would seem that H$_2$S and NH$_3$ react together to form a deep cloud of NH$_4$SH(s) (at pressures of $\sim$ 40 bar), which leaves H$_2$S alone to condense at a lower pressure, but greater than $\sim$ 3 bar \cite{irwin18,irwin19methane}. Given the known temperature profile and methane abundance, methane should condense at $\sim $1.5 bar in both planets, and hence it might be expected that a thick cloud of methane should similarly shroud both worlds. However, this does not seem to be the case, although upper tropospheric (200-600 mb) bright clouds are commonly seen in Neptune's atmosphere, and occasionally in Uranus's atmosphere. In addition, several dark spots have been seen in Neptune's atmosphere, most famously the Great Dark Spot observed by Voyager-2 in 1989 \cite{smith89}, but also in more recent Hubble Space Telescope observations \cite{hammel95,sromovsky11,wong18,hsu19}. These spots are of unknown origin, but seem only to be visible at wavelengths $<$ 700 nm. One dark spot has been reported in Uranus's atmosphere \cite{hammel09}, which was less dark, but appears to have been observable to wavelengths as long as 1.5 $\mu$m.

\begin{figure*}
\centering
\includegraphics[width=\textwidth]{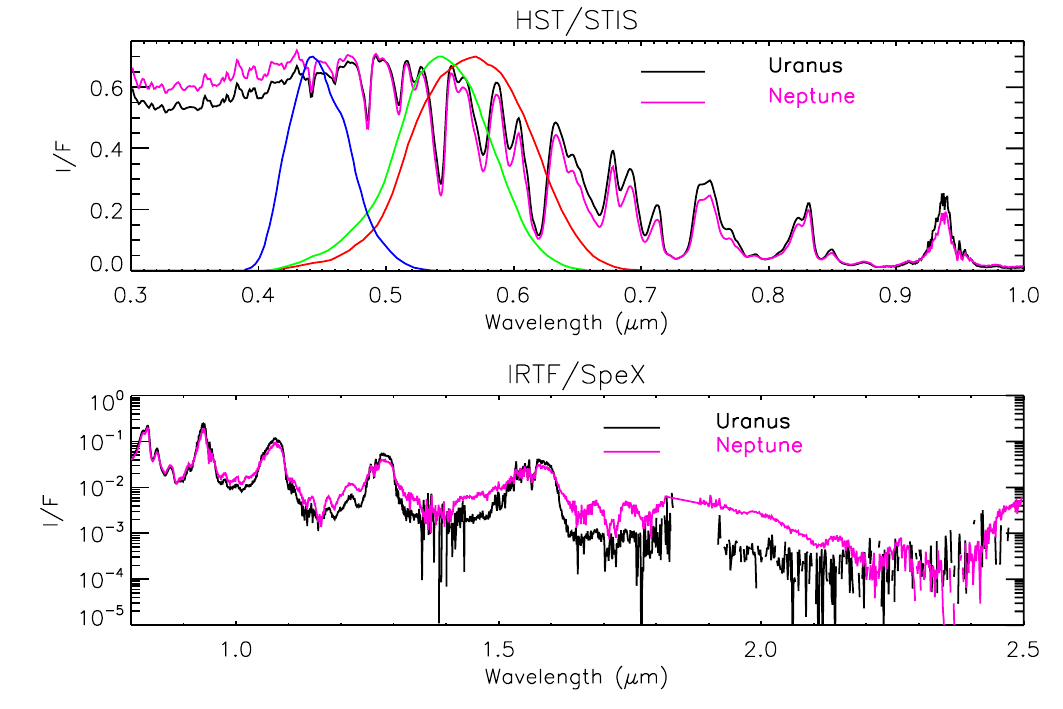}
\caption{Composite HST/STIS and IRTF/SpeX central-meridian-averaged I/F spectra of Uranus and Neptune compared with each other over the HST/STIS (0.3--1.0 $\mu$m) and IRTF/SpeX (0.8--2.5 $\mu$m) spectral ranges. Note that the HST/STIS data have been smoothed to the IRTF/SpeX resolution of 0.002 $\mu$m. Also overplotted, for reference, in the top panel are the red, green, blue sensitivities of the human eye \cite{stockmansharp00,stockman19}.
}
\label{fig:uranus_neptune_IFcompare}
\end{figure*}

At first glance, the visible/near-infrared spectra of these planets are very similar, as can be seen in Fig. \ref{fig:uranus_neptune_IFcompare}. However, it appears that at the peaks, Neptune is generally slightly less bright, while at longer methane-absorbing wavelengths, and in the UV, Neptune is more reflective. To simulate these spectra, a number of models of the vertical distribution of clouds and hazes have been proposed for Uranus \cite<e.g.,>{kark09,tice13,irwin15,irwin18,roman18,sromovsky19} and for Neptune \cite<e.g.,>{kark11,irwin11,luszcz16,irwin19h2s,irwin19methane}. All these models mostly have a thick aerosol layer at 2--4 bars (assumed to be photochemical haze and perhaps mixed with H$_2$S ice) and some sort of overlying haze. This aerosol layer has been modelled as a uniformly mixed (with height) haze \cite<e.g.,>{tice13, irwin15}, or as a discrete detached haze \cite<e.g.,>{irwin19methane}, or has been modelled \textit{ab initio} from microphysical models \cite<e.g.,>{toledo19, toledo20}. The reason that all these constituents are thought to be photochemically-produced hazes, rather than clouds, is that away from the discrete, bright upper tropospheric methane ice clouds, they are optically thin, appear to be spatially homogeneous and also need at some wavelengths to have  single-scattering albedos considerably less than unity \cite<e.g.,>{irwin11,irwin15} in order to be consistent with the observed limb-darkening/limb-brightening. The particles in the upper-tropospheric/lower-stratospheric haze, in particular, are seen to be rather dark in the 1--2 $\mu$m region \cite<e.g.,>{irwin11}, and the particles in the `main' 2--4-bar layer also seem to have significant absorption features. By contrast, freshly condensed methane (and presumably also H$_2$S ice) is expected to be nearly conservatively scattering.

Historically, cloud/haze retrieval studies have taken a set of observed spectra, covering a particular range of wavelengths, and found a cloud/haze model consistent with these observations using simple cloud/haze parameterisations and \textit{ad hoc} adjustments of single-scattering albedo and scattering phase function asymmetry. Different studies have used different wavelength ranges and different \textit{ad hoc} cloud/haze models, which has made it difficult to interpret these studies simultaneously to gain a deeper, simpler representation of the aerosol structures of these worlds. What makes such a holistic approach even more difficult is that we have no `ground truth' information on what the ice giant aerosols are made of and what their spectral properties are. Hence, the problem is extremely degenerate and multiple solutions exist.  Recently, attempts have been made to fit a wide range of wavelengths simultaneously using moderately self-consistent models. \citeA{irwin15} modelled IRTF/SpeX observations (0.8--1.8 $\mu$m) of Uranus, recorded near the disc centre, using a simple two-component model comprising a vertically thin aerosol layer based at $\sim 2$ bar, with variable opacity, base pressure and fractional scale height, combined with an extended haze of variable opacity, but fixed base pressure of 1.0 bar and fixed fractional scale height of 1.0 (i.e., uniformly mixed). This cloud/haze scheme was linked with a self-consistent cloud-scattering model that fitted for the imaginary refractive index spectra of the two constituents, reconstructing the real part using a Kramers-Kronig analysis and then calculated the extinction cross-sections, single-scattering albedos and phase functions using Mie theory.  It was found that the particles in the `main' $\sim$ 2-bar aerosol layer must be significantly more absorbing at wavelengths longer than 1 $\mu$m. Another study to find a model able to account for the observed reflectivity spectrum over a wide wavelength range is that of \citeA{sromovsky19}, who reanalysed  HST/STIS observations (0.3--1.0 $\mu$m) of  Uranus \cite{kark09}. Like \citeA{irwin15}, \citeA{sromovsky19} favoured a simple two-component cloud/haze structure to the more complex five-layer structure used in their previous analysis \cite{sromovsky14}, and like \citeA{irwin15} were able to make their cloud/haze model consistent out to 1.6 $\mu$m with IRTF/SpeX and Keck/NIRC2 observations by increasing the particles' imaginary refractive index at longer wavelengths. This increase of inferred $n_{imag}$ with wavelength has also been reported by \citeA{dekleer15}. \citeA{sromovsky19} noted that a plausible condensable species in the 1--3 bar region is H$_2$S, but since the spectral characteristics of this condensate are not currently known, the determined wavelength dependence of $n_{imag}$ could not be used to constrain its identity.  Finally, \citeA{sromovsky19} reported the possible detection of a second, deeper aerosol layer near $\sim$ 10 bar, which they tentatively suggested to be composed of NH$_4$SH. 

While attempts have been made to model Uranus spectra over a wider wavelength range, the same cannot be said for Neptune. Neptune is a significantly smaller target than Uranus and hence more difficult to spatially resolve, especially at longer wavelengths. It is also significantly more cloudy \cite<i.e., upper tropospheric methane ice clouds are widespread, for example>{irwin16}, which at lower spatial resolution makes it difficult to disentangle sunlight reflected from the background aerosol structure from that reflected from the discrete bright clouds. However, given these observational difficulties it is surprising that to our knowledge 
no attempt has previously been made to fit the spectra of both these planets with a single model, even though the observable atmospheres of these planets have similar tropospheric temperature and composition and have similar observable spectra, although we do note that the Uranus and Neptune models of \citeA{kark09} and \citeA{kark11} are very similar. In this paper we present a holistic aerosol model of both Uranus and Neptune and also propose a simple way of explaining the phenomenon of `dark spots', which are occasionally seen in these atmospheres, but whose nature has until now been a mystery.

\section{Observations} \label{observations}
The observations analysed in this study come from a variety of sources. The imaging observations from HST/WFC3 and Voyager-2/ISS (for Neptune) are described later, but here we give an overview of the spectral observations considered. 

\subsection{Uranus spectral observations}

We analysed HST/STIS observations of Uranus from 2002 \cite{kark09}, 2012 \cite{sromovsky14} and 2015 \cite{sromovsky19}. Since our primary intention in this paper was not to revisit the question of latitudinal variability of methane abundance or polar brightening, we concentrated on the 2002 data \cite{kark09}, when the disc of Uranus appeared particularly featureless, allowing us to combine the data from all latitudes together. This observation was made on August 19th 2002, between 01:43 and 10:57 UT. HST/STIS is actually a long slit spectrometer, but the slit was aligned parallel to the central meridian and then stepped from the central meridian to the edge of the planet to form a `cube' of half the planet, where at each location on the disc a complete spectrum covering 300.4 -- 1020.0 nm was recorded at a resolution of 1 nm, sampled every 0.4 nm. 

We also analysed an observation of Uranus made using IRTF/SpeX, another long-slit spectrometer. In this case,  the slit was aligned on the disc centre and the fluxes integrated along the central meridian. This standard reference spectrum is available on the IRTF Spectral Library website\footnote{\url{http://irtfweb.ifa.hawaii.edu/~spex/IRTF_Spectral_Library}}, reported by \citeA{rayner09}, and was made in SXD mode (0.8 -- 2.5 $\mu$m) on May 18th 2000. These data have a spectral resolution of 0.002 $\mu$m. 

Finally, we analysed observations of Uranus made with Gemini/NIFS in H-band (i.e., 1.47 -- 1.8 $\mu$m) in 2009 and 2010 \cite{irwin11a,irwin12}. Gemini/NIFS is an integral field-unit (IFU) spectrometer that simultaneously records spatial and spectral information, and where each pixel or `spaxel' is composed of a complete spectrum covering the targeted spectral range with a spectral resolving power of $R\sim 5200$. The actual cube used for our retrieval study was recorded on 2nd September 2009, which had good spatial resolution and was reasonably clear of discrete features. 

To enable easy intercomparison and simulation speed the HST/STIS and Gemini/NIFS data were smoothed to the spectral resolution of IRTF/SpeX of 0.002 $\mu$m (i.e., 2 nm), and sampled with a step of 0.001 $\mu$m.

During the time period spanned by these observations, Uranus moved through its orbit about the Sun and Table \ref{tab:obs_summary} lists the date, sub-Earth latitude (planetocentric) $\phi_E$, and also the apparent target-centered longitude of the Sun $L_s$ (which gives a convenient measure of season, with $0^\circ$ being northern spring equinox, $90^\circ$ being northern summer solstice, etc.) of the observations. In addition, we list the disc-integrated photometric magnitudes of Uranus as observed and reported by \citeA{lockwood19} in the $y$ (551 nm) and $b$ (472 nm) filters of the Str\"omgren photometric system, which gives a measure of the overall disc brightness and blueness. It can be seen that the IRTF/SpeX and HST/STIS observations were both made with $L_s \sim 335^\circ$ and $\phi_E \sim -25^\circ$. However, the Gemini/NIFS observation comes from just after the northern spring equinox in 2007 with $\phi_E = 7.9^\circ$. Hence, while the IRTF/SpeX and HST/STIS observations will be slightly more weighted to conditions in the southern hemisphere, the Gemini/NIFS observations will be more weighted to equatorial regions and in a slightly later season. However, neither set of observations will sample well the hazes in the polar regions. From 2000 to 2009 the disc-averaged magnitude of Uranus decreased by $\sim 0.03$ and became slightly more blue as the southern polar `hood' or `cap' \cite<e.g.,>{sromovsky08} became less visible. However, these changes are relatively small, enabling us to assume that all three sets of observations were observing approximately the same disc of Uranus. However, care should be taken when comparing this work with Voyager-2 studies of Uranus's aerosol structure, since Uranus was then near southern summer solstice and had a well developed `hood' and markedly higher albedo and thus brightness \cite{lockwood19}.

\begin{table}
\scriptsize
\caption{Summary of Spectral Observations of Uranus and Neptune}
\begin{tabular} {l l l l l l l l}
\hline

Uranus observations &  &  &  &  & & & \\
\hline
Date & Instrument & Wavelength range & $L_s$ & Sub-earth  & b-mag & y-mag & b-y \\
 &  &  &  & latitude $\phi_E$& 471 nm & 511 nm & \\
\hline
May 18th 2000 & IRTF/SpeX & 0.8 -- 2.5 $\mu$m & 330.4$^\circ$ & $-27.6^\circ$ & 5.769 & 5.560 & 0.209 \\
August 19th 2002 & HST/STIS & 300 -- 1020 nm & 339.3$^\circ$ & $-21.4^\circ$ & 5.779 & 5.577 & 0.202\\
September 2nd 2009 & Gemini/NIFS & 1.47 -- 1.8 $\mu$m & 6.8$^\circ$ & $+7.9^\circ$ & 5.785 & 5.592 & 0.193\\
\hline

Neptune observations &  &  &  &  & &  &\\
\hline
Date & Instrument & Wavelength range & $L_s$ & Sub-earth  & b-mag & y-mag & b-y \\
 &  &  &  & latitude $\phi_E$ & 471 nm & 511 nm & \\
\hline
June 30th 2000 & IRTF/SpeX & 0.8 -- 2.5 $\mu$m & 259.5$^\circ$ & $-28.7^\circ$ & 7.830 & 7.704 & 0.126\\
August 3rd 2003 & HST/STIS & 300 -- 1020 nm & 266.2$^\circ$ & $-29.1^\circ$ & 7.809 & 7.689 & 0.120\\
September 7th 2009 & Gemini/NIFS & 1.47 -- 1.8 $\mu$m & 279.5$^\circ$ & $-28.8^\circ$ & 7.819 & 7.694 & 0.125\\
\hline
\label{tab:obs_summary}
\end{tabular}
\end{table}

\subsection{Neptune spectral observations}

We analysed HST/STIS observations of Neptune \cite{kark11} made on August 3rd 2003, between 4:38 and 14:15 UT. As for the Uranus HST/STIS data these form a `cube' of half of Neptune's disc, covering 300.4 -- 1020.0 nm. 

We also analysed a reference observation of Neptune made using IRTF/SpeX and integrated along the central meridian, available on the IRTF Spectral Library \cite{rayner09}. The SXD component of this reference spectrum analysed here (0.8 -- 2.5 $\mu$m), was observed on June 30th 2000.

Finally, we analysed observations of Neptune made with Gemini/NIFS in 2009 (H-band) and 2011 (I, J, and H-band, i.e., 0.94 -- 1.16 $\mu$m, 1.14 -- 1.36 $\mu$m, and 1.47 -- 1.8 $\mu$m, respectively) \cite{irwin11,irwin16}. The actual cube used for our retrieval study was recorded in the H-band on 7th September 2009 and was chosen for its good spatial resolution and reasonably limited distribution of discrete cloud features. 

During the time period spanned by these observations, Neptune also moved through its orbit about the Sun and Table \ref{tab:obs_summary} again lists the date, sub-Earth latitude $\phi_E$, apparent target-centered longitude of the Sun $L_s$, and the disc-integrated photometric magnitudes \cite{lockwood19} of the observations. Neptune's southern summer solstice occurred in 2005, but Neptune's slower orbit about the Sun means that $L_s$ and $\phi_E$ differed little during the total elapsed period, and can be assumed to be $\sim 270^\circ$ and $\sim -29^\circ$, respectively. However, it can be seen that all three data sets will be weighted towards conditions in the southern hemisphere, with the south polar region clearly visible. From 2000 to 2009 the disc-averaged magnitudes and colour of Neptune were found not to vary significantly \cite{lockwood19}, again enabling us to assume that all three sets of observations were observing approximately the same disc of Neptune. However, care should again be taken when comparing this work with Voyager-2 studies of Neptune's aerosol structure, and indeed the Voyager-2/ISS observations described later in this paper, since Neptune was then still approaching southern summer solstice and was noticeably darker \cite{lockwood19}.

Once again, to enable easy intercomparison the HST/STIS and Gemini/NIFS  data were smoothed to IRTF/SpeX spectral resolution.

\section{Analysis}\label{analysis}
 
 \subsection{Atmospheric and Radiative Transfer Modelling}
 
 We modelled these observations using the radiative transfer and retrieval tool, NEMESIS \cite{irwin08,irwin22a,irwin22b}.
 To account for multiple scattering, NEMESIS employs a plane-parallel matrix operator model \cite{plass73}, where integration over zenith angle is performed with a Gauss-Lobatto quadrature scheme, while the azimuth integration is done with Fourier decomposition. We have found that five zenith angles are usually sufficient to model the giant planet I/F spectra and the number of azimuth Fourier components is set from the viewing and solar zenith angles as $N_F=\left \lfloor{\Theta/3}\right \rfloor$, where $\Theta = \max[\theta,\theta_0]$ and where $\theta$ is the observation zenith angle (in degrees) and $\theta_0$ is the solar zenith angle (in degrees). This scaling of $N_F$ with $\Theta$ is necessary to reconstruct reliably the scattering functions at higher zenith angles, but comes at a considerable computational cost. Calculations at intermediate zenith angles are interpolated between the matrix operator calculations at the nearest tabulated viewing and solar zenith angles. The five-angle zenith angle quadrature scheme used here is listed in Table 1 of \citeA{irwin21}. 
 
 NEMESIS was run in correlated-$k$ mode, with the methane $k$-tables generated from a number of different sources as described in Section \ref{sec:model}. For H$_2$--H$_2$ and H$_2$--He collision-induced absorption we used the coefficients of \citeA{borysow89a,borysow89b,borysow00}, assuming a thermally-equilibriated ortho:para hydrogen ratio for both Uranus and Neptune. Rayleigh scattering was included as described in \citeA{irwin19methane}, and the effects of polarization and Raman scattering were included as described in Section \ref{sec:model}. We used the solar spectrum of \citeA{chance10}, extrapolated to longer/shorter wavelengths with the solar spectrum of \citeA{kurucz93}. This combined solar spectrum was smoothed with a triangular line shape of FWHM = 0.002 $\mu$m for consistency with the Uranus and Neptune resampled spectra.
 
 \subsection{Atmospheric Models}
 
 The reference 
 temperature and mole fraction profile for Uranus is the same as that used by \citeA{irwin18} and is based on the `F1’ temperature profile determined from Voyager 2 and HST/STIS observations \cite{sromovsky11}, with He:H$_2$=0.131 and including 0.04\% mole fraction of neon. We adopted a simple `step' model for the methane profile with variable deep mole fraction (\textit{a priori} 4\%), variable relative humidity above the condensation level and a limiting stratospheric mole fraction of $1\times 10^{-4}$ \cite{encrenaz98, orton14}. For the atmospheric static stability study discussed later, we also added saturation-limited profiles of H$_2$S and NH$_3$. When dealing with condensing gases such as CH$_4$, once the profile for that gas was determined, the abundance of H$_2$ and He was scaled at each level, keeping He:H$_2$=0.131, to ensure that the mole fractions added up to 1.0
 
 The reference temperature and mole fraction profile for Neptune is the same as that used by \citeA{irwin19methane} and \citeA{irwin21} and is based on the  `N' profile determined by Voyager-2 radio-occultation measurements \cite{lindal92}, with He:H$_2$ = 0.177 (15:85) and including 0.3\% mole fraction of N$_2$. 
 We again adopted a simple `step' methane model with a variable deep mole fraction, variable relative humidity above the condensation level and limiting the stratospheric mole fraction to $1.5\times 10^{-3}$ \cite{lellouch10}. Again, the abundances of H$_2$ and He were adjusted at each level to ensure the mole fractions added up to 1.0, keeping He:H$_2$ = 0.177.

 \subsection{Spectral Modelling} \label{sec:model}
 
 The HST/STIS and IRTF/SpeX observations of both Uranus and Neptune were first combined to give composite central-meridian-averaged spectra of both planets, which are shown in 
 Fig. \ref{fig:uranus_neptune_IFcompare}.
 To create these combined spectra, we took the HST/STIS observations of Uranus and Neptune from 2002 and 2003, respectively, averaged them along the central meridian (masking discrete cloud features in the Neptune STIS data) and then scaled the IRTF/SpeX Uranus and Neptune spectra to match at overlapping wavelengths from 0.85 to 1.0 $\mu$m. This scaling was necessary as the IRTF/SpeX data are in units of line-integrated total flux and needed to be converted to I/F. Given uncertainty in the slit size and position it was not possible to do this \textit{ab initio} with sufficient accuracy and hence re-scaling was necessary to ensure consistency with HST/STIS data. The Gemini/NIFS data, although not used in this initial analysis, were also averaged along the central meridian and their reflectivities adjusted to match the scaled IRTF/SpeX spectrum at overlapping wavelengths for later use. 

\begin{figure*} 
\centering
\includegraphics[width=\textwidth]{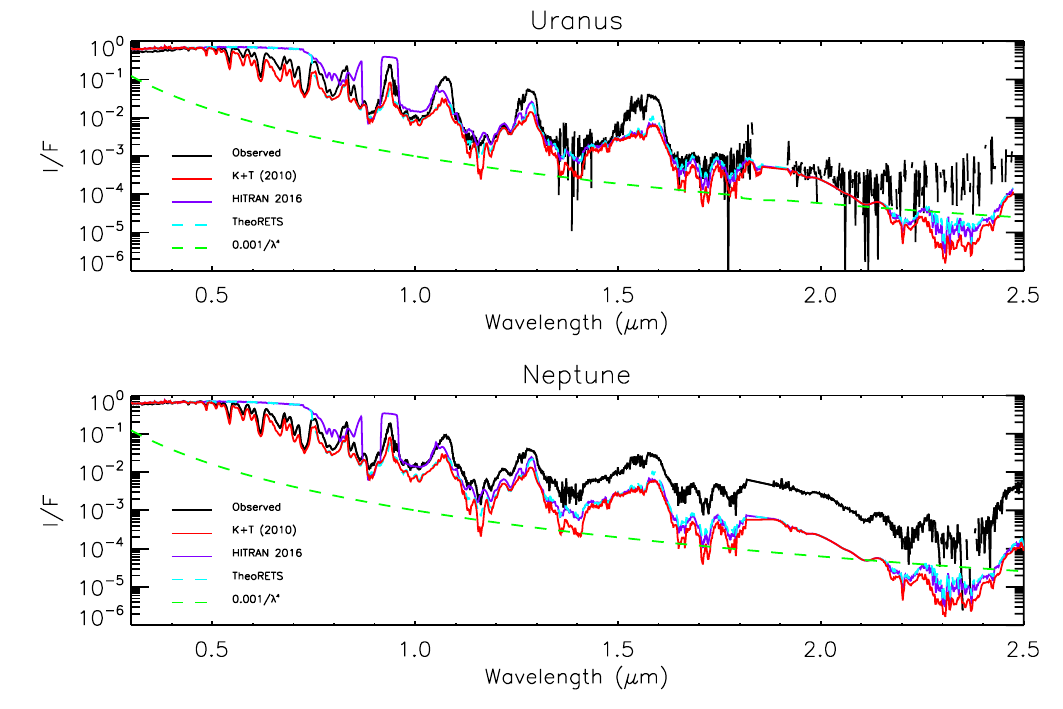}
\caption{Combined HST/STIS and IRTF/SpeX central-meridian-averaged I/F spectra of Uranus and Neptune compared with calculations from an aerosol-free atmosphere including only Rayleigh/Raman scattering and gaseous absorption from methane and hydrogen-helium collision-induced absorption. The observations are plotted in black, while calculations using different sources of methane absorption from band data \cite{kark10} and line datasets HITRAN2016 \cite{hitran16} and TheoRETS \cite{Rey18} are over-plotted in red, purple and cyan, respectively. Also plotted in green in both panels are simple $1/\lambda^4$ curves, showing the general trend of the combined spectra. N.B., the HITRAN2016 and TheoRETS do not cover the entire range and so at shorter wavelengths the calculations revert to Rayleigh/Raman scattering only.}
\label{fig:uranus_neptune_IF1}
\end{figure*}
 
 Until now, our radiative transfer model NEMESIS has not incorporated Raman scattering, but since the HST/STIS observations extend to 0.3 $\mu$m, where  \citeA{sromovsky05} shows that Raman scattering is significant, it was  necessary here to incorporate this effect. To do this, we followed the approach of \citeA{sromovsky05} and considered only the $S(0)$, $S(1)$ and combined `\textit{Q}' transitions, using the cross-section absorption coefficients of  \citeA{fordbrowne73}. Radiation scattered from short to longer wavelengths was introduced at the longer wavelengths as an additional pseudo-thermal-emission term, using the assumption of \citeA{sromovsky05} that this Raman re-emitted radiation is effectively isotropic, with wavenumber shifts of 354.69 cm$^{-1}$, 587.07 cm$^{-1}$ and 4160.00 cm$^{-1}$, respectively for the $S(0)$, $S(1)$ and combined `\textit{Q}' transitions. In addition to including Raman scattering, we also further revised NEMESIS to incorporate a correction for polarisation effects as described by \citeA{sromovsky05pol}.
 
 Using our upgraded NEMESIS model we show in Fig. \ref{fig:uranus_neptune_IF1} synthetic spectra calculated from our assumed Uranus and Neptune atmospheres for aerosol-free conditions, including Rayleigh scattering, Raman scattering, H$_2$-H$_2$ and H$_2$-He collision-induced absorption (CIA), and absorption by gaseous methane. As can be seen, the  measured reflectivity (I/F) of both Uranus and Neptune decreases rapidly with wavelength and is well matched overall by a $1/\lambda^4$ curve, suggesting that any aerosols which are present are small and of low integrated opacity. In Fig. \ref{fig:uranus_neptune_IF1} we compare calculations made using three different available sources of methane absorption data. The band data of \citeA{kark10}, converted to k-tables by \citeA{irwin11}, covers the entire wavelength range and reproduces most of the observed features well. However, at longer wavelengths we can also use k-tables generated from HITRAN2016 methane line data \cite{hitran16}, and a more recent compilation of methane line data from the TheoRETS project \cite{Rey18}. The HITRAN2016 data can be seen to be in good agreement with the band/k-data at wavelengths longer than 1.0 $\mu$m, while the TheoRETS data extends reasonably well to even shorter wavelengths of $0.75$ $\mu$m. However, neither set extends to visible wavelengths and since our aim in this study is to find a holistic aerosol model of Uranus and Neptune that matches all wavelengths covered by HST/STIS, IRTF/SpeX and Gemini/NIFS \textbf{simultaneously} we used the band/k-data of \citeA{kark10}/\citeA{irwin11} in this study. It is also apparent from Figs.  \ref{fig:uranus_neptune_IFcompare}--\ref{fig:uranus_neptune_IF1} that Neptune 
 is more reflective than Uranus at 
 methane-absorbing wavelengths longer than $\sim 1$ $\mu$m, indicating that Neptune has a higher opacity of upper-atmosphere aerosols, which must be of a size greater than $\sim$ 1 $\mu$m in order to be visible at these longer wavelengths. Furthermore, it would appear that any aerosols in Uranus's upper atmosphere must be of very low opacity, since a pure Rayleigh/Raman scattering calculation already matches the observed reflectivity at methane-absorbing wavelengths reasonably well.
 
\begin{figure*} 
\centering
\includegraphics[width=\textwidth]{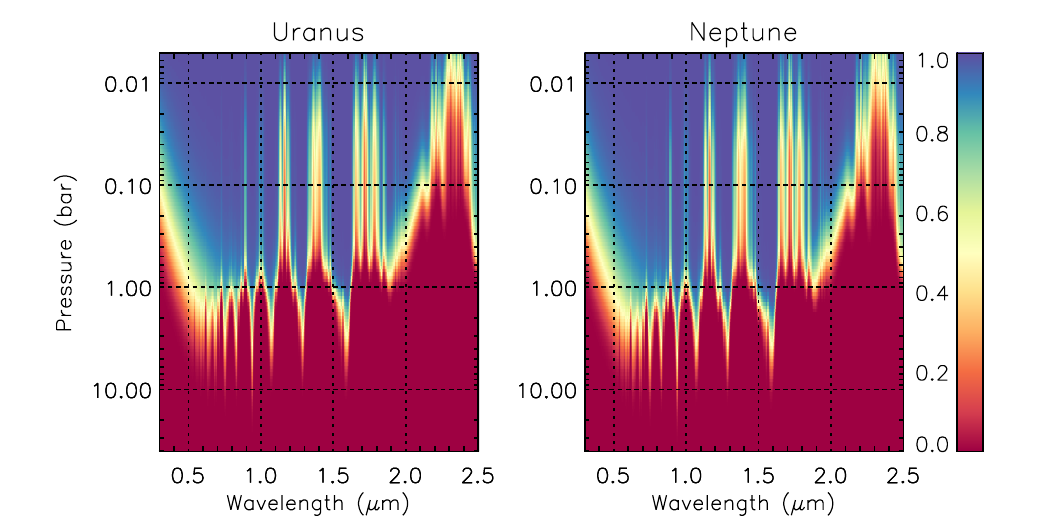}
\caption{Two-way transmission from space, for a vertical path, to different levels in our standard Uranus and Neptune atmospheres for aerosol-free conditions. These calculations include Rayleigh scattering, Raman scattering and absorption from gaseous methane and hydrogen-helium collision-induced absorption.} \label{fig:uranusneptune_reflatm}
\end{figure*}

In order to analyse the observations, we needed to determine the depths to which sunlight can penetrate at different wavelengths. In Fig. \ref{fig:uranusneptune_reflatm} we show contour plots of the two-way transmission from space, for a vertical path, to different levels in the atmospheres of Uranus and Neptune for aerosol-free conditions, including Rayleigh scattering, Raman scattering, methane absorption and hydrogen-helium collision-induced absorption. Here it can be seen that the penetration depth is very similar for Uranus and Neptune, but that a radiative transfer model wishing to simulate the observations needs potentially to cover a very wide range of pressure levels. Hence, the pressure range for both Uranus and Neptune models was set to cover 40 -- 0.001 bar and the atmosphere split into 39 layers (equally spaced in log pressure) for our radiative transfer calculations.

To best analyse the combined HST/STIS, IRTF/SpeX and Gemini/NIFS observations, and place strong constraint on the vertical structure and spectral dependence of particle scattering properties, we needed to develop an efficient way of modelling the observed centre-to-limb variations of the HST/STIS and Gemini/NIFS data and combine these with the wider wavelength-range central-meridian-averaged IRTF/SpeX observations, which provide important constraints on particle size and better probe lower pressure levels at the longer wavelengths. To do this we employed the Minnaert limb-darkening approximation \cite{minnaert41}, where for an observation at a particular wavelength, the reflectivity $R$ is approximated as:
 
\begin{linenomath*}
\begin{equation}\label{eq:minnaert}
R = R_{0} \mu_{0}^{k} \mu^{k-1}.
\end{equation}
\end{linenomath*}

Here, $\mu$ and $\mu_0$ are the cosines of the viewing and solar zenith angles, respectively,
$R_0$ is the fitted nadir reflectance, and $k$ is the fitted limb-darkening parameter. \citeA{irwin21} applied this model to VLT/MUSE observations of Neptune made in 2018 and found it to give a very good approximation to the observations and was also well reproduced by the NEMESIS radiative transfer model. Hence, in this study we used the Minnaert approximation to simplify our calculation of the disc-averaged limb-darkening and central-meridian averaged spectra following the scheme:

\begin{enumerate}
    \item As previously described, the HST/STIS and Gemini/NIFS observations were first averaged to the IRTF/SpeX resolution of 0.002 $\mu$m and sampled on a regular grid of spacing 0.001 $\mu$m. The  masked  HST/STIS  data  were  then averaged  along  the  central  meridian  and  used to  scale  the  IRTF/SpeX  observations  to  make  the  combined  HST/IRTF  processed observations self-consistent. The Gemini/NIFS data were also averaged along the central meridian (masking out discrete cloud features) and themselves scaled to be consistent with the combined HST/STIS and IRTF/SpeX observations.
    
    \item The averaged HST/STIS and Gemini/NIFS `cubes' were then masked to exclude discrete cloud features and the remaining observations at all latitudes Minnaert-analysed to derive spectra of the disc-averaged $R_0(\lambda)$ and $k(\lambda)$. This was simplified in this case since Uranus and Neptune are so distant that the solar zenith angle and viewing zenith angle can be approximated to be the same (i.e., $\mu = \mu_0$) and hence $R = R_{0} \mu^{2k-1}$.
    
    \item The fitted Minnaert parameters $R_0(\lambda)$ and $k(\lambda)$ were used to generate reconstructed spectra for HST/STIS and Gemini/NIFS at two zenith angles ($0^\circ$, $61.45^\circ$), corresponding to two of the five zenith angles of the Gauss-Lobatto multiple-scattering radiative transfer model used in NEMESIS. The higher angle is large enough to ensure we fully capture the observed limb-darkening/limb-brightening, and is coincident with one of the Gauss-Lobatto quadrature angles, thus obviating the need for any interpolation \cite{irwin21}. However, it is not so large that the computational cost becomes excessive. 
    
\end{enumerate}

The two resulting reconstructed HST/STIS spectra, two reconstructed Gemini/NIFS spectra and the measured IRTF/SpeX central-meridian line-averaged spectrum were then used as a set of `measured' observations to NEMESIS. By fitting an aerosol model to the HST/STIS and Gemini/NIFS processed spectra at $0^\circ$ and $61.45^\circ$ zenith angles, and extracting our best fits to the Minnaert parameters $R_0(\lambda)$ and $k(\lambda)$, we could then simply reproduce the observations at any other angle
from Eq. \ref{eq:minnaert}, assuming that the Minnaert approximation holds at all the other zenith angles of our zenith-angle quadrature scheme, which \citeA{irwin21} found to be a good approximation for their analysis of VLT/MUSE Neptune observations. However, we still needed an efficient way of simulating the central-meridian-averaged IRTF/SpeX spectra. We could have computed spectra at multiple locations along the central meridian and averaged these, but this would have been slow, especially near the disc edges. Instead, for each iteration we calculated spectra at two zenith angles ($0^\circ$ and $42.47^\circ$\footnote{The reason we chose the middle zenith angle of our quadrature scheme ($42.47^\circ$) here, instead of the second largest ($61.45^\circ$) as we did for the HST/STIS observations, is that calculations at lower zenith angles require fewer Fourier azimuth components and are more rapid; since we only needed to approximate the line-average in this case, rather than limb-darkening, the lower angle was found to be sufficiently accurate.}), extracted Minnaert parameters $R_0$ and $k$ at each wavelength (using $k=(1+\log(R_\mu/R_0)/\log(\mu))/2$, where $R_0$ is the nadir-calculated radiance and $R_\mu$ is the radiance calculated at $\mu = \cos(42.47^\circ)$)
and used these to compute the central-meridian line-average as follows. In \ref{App:B} we show that the mean radiance integrated along the central meridian of a  planet whose limb-darkening is well represented by the Minnaert approximation can be written (assuming $\mu = \mu_0$) as:
\begin{equation} \label{eq:line}
    \overline{R}_{line} = R_0 \int_0^{\pi/2}  (\cos\theta)^{2k} \dd \theta.
\end{equation}

Hence, if we have an estimate of the Minnaert limb-darkening coefficient, $k$, and nadir radiance, $R_0$, we can easily compute the corresponding central meridian line-averaged radiance. This function cannot be integrated analytically and so we pre-computed a table of $\overline{R}_{line}/R_0$ versus $k$ and then interpolated this to the value of $k$ derived at each wavelength. Hence, using calculations at just two angles to determine $k$ we were able to accurately simulate the central-meridian line-averaged IRTF/SpeX observations using Eq. \ref{eq:line}. 
For the HST/STIS and Gemini/NIFS data, we further verified that the central-meridian-line-average calculated from Eq. \ref{eq:line} using the disc-averaged Minnaert parameters was consistent with the measured IRTF/SpeX observations. For the Neptune Gemini/NIFS observations, this test necessitated some iteration on the degree to which discrete clouds were masked in order to generate the best overall  set of self-consistent Minnaert coefficients. All five spectra could then be fitted simultaneously, or individual spectra fitted separately as necessary.

Our starting assumption in this analysis is that the cloud/haze structures of Uranus and Neptune are substantially the same, but that Neptune has thicker upper-tropospheric/lower-stratospheric haze. Hence, to begin with we analysed the simpler case of the Uranus observations and extended to Neptune observations later.

\begin{figure*} 
\centering
\includegraphics[width=\textwidth]{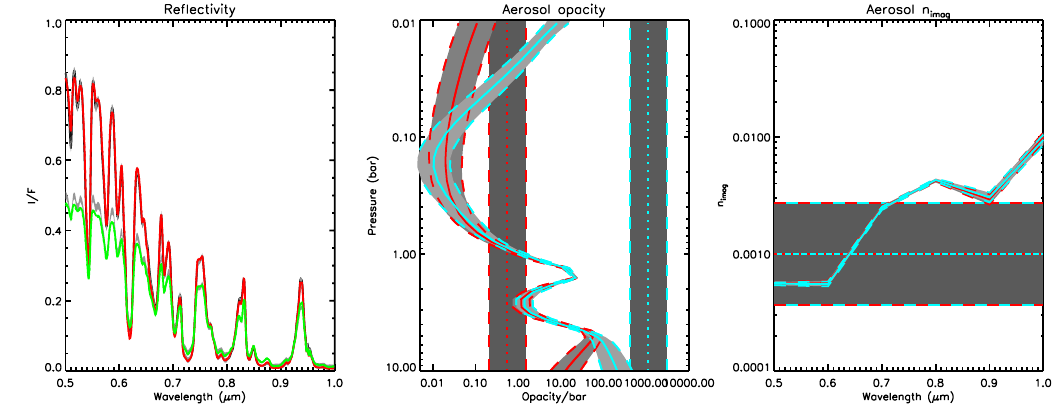}
\caption{Example `bracketed' retrieval fit to reconstructed HST/STIS spectra of Uranus from 0.5--1.0 $\mu$m at $0^\circ$ and $61.45^\circ$, using two continuous \textit{a priori} distributions of aerosols of radius 0.1 $\mu$m and variable $n_{imag}$ spectra. The left-hand panel compares the modelled spectra (red for $0^\circ$ and green for $61.45^\circ$) to the observations (grey) (N.B. The modelled fits for the two \textit{a priori} cases are indistinguishable at this scale). The middle panel shows the retrieved aerosol profiles (in terms of opacity/bar at the reference wavelength of 0.8 $\mu$m) for the two widely separated \textit{a priori} profiles, indicating two aerosol layers at depth. Here, the solid lines (red or cyan) show the retrieved profiles, while the dotted lines show the \textit{a priori} profiles; the \textit{a priori} and  retrieved errors are indicated by the shaded regions and dashed lines. It can be seen that the retrievals tend back to \textit{a priori} above and below the region of maximum sensitivity between 1 and $\sim$5 bar, and thus that although we can detect a second aerosol layer at $p>\sim$4 bar, we cannot detect whether if it has a base in the 5--10-bar range. The right-hand panel shows the two fitted $n_{imag}$ spectra, with the dotted line indicating the \textit{a priori} value  for both cases.} \label{fig:HSTcontinuous}
\end{figure*}

\subsection{Analysis of HST/STIS Uranus observations}

We first analysed the HST/STIS reconstructed observations of Uranus at zenith angles of $0^\circ$ and $61.45^\circ$ from 0.5--1.0 $\mu$m. We conducted a first-pass retrieval using a continuous distribution of haze
particles, in order to let the data (rather than \textit{a priori} assumptions) determine
the location of aerosol layers. In this first pass, the particle sizes were assumed to follow a
Gamma distribution with mean radius 0.1 $\mu$m and variance $\sigma = 0.3$. We also fitted for the imaginary refractive index spectrum (\textit{a priori} value set to 0.001) of these particles at six wavelengths from 0.5 to 1.0 $\mu$m, step 0.1 $\mu$m, and then reconstructed the real refractive index spectrum using the Kramers-Kronig relation, assuming the real refractive index at 0.8 $\mu$m was fixed to a value of 1.4. The resulting complex refractive index spectrum was then used to calculate, using Mie theory, the extinction cross-section, single-scattering albedo, and phase function spectra. However, since we expect the particles (ice and haze) to be non-spherical, the phase function spectra were approximated with combined Henyey-Greenstein phase function parameters to average over features peculiar to spherical particles such as the `rainbow' and `glory'. For the methane profile we assumed a `step' function with retrievable deep abundance up to the condensation level, but with the relative humidity above the condensation level fixed to 100\% 
and the stratospheric abundance limited to not exceed $1\times 10^{-4}$. The retrieval was attempted with two widely spaced \textit{a priori} aerosol abundance profiles (assumed to have constant opacity/bar at all pressure levels) to achieve a `bracketed' retrieval, where the retrieved profiles will overlap where they are well constrained by the measurements and tend back to their respective  \textit{a priori} values at pressures where there is little sensitivity. Fig. \ref{fig:HSTcontinuous} shows the fits to the reconstructed spectra together with the fitted aerosol abundance profiles. As can be seen, the fit struggles slightly at short wavelengths, but otherwise we find that the HST/STIS observations are well fitted by an aerosol structure with a thin, well defined peak at $\sim 1.5$ bar and a second layer at pressures greater than $\sim$4 bar. Although we can clearly detect the top of this deeper layer, we cannot constrain its base pressure as the bracketed retrievals tend back to their respective \textit{a priori} values at pressures exceeding approximately 5--7 bar; this makes sense given that the two-way transmission to space for a vertical path rapidly falls to zero in this pressure region, even for an aerosol-free atmosphere (Fig. \ref{fig:uranusneptune_reflatm}).  This indication of a double-peaked cloud/haze structure was also noted by \citeA{sromovsky19}, as described earlier.

\begin{figure*} 
\centering
\includegraphics[width=1.0\textwidth]{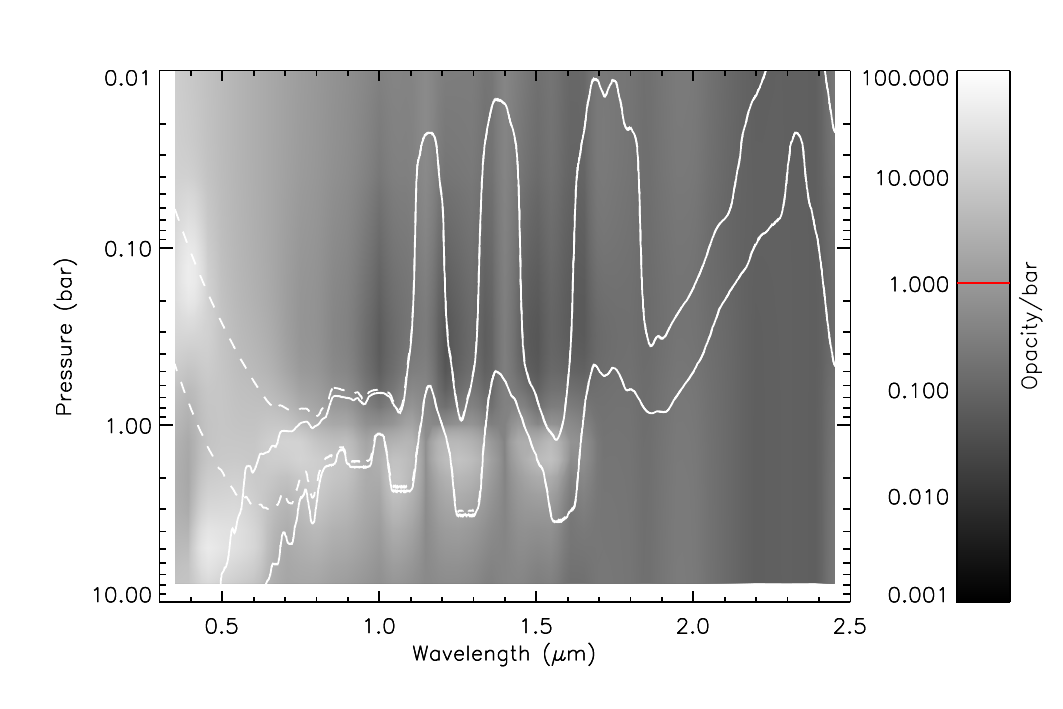}
\caption{Contour plot of vertical aerosol structure (opacity/bar) inferred from `snippet' Uranus retrievals, where for each wavelength the aerosol structure is retrieved from the wavelengths in a bin of width 0.1 $\mu$m centred on that wavelength. The contour levels are indicated in the bar on the right. From 0.3 to 0.8 $\mu$m the aerosol structure is infered from the HST/STIS nadir spectrum, while from 0.85 to 2.5 $\mu$m the IRTF/SpeX line-averaged spectrum is used. 
Overplotted are white lines showing the two-way nadir transmission to space of 0.2 and 0.8 for an aerosol-free atmosphere, either excluding (solid lines) or including (dashed lines) Rayleigh-Raman scattering. The retrieved aerosol opacity/bar shown is that derived with the full Rayleigh-Raman scattering taken into account. The red line in the opacity/bar key is the assumed \textit{a priori} value.
} \label{fig:snippet_uranus}
\end{figure*}

\subsection{Analysis of combined HST/STIS and IRTF/SpeX Uranus observations - snippet analysis}
Having noted that there may be two aerosol layers at depth in Uranus's atmosphere, we attempted to fit the combined HST/STIS and IRTF/SpeX data with three aerosol layers: one based at $p > $ 5--7 bar; one based at 1--2 bar; and an additional extended haze in the upper troposphere/lower stratosphere. 
We allowed the mean particle sizes of these constituents to vary and also fitted for the imaginary refractive index spectra of each particle type. We found this to be a massively degenerate problem, with multiple combinations of parameters able to match the observations equally well. To attempt to prevent the solutions becoming stuck in local minima, we performed multiple retrievals with randomly varied \textit{a priori} values of the pressure levels of these aerosol components and also their imaginary and real refractive indices, but were still unable to
find a reliable model that was clearly better than any of the others. An additional problem is that since the HST/STIS data sets do not include estimated errors, we needed a way to set appropriate error estimates on these data that were compatible with the published errors on the IRTF/SpeX observations and so provided appropriate weight between the two data sets. Most of all, though, we needed a way to decouple the vertical structure of the best-fitting aerosol profile from the
wavelength-dependent scattering and absorption properties of the particles necessary to model the spectra over a wide wavelength range. 

The method we settled on to do this was to split the HST/STIS and IRTF/SpeX spectra into a series of narrow wavelength-range `snippets', each covering a wavelength bin of width 0.1 $\mu$m (narrow, but in most cases covering a reasonably large range of methane absorption strengths) and stepped by 0.05 $\mu$m to achieve Nyquist sampling. Then for each spectral snippet, we adopted a continuous distribution of
aerosols with 
$r_{mean} = 0.1$ $\mu$m and $\sigma=0.05$, assumed that $n_{real}$ = 1.4, and fitted for the imaginary refractive index, which we assumed to be constant across the moderately narrow wavelength bins chosen. For the snippet covering 0.8 -- 0.9 $\mu$m, which includes CIA bands of H$_2$--H$_2$ and H$_2$--He, we also fitted for the deep CH$_4$ abundance, which we then kept fixed for all other snippets. Hence, this snippet had to be fitted first.

For the HST/STIS `snippets' we only fitted to the nadir spectrum to keep the retrievals fast, while for the IRTF/SpeX `snippets' we fitted to the central-meridian line average as described earlier, modelling spectra at two zenith angles (0$^\circ$ and 42.47$^\circ$) and combining using Eq. \ref{eq:line}. 
At this point we considered what errors would be appropriate to assume for the data. Looking at the quality of the fits, we found we were able to fit the HST/STIS reflectivity spectra to a precision of $\chi^2/n \sim 1$ if we assumed the errors to be equivalent to 1/50 of the peak reflectivity in each bin. 
This estimated error covers unknown systematic uncertainties from
sources such as the Lucy-Richardson spatial deconvolution applied to the
HST/STIS data, the inhomogeneous aerosol structure, and the methane absorption
k-table parameters.
While the HST/STIS data do not include error estimates, the IRTF/SpeX data do contain these, and so we added the systematic R/50 errors in quadrature with the measured/scaled errors. Our approach meant that: 1) we achieved a good balance between the STIS and SpeX data; 2) we did not overfit the peak radiance wavelengths;  and 3) in regions of low reflectance for the SpeX data we did not attempt to fit noise.

The fitted aerosol profiles for each snippet were then combined together and plotted as a contour function of aerosol opacity/bar against the mean bin wavelength as shown in Fig. \ref{fig:snippet_uranus}. Also plotted on Fig. \ref{fig:snippet_uranus} are the depths for two-way nadir transmission of 0.2 and 0.8, respectively, for an aerosol-free atmosphere and including or excluding Rayleigh and Raman scattering to give an idea of how deeply we can see at different wavelengths. 
We can draw a number of conclusions from Fig. \ref{fig:snippet_uranus}. First of all these `snippet' retrievals favour a vertically confined aerosol layer at $\sim$ 1.5--2.5 bar for all wavelengths that penetrate deeply, except 0.4--0.6 $\mu$m, where a much deeper aerosol layer at $p>\sim$5 bar is indicated. The apparent base of this deeper layer is, as in the low \textit{a priori} example shown in Fig. \ref{fig:HSTcontinuous}, caused by the retrieved profile tending back to \textit{a priori} at depth. The highest opacity/bar values of these constituents are generally found between two-way transmission levels of 0.2 and 0.8, where the retrieval has most sensitivity. 
It is likely that the deeper $p>$ 5--7 bar aerosol layer is present at all wavelengths, but cannot be detected at longer wavelengths due to 
lower reflectivity, increased gas absorption, and increased opacity of the
overlying 1.5--2.5-bar aerosol layer. It should also be noted that the aerosol layer at $\sim 2$ bar has a much lower integrated opacity than the deeper layer, since although both have similar peak opacity/bar values, the 2-bar layer spans a smaller range of pressure. The narrow wavelength range of detectability makes it difficult to say much about the particle size of aerosols in the deepest layer. However, for the 1.5--2.5 bar layer the opacity drops slowly with wavelength, suggesting particle sizes of the order of 1 $\mu$m.  Finally, it can be seen that at short wavelengths (i.e., 0.3--1.0 $\mu$m) there is general indication of an extended upper tropospheric/lower stratospheric haze, where the opacity/bar increases with altitude at pressures less than $\sim 1$ bar, but where the opacity drops with increasing wavelength. The retrieved wavelength dependence of this constituent of $\sim 1/\lambda^4$ suggests a very small particle size, while the increasing opacity/bar with altitude is consistent with microphysical modelling of haze production in Ice Giant atmospheres of \citeA{toledo19,toledo20}, who predict a haze distribution with a fractional scale height of 2--4.

\begin{figure*} 
\centering
\includegraphics[width=1.0\textwidth]{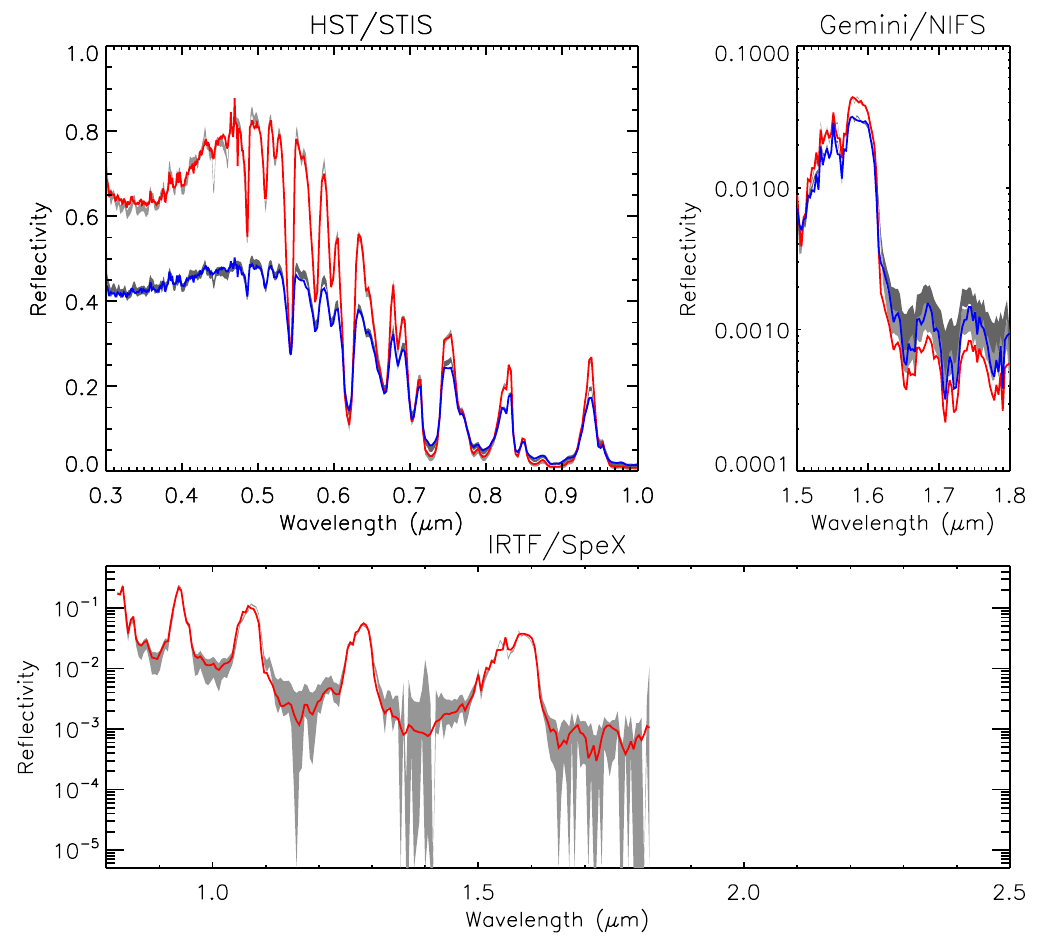}
\caption{Example of one of the best fitted spectra for Uranus, with $\chi^2/n = 3.76$ for our assumed errors. Here, the measured spectra and assumed error range are shown in grey and the fits at 0$^\circ$ and 61.45$^\circ$ are shown in red and blue, respectively, for the HST/STIS and Gemini/NIFS observations. For the IRTF/SpeX observation, the simulated central-meridian line average is shown in red. It can be seen that we achieve a very good fit to the data for all three data sets.} \label{fig:spectrum_uranus}
\end{figure*}

\subsection{Simultaneous retrieval of combined HST/STIS, IRTF/SpeX, and Gemini/NIFS Uranus observations}
Having established the general vertical aerosol structure of Uranus from the first-pass retrieval, and
the wavelength-dependent absorption from the snippet analysis, we moved to a
parameterized model for the final retrieval. The parameterized model for Uranus
included three aerosol layers: 
1) a deep cloud/haze layer, `Aerosol-1' fixed with a base at 10 bar (i.e., well below where we are sensitive to), with variable fractional scale height; 2) a vertically-thin cloud/haze at 1--2 bar, `Aerosol-2', modelled as a Gaussian distribution with fixed width of $\sim 0.1$ scale heights; and 3) an extended tropospheric/stratospheric haze, `Aerosol-3', based at 1--2 bar with a fixed fractional scale height of 2.0. Since the Aerosol-1 layer is effectively infinitely deep compared with our sensitivity functions, in this study we only quote its integrated opacity from space down to the level at which this sensitivity diminishes, which can be seen from Fig. \ref{fig:HSTcontinuous} to be $\sim$5 bar. Also, since the deep Aerosol-1 layer is only of importance at short wavelengths we \textbf{assumed} it to be composed predominantly of very small particles and assumed a Gamma distribution of sizes with $r_{mean} =0.05$ $\mu$m and variance, $\sigma=0.05$. We assumed the same particle size distribution for the tropospheric/stratospheric haze particles (Aerosol-3) since the opacity of this component needs to fall as $1/\lambda^4$. For the intermediate cloud/haze layer at 1--2 bar (Aerosol-2) we tried a number of mean particle sizes of size $\sim$ 1 $\mu$m, with variance $\sigma=0.3$. Given the degenerate nature of this problem, we first fitted to the HST/STIS data alone in the wavelength range 0.8--0.9 $\mu$m to constrain the deep methane abundance and pressure of the 1--2 bar Aerosol-2 component, which we then used as the \textit{a priori} in subsequent all-wavelength retrievals. To do this, we generated a set of 30 retrievals with randomly chosen values of the Aerosol-1 fractional scale heights (in range 0.05--0.15), randomly chosen Aerosol-2 mean particle radii (in range 0.1--4.0 $\mu$m) and then fitted for the opacities of all three constituents, their mean imaginary refractive indices (assumed constant in 0.8--0.9 $\mu$m range), together with the deep methane abundance, setting the relative humidity above the condensation level to 100\% and limiting the stratospheric mole fraction to not exceed $10^{-4}$. The real refractive index of all components was assumed to be 1.4\footnote{Other values for the real refractive index were tried, but were not found to more or less appropriate}. Then, for the best-fitting aerosol/methane solutions to the 0.8--0.9 $\mu$m region (achieving $\chi^2/n < 1)$, we fitted the entire combined STIS/SpeX/NIFS dataset using these retrievals as the \textit{a priori}, keeping the mean Aerosol-2 particle size fixed, but fitting again for all other properties and also for the imaginary refractive index spectra from 0.2--1.9 $\mu$m, with step 0.1 $\mu$m and assuming $n_{real}=1.4$ at 0.8 $\mu$m. We  again reconstructed $n_{real}$ at other wavelengths using the Kramers-Kronig relation and then generated cross-section spectra, single-scattering albedo spectra and phase function spectra from Mie theory, approximating the phase functions with double Henyey-Greenstein functions. 

\begin{figure*} 
\centering
\includegraphics[width=1.0\textwidth]{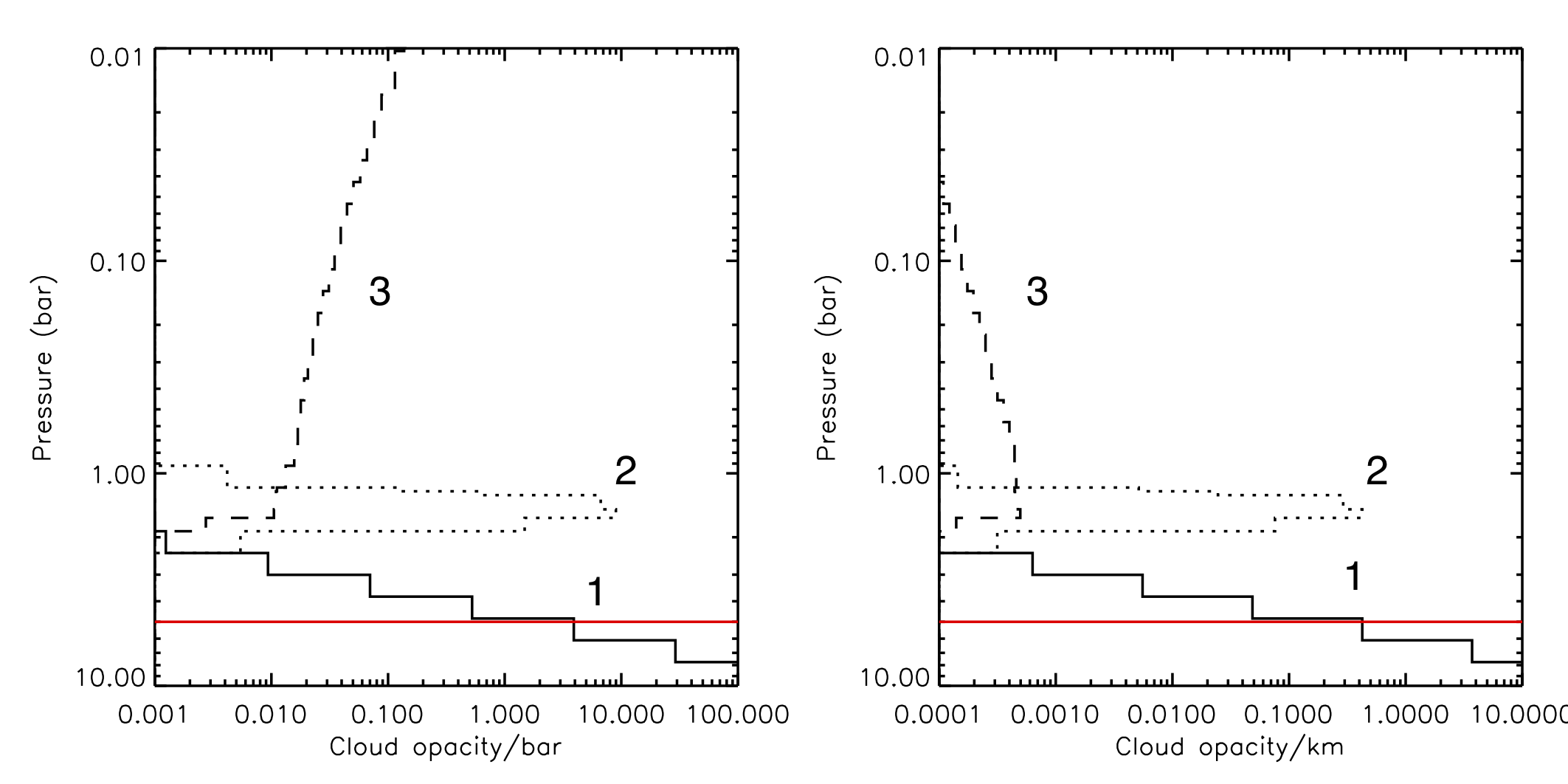}
\caption{Fitted vertical aerosol structure for Uranus's atmosphere from our example best-fit retrieval, presented in terms of opacity/bar (left) and opacity/km (right) at 0.8 $\mu$m. The deep aerosol layer (Aerosol-1) is shown as a solid line, the middle aerosol layer (Aerosol-2) as a dotted line, and the extended tropospheric/stratospheric haze (Aerosol-3) as a dashed line. The red line marks the 5-bar level, which is roughly the level where our sensitivity runs out.} \label{fig:cloud_uranus}
\end{figure*}

For the HST/STIS spectra at zenith angles of 0$^\circ$ and 61.45$^\circ$ we modelled 500 points from 0.2--1.0 $\mu$m, with 300 points from 0.2 to 0.5 $\mu$m, spaced every 0.001 $\mu$m to fully capture the Raman-scattering component, and the remaining 200 points equally spaced from 0.5 to 1.0 $\mu$m. Although the HST/STIS data do not extend below 0.3 $\mu$m, we still needed to model the spectra from 0.2 $\mu$m  upwards in order to compute the effect of sunlight being scattered to the longer wavelengths observed by STIS. We generated synthetic `measured' radiances at $\lambda < 0.3$ $\mu$m having the same reflectivity as observed at 0.3 $\mu$m, but with very large errors in order that NEMESIS did not try to fit to them, but did include their Raman-scattering effect on the measured wavelengths. For wavelengths longer than 0.3 $\mu$m, we set the reflectivity errors to be equivalent to 1/50 of the reflectivity of the nearest peak, but in addition prevented the reflectivity error from exceeding a value of 0.01 in order to force the model to fit the data well near $\sim 0.5$ $\mu$m, where the observed reflectances peak. We also halved the errors in the 0.8--0.9 $\mu$m range to ensure good fit at the wavelengths best able to differentiate between methane abundance and 1--2-bar aerosol layer (Aerosol-2) pressure.

For the Gemini/NIFS data, we selected 100 wavelengths equally spaced between 1.5 and 1.8 $\mu$m, again at 0$^\circ$ and 61.45$^\circ$ zenith angles. The errors were set to 1/100 of the maximum reflectivity of the 1.55 $\mu$m peak to give these data slightly more weight in the final fit than the IRTF/SpeX data since they contain clearer limb-darkening information. 

Finally, for the IRTF/SpeX data we selected 200 wavelengths, equally spaced between 0.8 and 1.9 $\mu$m (longer wavelengths were discounted as they can be seen in Fig. \ref{fig:uranus_neptune_IFcompare} to have low SNR), and set the errors to be 1/50 of the nearest peak reflectance, added in quadrature with the published errors. 

\begin{figure*} 
\centering
\includegraphics[width=1.0\textwidth]{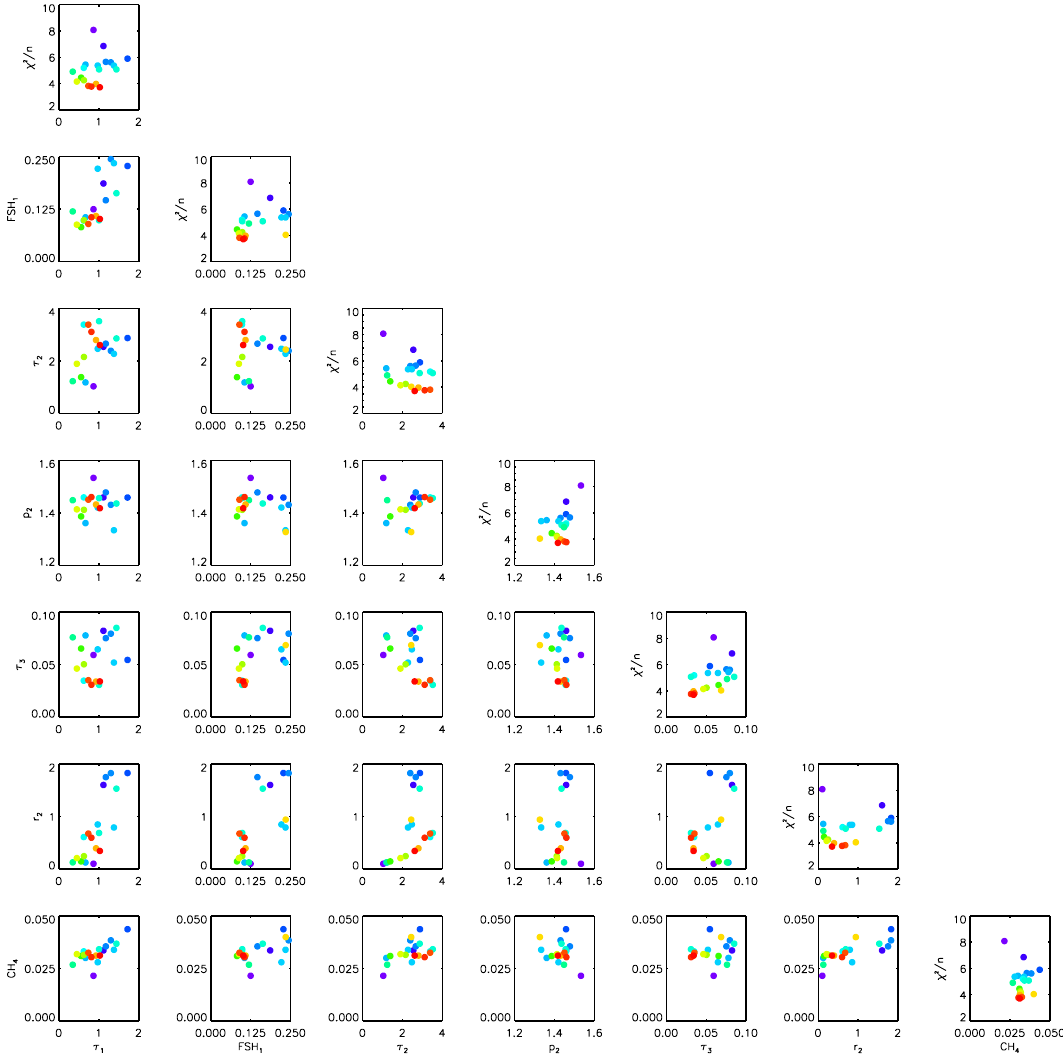}
\caption{`Corner' plot of 30 retrievals of the combined HST/STIS, IRTF/SpeX and Gemini/NIFS Uranus dataset with $\chi^2/n < 10$. Note that for the opacity of the Aerosol-1 layer, $\tau_1$, the opacity plotted is that integrated from space down to the 5-bar pressure level. The data points are colour-coded by the $\chi^2/n$ of the fit, with red fitting best and purple worst. Along the leading diagonal, instead of plotting the marginalised errors as would be usual for such plots, we plot $\chi^2/n$ as a function of fitted parameter, again colour-coded with red points having the lowest $\chi^2/n$ and purple points the highest. Although some parameters are better constrained than others, it can be seen that there is little cross-correlation between any of the fitted parameters.} \label{fig:corner_uranus}
\end{figure*}

Figure \ref{fig:spectrum_uranus} shows one of our best fits to the combined data set where we can see that we match the observed reflectivity spectra very well. The vertical cloud/haze structure retrieved in this case is shown in Fig. \ref{fig:cloud_uranus}, both in terms of opacity/bar and opacity/km at 0.8 $\mu$m (see \ref{App:C} for an explanation of the different opacity units), which is comprised of a deep aerosol layer (Aerosol-1) based at 10 bar with fractional scale height 0.11 and integrated opacity to 5 bar (at 0.8 $\mu$m) 0.81, a middle vertically-thin aerosol layer (Aerosol-2) at 1.48 bar and opacity 3.1, composed of particles of mean radius 0.6 $\mu$m, and an extended haze (Aerosol-3), based at 1.6 bar with fixed fractional scale height 2.0 and retrieved opacity 0.03. The retrieved methane mole fraction is 3.0\% and the relative humidity above the condensation level was fixed to 100\%. It should be stressed here that there is no single solution that clearly fits better than all others. As we have performed multiple slightly perturbed retrievals there is a general class of solutions that fit best, of which this is one representative example. The complete set of retrieved parameters is shown in Fig. \ref{fig:corner_uranus} and the range of solutions and errors are listed in Table \ref{tab:retrieval_results}.

\begin{table}
\scriptsize
\caption{Fitted model parameters to combined HST/STIS, IRTF/SpeX and Gemini/NIFS data. Note that the fractional scale height of the Gaussian-shaped Aerosol-2 layer (and Aerosol-4 layer for Neptune) was fixed at 0.1 and the fractional scale height of the Aerosol-3 layer was fixed at 2.0. Note also that the quoted opacities for the Aerosol-1 layer, $\tau_1$ are integrated from space down to a pressure level of 5 bar.}
\begin{tabular} {l l l}
\hline
Parameter & Uranus & Neptune \\
\hline
Opacity $\tau_1$ (at 0.8 $\mu$m) & 0.8 -- 1.1 & 0.5 -- 0.8 \\
Frac. Sc. Ht. $fsh_1$ & 0.08 -- 0.12 & 0.14 -- 0.18 \\
Radius $r_1$  & 0.05 $\mu$m (fixed) & 0.05 $\mu$m (fixed)\\
\hline
Pressure $p_2$ & 1.4 -- 1.5 bar & 2.0 -- 2.1 bar\\
Opacity $\tau_2$ (at 0.8 $\mu$m) & 2.0 -- 3.5 & 1 -- 2 \\
Radius $r_2$  & 0.3 -- 0.8 $\mu$m & 0.3 -- 0.8 $\mu$m\\
\hline
Pressure $p_3$ & 1.6 bar (fixed) & 1.6 bar (fixed) \\
Opacity $\tau_3$ (at 0.8 $\mu$m) & $0.03 \pm 0.01$ & $0.04 \pm 0.01$\\
Radius $r_3$  & 0.05 $\mu$m (fixed) & 0.05 $\mu$m (fixed)\\
\hline
Pressure $p_4$ & n/a & 0.2 bar (fixed) \\
Opacity $\tau_4$ (at 0.8 $\mu$m) & n/a & $0.030 \pm 0.005$\\
Radius $r_4$  & n/a & $2.5 \pm 0.5$ $\mu$m \\
\hline
Deep CH$_4$ mole fraction & $3 \pm 1$ \% & $7 \pm 1$ \% \\
Tropopause CH$_4$ RH & 100\% (fixed) & $35 \pm 5$ \% \\
\hline
\label{tab:retrieval_results}
\end{tabular}
\end{table}

\begin{figure*} 
\centering
\includegraphics[width=1.0\textwidth]{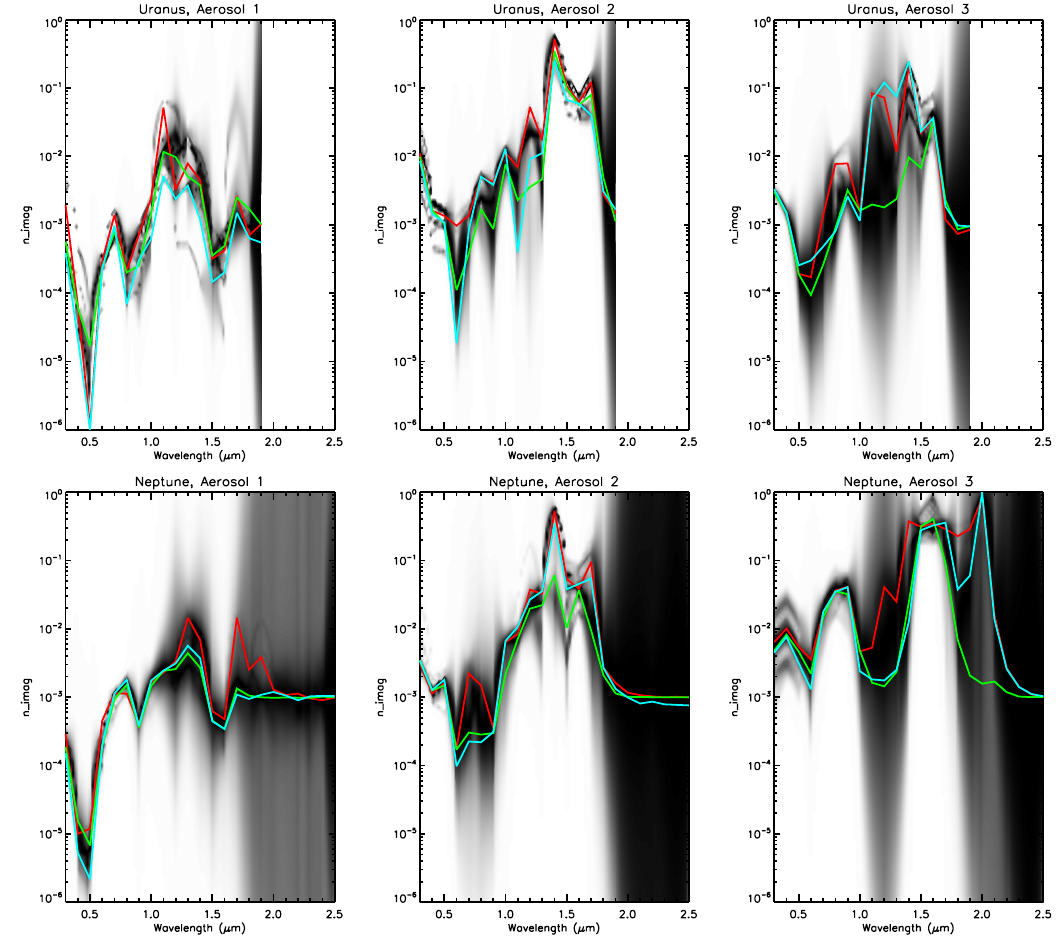}
\caption{Fitted $n_{imag}$ spectra for Uranus (top row) and Neptune (bottom row) for the lower cloud/haze at 10 bar (Aerosol-1), the middle cloud/haze at 1--2 bar (Aerosol-2) and the vertically-extended upper tropospheric/lower stratospheric haze (Aerosol-3). The filled contour plots show the linear addition of the best-fitting imaginary index distributions. Over-plotted on these distributions are the extracted mean $n_{imag}$ spectra, where red are the weighted averages of all best-fitting retrieved spectra, green are the contour-map-weighted averages, and cyan are the retrieved spectra from the representative retrieval case shown for each planet. These spectra are listed in Tables \ref{tab:refuranus} and Tables \ref{tab:refneptune}. N.B., the \textit{a priori} value of $n_{imag}$ was set to 0.001 and this is the value the retrievals tend back to when the data are no longer constraining. The IRTF/SpeX data for Uranus were truncated at 1.9 $\mu$m as it can be seen  from Fig. \ref{fig:uranus_neptune_IFcompare} that the observations are too noisy to use at longer wavelengths.} \label{fig:nimag_combined}
\end{figure*}

The estimated $n_{imag}$ spectra for the three constituents are shown in Fig. \ref{fig:nimag_combined}. Here, we have taken the fitted $n_{imag}$ spectra for the best-fitting cases and combined them together to give an idea of the range of acceptable solutions, together with two estimates of the optimal solution and the fitted spectra from our example best-fitting case, which are listed in Table \ref{tab:refuranus}. As can be seen there are certain features common to all the best-fitting retrieved imaginary refractive index spectra, namely:

\begin{enumerate}
\item The lowest aerosol layer (Aerosol-1) has very low $n_{imag}$ at 0.4--0.5 $\mu$m, which increases either side, but never gets very big. This means this layer is bright and highly scattering at 0.5 $\mu$m, but less so either side. Since we chose this layer to be composed of very small particles its contribution diminishes rapidly at longer wavelengths.

\item The 1--2-bar aerosol layer (Aerosol-2) has lowest $n_{imag}$ from 0.4--0.6 $\mu$m, which then increases sharply on either side. This increase is necessary to give sufficient ultraviolet (UV) absorption and also to reproduce the observed limb-darkening/limb-brightening seen at longer wavelengths. The retrieved mean radius of the particles in this layer of $\sim 1$ $\mu$m, makes them moderately forward-scattering at 0.5 $\mu$m, lowering their back-scatter and thus contribution to the reflected spectra, and increasing their `transparency' to allow light to be reflected from the deeper $p > $ 5--7 bar Aerosol-1 layer. 

\item The tropospheric/stratospheric haze (Aerosol-3) $n_{imag}$ spectrum is moderately similar to that of the Aerosol-2 layer, but is less well constrained. 
\end{enumerate}

\begin{table*}
\scriptsize
\caption{Estimated $n_{imag}$ spectra for Uranus aerosols. Here, for each aerosol type, `Mean 1' is the weighted average for all best-fitting retrieved spectra, while `Mean 2' are the averages of the contour maps shown in Fig. \ref{fig:nimag_combined}. `Sample' is the retrieved spectra from the representative best-fitting sample case shown for each planet.}
\centering
\begin{tabular} {l l l l l l l l l l}
\hline
 & Aerosol-1  & ($>$ 5--7 bar) & & Aerosol-2  & (1--2 bar) & & Aerosol-3  & (extended) & \\
$\lambda$($\mu$m) &  Mean 1 & Mean 2 & Sample &  Mean 1 & Mean 2 & Sample &  Mean 1 & Mean 2 & Sample \\
\hline
0.3 & 1.95$\times 10^{-3}$ & 5.62$\times 10^{-4}$ & 3.88$\times 10^{-4}$ & 1.10$\times 10^{-2}$ & 9.90$\times 10^{-3}$ & 8.59$\times 10^{-3}$ & 3.07$\times 10^{-3}$ & 2.98$\times 10^{-3}$ & 3.31$\times 10^{-3}$ \\
0.4 & 6.73$\times 10^{-5}$ & 6.17$\times 10^{-5}$ & 2.19$\times 10^{-5}$ & 1.61$\times 10^{-3}$ & 1.64$\times 10^{-3}$ & 1.44$\times 10^{-3}$ & 1.50$\times 10^{-3}$ & 1.48$\times 10^{-3}$ & 1.51$\times 10^{-3}$ \\
0.5 & 1.10$\times 10^{-6}$ & 1.66$\times 10^{-5}$ & 1.00$\times 10^{-6}$ & 1.35$\times 10^{-3}$ & 1.03$\times 10^{-3}$ & 1.07$\times 10^{-3}$ & 1.91$\times 10^{-4}$ & 1.90$\times 10^{-4}$ & 2.54$\times 10^{-4}$ \\
0.6 & 2.96$\times 10^{-4}$ & 2.36$\times 10^{-4}$ & 2.30$\times 10^{-4}$ & 9.65$\times 10^{-4}$ & 1.10$\times 10^{-4}$ & 1.86$\times 10^{-5}$ & 1.70$\times 10^{-4}$ & 9.34$\times 10^{-5}$ & 2.99$\times 10^{-4}$ \\
0.7 & 1.39$\times 10^{-3}$ & 7.88$\times 10^{-4}$ & 9.66$\times 10^{-4}$ & 1.37$\times 10^{-3}$ & 3.78$\times 10^{-4}$ & 8.37$\times 10^{-4}$ & 1.38$\times 10^{-3}$ & 2.59$\times 10^{-4}$ & 4.79$\times 10^{-4}$ \\
0.8 & 2.12$\times 10^{-4}$ & 1.99$\times 10^{-4}$ & 6.88$\times 10^{-5}$ & 4.89$\times 10^{-3}$ & 1.69$\times 10^{-3}$ & 5.03$\times 10^{-3}$ & 7.74$\times 10^{-3}$ & 8.92$\times 10^{-4}$ & 7.96$\times 10^{-4}$ \\
0.9 & 7.21$\times 10^{-4}$ & 2.51$\times 10^{-4}$ & 3.04$\times 10^{-4}$ & 4.24$\times 10^{-3}$ & 8.63$\times 10^{-4}$ & 3.84$\times 10^{-3}$ & 7.87$\times 10^{-3}$ & 3.21$\times 10^{-3}$ & 2.60$\times 10^{-3}$ \\
1.0 & 2.46$\times 10^{-3}$ & 1.52$\times 10^{-3}$ & 6.26$\times 10^{-4}$ & 1.13$\times 10^{-2}$ & 7.70$\times 10^{-3}$ & 1.29$\times 10^{-2}$ & 1.52$\times 10^{-3}$ & 1.60$\times 10^{-3}$ & 1.14$\times 10^{-3}$ \\
1.1 & 5.18$\times 10^{-2}$ & 1.16$\times 10^{-2}$ & 5.08$\times 10^{-3}$ & 6.91$\times 10^{-3}$ & 2.29$\times 10^{-3}$ & 3.92$\times 10^{-4}$ & 8.38$\times 10^{-2}$ & 1.97$\times 10^{-3}$ & 6.73$\times 10^{-2}$ \\
1.2 & 3.23$\times 10^{-3}$ & 9.75$\times 10^{-3}$ & 2.37$\times 10^{-3}$ & 5.25$\times 10^{-2}$ & 3.55$\times 10^{-3}$ & 9.09$\times 10^{-3}$ & 7.20$\times 10^{-2}$ & 1.79$\times 10^{-3}$ & 1.20$\times 10^{-1}$ \\
1.3 & 7.82$\times 10^{-3}$ & 5.05$\times 10^{-3}$ & 3.72$\times 10^{-3}$ & 1.72$\times 10^{-2}$ & 4.62$\times 10^{-3}$ & 1.12$\times 10^{-2}$ & 1.16$\times 10^{-2}$ & 2.37$\times 10^{-3}$ & 7.53$\times 10^{-2}$ \\
1.4 & 4.46$\times 10^{-3}$ & 3.78$\times 10^{-3}$ & 1.18$\times 10^{-3}$ & 5.20$\times 10^{-1}$ & 3.46$\times 10^{-1}$ & 2.52$\times 10^{-1}$ & 2.16$\times 10^{-1}$ & 9.62$\times 10^{-3}$ & 2.46$\times 10^{-1}$ \\
1.5 & 3.20$\times 10^{-4}$ & 3.52$\times 10^{-4}$ & 1.45$\times 10^{-4}$ & 1.06$\times 10^{-1}$ & 9.83$\times 10^{-2}$ & 6.72$\times 10^{-2}$ & 2.19$\times 10^{-2}$ & 6.77$\times 10^{-3}$ & 2.31$\times 10^{-2}$ \\
1.6 & 4.19$\times 10^{-4}$ & 4.89$\times 10^{-4}$ & 2.01$\times 10^{-4}$ & 6.21$\times 10^{-2}$ & 5.65$\times 10^{-2}$ & 5.87$\times 10^{-2}$ & 3.53$\times 10^{-2}$ & 3.34$\times 10^{-2}$ & 3.62$\times 10^{-2}$ \\
1.7 & 2.64$\times 10^{-3}$ & 2.49$\times 10^{-3}$ & 1.49$\times 10^{-3}$ & 1.23$\times 10^{-1}$ & 7.95$\times 10^{-2}$ & 4.01$\times 10^{-2}$ & 1.19$\times 10^{-3}$ & 2.28$\times 10^{-3}$ & 1.85$\times 10^{-3}$ \\
1.8 & 6.98$\times 10^{-4}$ & 1.70$\times 10^{-3}$ & 6.35$\times 10^{-4}$ & 2.75$\times 10^{-3}$ & 4.89$\times 10^{-3}$ & 3.07$\times 10^{-3}$ & 7.35$\times 10^{-4}$ & 8.57$\times 10^{-4}$ & 9.75$\times 10^{-4}$ \\
1.9 & 1.02$\times 10^{-3}$ & 1.00$\times 10^{-3}$ & 5.47$\times 10^{-4}$ & 1.47$\times 10^{-3}$ & 1.12$\times 10^{-3}$ & 1.67$\times 10^{-3}$ & 8.44$\times 10^{-4}$ & 9.53$\times 10^{-4}$ & 9.42$\times 10^{-4}$ \\
\hline
\label{tab:refuranus}
\end{tabular}
\end{table*}

\begin{table*}
\scriptsize
\caption{Estimated $n_{imag}$ spectra for Neptune aerosols. Again, for each aerosol type, `Mean 1' is the weighted average for all best-fitting retrieved spectra, while `Mean 2' are the averages of the contour maps shown in Fig. \ref{fig:nimag_combined}. `Sample' is the retrieved spectra from the representative best-fitting sample case shown for each planet.}
\centering
\begin{tabular} {l l l l l l l l l l}
\hline
 & Aerosol-1  & ($>$ 5--7 bar) & & Aerosol-2  & (1--2 bar) & & Aerosol-3  & (extended) & \\
$\lambda$($\mu$m) &  Mean 1 & Mean 2 & Sample &  Mean 1 & Mean 2 & Sample &  Mean 1 & Mean 2 & Sample \\
\hline

0.3 & 2.89$\times 10^{-4}$ & 1.87$\times 10^{-4}$ & 1.52$\times 10^{-4}$ & 3.13$\times 10^{-3}$ & 3.46$\times 10^{-3}$ & 3.40$\times 10^{-3}$ & 6.39$\times 10^{-3}$ & 4.78$\times 10^{-3}$ & 4.46$\times 10^{-3}$ \\
0.4 & 1.00$\times 10^{-5}$ & 1.57$\times 10^{-5}$ & 5.27$\times 10^{-6}$ & 1.13$\times 10^{-3}$ & 1.26$\times 10^{-3}$ & 1.27$\times 10^{-3}$ & 1.01$\times 10^{-2}$ & 8.23$\times 10^{-3}$ & 7.62$\times 10^{-3}$ \\
0.5 & 1.19$\times 10^{-5}$ & 6.76$\times 10^{-6}$ & 2.16$\times 10^{-6}$ & 1.61$\times 10^{-3}$ & 1.49$\times 10^{-3}$ & 1.79$\times 10^{-3}$ & 5.47$\times 10^{-3}$ & 4.66$\times 10^{-3}$ & 3.14$\times 10^{-3}$ \\
0.6 & 4.54$\times 10^{-4}$ & 1.91$\times 10^{-4}$ & 2.02$\times 10^{-4}$ & 1.76$\times 10^{-4}$ & 1.71$\times 10^{-4}$ & 9.71$\times 10^{-5}$ & 3.56$\times 10^{-3}$ & 2.19$\times 10^{-3}$ & 1.31$\times 10^{-3}$ \\
0.7 & 1.13$\times 10^{-3}$ & 9.28$\times 10^{-4}$ & 1.14$\times 10^{-3}$ & 2.22$\times 10^{-3}$ & 3.03$\times 10^{-4}$ & 2.24$\times 10^{-4}$ & 1.71$\times 10^{-2}$ & 1.46$\times 10^{-2}$ & 1.77$\times 10^{-2}$ \\
0.8 & 1.13$\times 10^{-3}$ & 1.51$\times 10^{-3}$ & 1.79$\times 10^{-3}$ & 1.48$\times 10^{-3}$ & 2.83$\times 10^{-4}$ & 2.19$\times 10^{-4}$ & 3.61$\times 10^{-2}$ & 3.53$\times 10^{-2}$ & 3.48$\times 10^{-2}$ \\
0.9 & 4.00$\times 10^{-4}$ & 3.62$\times 10^{-4}$ & 3.83$\times 10^{-4}$ & 3.52$\times 10^{-4}$ & 2.97$\times 10^{-4}$ & 3.09$\times 10^{-4}$ & 4.13$\times 10^{-2}$ & 3.21$\times 10^{-2}$ & 4.06$\times 10^{-2}$ \\
1.0 & 1.64$\times 10^{-3}$ & 1.46$\times 10^{-3}$ & 1.75$\times 10^{-3}$ & 6.36$\times 10^{-3}$ & 2.16$\times 10^{-3}$ & 6.65$\times 10^{-3}$ & 4.66$\times 10^{-3}$ & 4.80$\times 10^{-3}$ & 2.39$\times 10^{-3}$ \\
1.1 & 2.28$\times 10^{-3}$ & 2.43$\times 10^{-3}$ & 2.43$\times 10^{-3}$ & 8.36$\times 10^{-3}$ & 7.09$\times 10^{-3}$ & 1.07$\times 10^{-2}$ & 5.29$\times 10^{-3}$ & 1.62$\times 10^{-3}$ & 1.82$\times 10^{-3}$ \\
1.2 & 3.35$\times 10^{-3}$ & 2.57$\times 10^{-3}$ & 3.11$\times 10^{-3}$ & 3.73$\times 10^{-2}$ & 1.97$\times 10^{-2}$ & 2.68$\times 10^{-2}$ & 4.05$\times 10^{-2}$ & 1.44$\times 10^{-3}$ & 1.75$\times 10^{-3}$ \\
1.3 & 1.45$\times 10^{-2}$ & 4.40$\times 10^{-3}$ & 5.66$\times 10^{-3}$ & 3.28$\times 10^{-2}$ & 2.21$\times 10^{-2}$ & 3.55$\times 10^{-2}$ & 2.49$\times 10^{-2}$ & 2.34$\times 10^{-3}$ & 2.50$\times 10^{-3}$ \\
1.4 & 6.87$\times 10^{-3}$ & 2.69$\times 10^{-3}$ & 3.71$\times 10^{-3}$ & 5.31$\times 10^{-1}$ & 6.08$\times 10^{-2}$ & 3.50$\times 10^{-1}$ & 3.73$\times 10^{-1}$ & 2.45$\times 10^{-2}$ & 1.26$\times 10^{-2}$ \\
1.5 & 6.23$\times 10^{-4}$ & 4.53$\times 10^{-4}$ & 4.43$\times 10^{-4}$ & 5.33$\times 10^{-2}$ & 1.03$\times 10^{-2}$ & 3.79$\times 10^{-2}$ & 3.10$\times 10^{-1}$ & 3.16$\times 10^{-1}$ & 2.77$\times 10^{-1}$ \\
1.6 & 4.76$\times 10^{-4}$ & 3.41$\times 10^{-4}$ & 3.41$\times 10^{-4}$ & 3.85$\times 10^{-2}$ & 3.66$\times 10^{-2}$ & 4.54$\times 10^{-2}$ & 3.46$\times 10^{-1}$ & 4.04$\times 10^{-1}$ & 3.26$\times 10^{-1}$ \\
1.7 & 1.46$\times 10^{-2}$ & 1.34$\times 10^{-3}$ & 1.09$\times 10^{-3}$ & 9.55$\times 10^{-2}$ & 9.45$\times 10^{-3}$ & 5.42$\times 10^{-2}$ & 2.89$\times 10^{-1}$ & 9.08$\times 10^{-2}$ & 3.57$\times 10^{-1}$ \\
1.8 & 2.53$\times 10^{-3}$ & 1.05$\times 10^{-3}$ & 9.41$\times 10^{-4}$ & 2.69$\times 10^{-3}$ & 2.10$\times 10^{-3}$ & 2.70$\times 10^{-3}$ & 2.27$\times 10^{-1}$ & 6.91$\times 10^{-3}$ & 3.76$\times 10^{-2}$ \\
1.9 & 3.84$\times 10^{-3}$ & 9.95$\times 10^{-4}$ & 1.08$\times 10^{-3}$ & 1.61$\times 10^{-3}$ & 1.13$\times 10^{-3}$ & 1.32$\times 10^{-3}$ & 2.88$\times 10^{-1}$ & 2.07$\times 10^{-3}$ & 6.00$\times 10^{-2}$ \\
2.0 & 1.28$\times 10^{-3}$ & 9.79$\times 10^{-4}$ & 1.19$\times 10^{-3}$ & 1.16$\times 10^{-3}$ & 1.01$\times 10^{-3}$ & 9.86$\times 10^{-4}$ & 8.22$\times 10^{-1}$ & 1.57$\times 10^{-3}$ & 9.92$\times 10^{-1}$ \\
2.1 & 1.08$\times 10^{-3}$ & 9.89$\times 10^{-4}$ & 1.06$\times 10^{-3}$ & 1.07$\times 10^{-3}$ & 1.00$\times 10^{-3}$ & 8.06$\times 10^{-4}$ & 1.59$\times 10^{-2}$ & 1.69$\times 10^{-3}$ & 1.38$\times 10^{-2}$ \\
2.2 & 1.12$\times 10^{-3}$ & 9.82$\times 10^{-4}$ & 9.00$\times 10^{-4}$ & 1.03$\times 10^{-3}$ & 1.00$\times 10^{-3}$ & 8.60$\times 10^{-4}$ & 2.68$\times 10^{-3}$ & 1.18$\times 10^{-3}$ & 2.59$\times 10^{-3}$ \\
2.3 & 9.54$\times 10^{-4}$ & 9.76$\times 10^{-4}$ & 1.03$\times 10^{-3}$ & 9.92$\times 10^{-4}$ & 1.00$\times 10^{-3}$ & 7.84$\times 10^{-4}$ & 1.42$\times 10^{-3}$ & 1.02$\times 10^{-3}$ & 1.41$\times 10^{-3}$ \\
2.4 & 9.04$\times 10^{-4}$ & 1.02$\times 10^{-3}$ & 1.04$\times 10^{-3}$ & 1.01$\times 10^{-3}$ & 1.00$\times 10^{-3}$ & 7.71$\times 10^{-4}$ & 1.10$\times 10^{-3}$ & 1.00$\times 10^{-3}$ & 1.12$\times 10^{-3}$ \\
\hline
\label{tab:refneptune}
\end{tabular}
\end{table*}

It should be noted that the fitted $n_{imag}$ spectra can alias all sorts of other deficiencies in the forward model, such as inaccuracies in the methane absorption data, assumed/estimated particle size, or vertical cloud/haze parameterisation, and so not all features in these spectra may necessarily reflect what we would measure were we able to make \textit{in situ} observations of the particle scattering properties. However, the similarity between the retrieved Aerosol-2 and Aerosol-3 spectra and the generally higher values of $n_{imag}$ leads us to posit that the middle Aerosol-2 layer is not a pure condensed cloud, but instead is a region of photochemically-produced haze transported from the upper atmosphere and perhaps mixed with some methane ice. This is a point we shall return to later. For the lower Aerosol-1 layer, this has a lower $n_{imag}$ and we posit that this is actually the H$_2$S condensation layer (rather than an NH$_4$SH layer, suggested by \citeA{sromovsky19}), condensing onto dark haze particles and thus not conservatively scattering. We did attempt to model these data with a conservatively-scattering cloud/haze, but found that there was a detectable contribution from this component at longer visible wavelengths, which from the `snippet' analysis we think is unlikely. Instead, this Aerosol-1 layer is only visible from 0.4 to 0.7 $\mu$m due to: a) this layer being most reflective at these wavelengths; b) at longer (and shorter) wavelengths it is obscured by the thicker, and increasingly less scattering overlying 1--2-bar Aerosol-2 layer; and c) we have assumed a small particle size to make sure the extinction cross-section falls as $1/\lambda^4$. Of course it is entirely possible that the particles in the Aerosol-1 layer are actually larger than the value of 0.05 $\mu$m we have assumed here, in which case the retrieval model would likely have estimated larger $n_{imag}$ at longer wavelengths in order to reduce the reflectivity to a sufficiently small values. Typical sizes of droplets in clouds on Earth are many orders of magnitude larger than value of 0.05 $\mu$m assumed here, although such clouds are formed of different condensate at very different pressures and temperatures so may not be representative. However, it is worth emphasising that the actual mean radius of the Aerosol-1 particles is very uncertain and is not unambiguously constrained by these data. 

One final point to make here is that the computed opacity/km of the Aerosol-2 layer at 0.8 $\mu$m is rather low, peaking at about 1.0--10 km$^{-1}$. This underlines our suggestion that, assuming these are layers are homogeneous, they are extended photochemically-produced haze layers, rather than opaque cloud decks, since we would be able to see many hundreds of metres in all directions were we to be present in these layers.

\begin{figure*} 
\centering
\includegraphics[width=\textwidth]{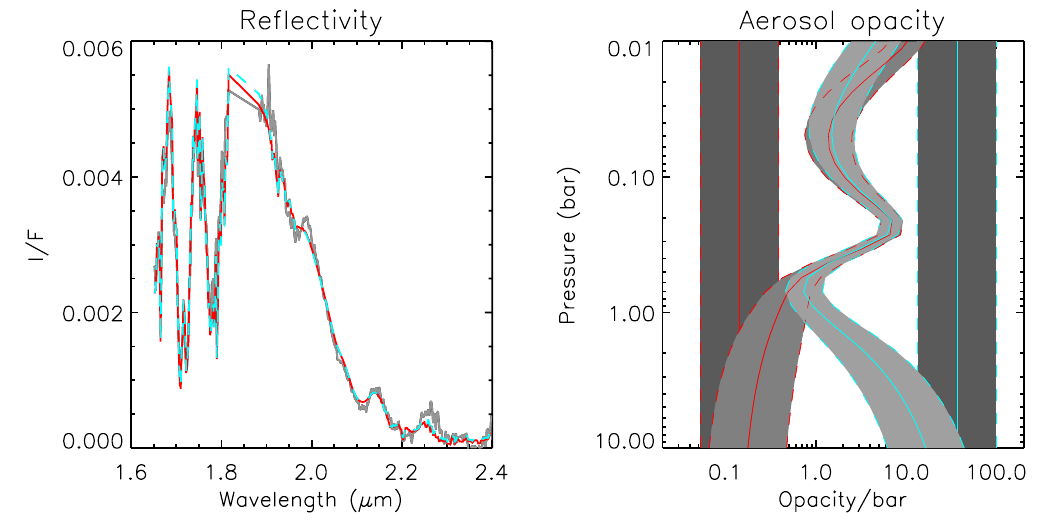}
\caption{Example `bracketed' fits to the longwave part of the IRTF/SpeX spectrum of Neptune using two different continuous \textit{a priori} distributions of aerosol particles of radius 0.1 $\mu$m and and fixed $n_{imag}=0.1$. The left panel compares the modelled spectra (red and cyan-dashed for two different priors) to that observed (grey), while the right panel shows the retrieved aerosol profiles for the two different \textit{a priori}, indicated in red and cyan respectively, showing that the spectrum is best fit in both cases with a layer that has peak opacity just below the tropopause near 0.2 bar. Note that in the right-hand panel the \textit{a priori} and retrieved error ranges are shaded in grey and edged by dashed coloured lines.} \label{fig:IRTFcontinuous}
\end{figure*}

\subsection{Analysis of IRTF/SpeX Neptune H+K band observations}
Having found a good fit to our Uranus observations, we then turned our attention to the Neptune observations. Before analysing the combined HST/STIS, IRTF/SpeX and Gemini/NIFS data sets we first of all sought to understand the gross differences between the Uranus and Neptune IRTF/SpeX observations, in particular the high reflectivities seen at methane absorption wavelengths longer than 1 $\mu$m for Neptune, but not seen for Uranus. 

The 1.65--2.4 $\mu$m region is particularly good for determining upper atmospheric aerosol density since the strong methane absorption band at 1.7 $\mu$m lies next to  strong H$_2$-H$_2$/H$_2$-He collision-induced absorption bands centred at 2.1 $\mu$m. The weighting functions of the methane-absorbing and hydrogen/helium-CIA-absorbing regions cover a similar range of pressure levels in the upper troposphere (0.6--0.1 bar) and thus this spectral region can be used to resolve the degeneracy between aerosol opacity and upper tropospheric methane abundance and retrieve reliable estimates of both \cite{roman18}.

Fig. \ref{fig:IRTFcontinuous} shows a `bracketed' retrieval to the IRTF/SpeX spectrum in the wavelength range 1.65--2.4 $\mu$m, assuming two widely-separated continuous distributions of small aerosols of mean radius 0.1 $\mu$m, variance $\sigma=0.3$, and fixed $n_{imag}=0.1$. As can be seen, a sharply peaked aerosol profile is favoured, peaking at $\sim$0.2 bar, with the local methane relative humidity retrieved to be $\sim$32\%. 
At first glance this fitted aerosol structure seems contrary the results of microphysical modelling \cite{toledo19,toledo20} where, since photochemically-produced haze particles are produced in the stratosphere and then progressively mixed to deeper and deeper pressures, we expect a distribution with decreasing opacity/bar with increasing pressure. To understand what was going on here, we needed to examine the reflectivities over a wider wavelength range, which, as we did for Uranus, we achieved with a Neptune `snippet' analysis, which we outline in the next section.

\subsection{Analysis of combined HST/STIS and IRTF/SpeX Neptune observations - snippet analysis}
\begin{figure*} 
\centering
\includegraphics[width=1.0\textwidth]{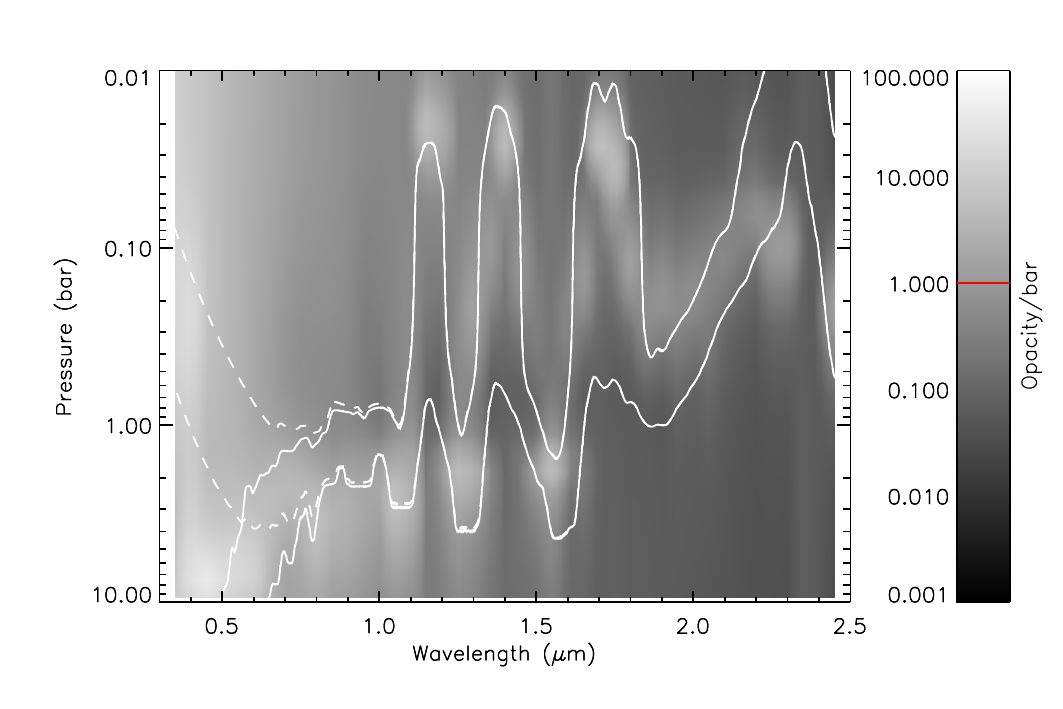}
\caption{As Fig. \ref{fig:snippet_uranus}, but for Neptune, showing a contour plot of vertical aerosol structure (opacity/bar) inferred from our `snippet' Neptune retrievals, where for each wavelength the aerosol structure is retrieved from the wavelengths in a bin of width 0.1 $\mu$m centred on that wavelength. The red line in the opacity/bar key is again the assumed \textit{a priori} value.} \label{fig:snippet_neptune}
\end{figure*}

As for our Uranus analysis, we split the HST/STIS and IRTF/SpeX observations into small, manageable `snippets' of wavelength width 0.1 $\mu$m, spaced every 0.05 $\mu$m and retrieved the vertical aerosol profile for a Gamma distribution of particles with mean radius 0.1 $\mu$m and variance $\sigma = 0.05$, varying also the imaginary refractive index $n_{imag}$ and keeping $n_{real}=1.4$. The resulting retrieved aerosol vertical/wavelength structure is shown in Fig. \ref{fig:snippet_neptune}. As can be seen, the aerosol structure shares many similarities with the equivalent Uranus results (Fig. \ref{fig:snippet_uranus}), with a deep Aerosol-1 layer seen at $p>$ $\sim$4 bar and a middle Aerosol-2 layer at $\sim$2 bar (slightly deeper than for Uranus). 
At short wavelengths it can also be seen that the opacity/bar increases with height in the upper troposphere/lower stratosphere, and decays rapidly with wavelength, consistent with our expectations with small, photochemically-produced haze particles. However, for Neptune there is clearly also a component of vertically confined particles centred at $\sim$0.2 bar, which must have a larger mean size in order that their opacity does not drop noticeably with wavelength. We believe here that we are detecting in the IRTF/SpeX data the signature of an additional component of larger-sized particles at $\sim$0.2 bar, which we surmise to be condensed methane ice. Such upper tropospheric methane condensation clouds are often seen in Neptune observations, especially in the regions at 20--40$^\circ$N and 20--40$^\circ$S and are seen to be highly temporally and spatially variable. We have tried to mask out such regions in the HST/STIS and Gemini/NIFS data. However, since we do not have an image of Neptune observed simultaneously with the IRTF/SpeX central-meridian line average, we do not know how significant this component of  upper tropospheric methane clouds is in the recorded IRTF/SpeX spectrum. Hence, we need to be careful when interpreting the IRTF/SpeX Neptune data, especially when used in combination with the other data sets. However, the good correspondence between the scaled IRTF/SpeX and Gemini/NIFS data sets, achieved, as described earlier, by optimising the cut-off level for bright clouds in the Gemini/NIFS data, reassures us that these data are reasonably self-consistent.

\begin{figure*} 
\centering
\includegraphics[width=1.0\textwidth]{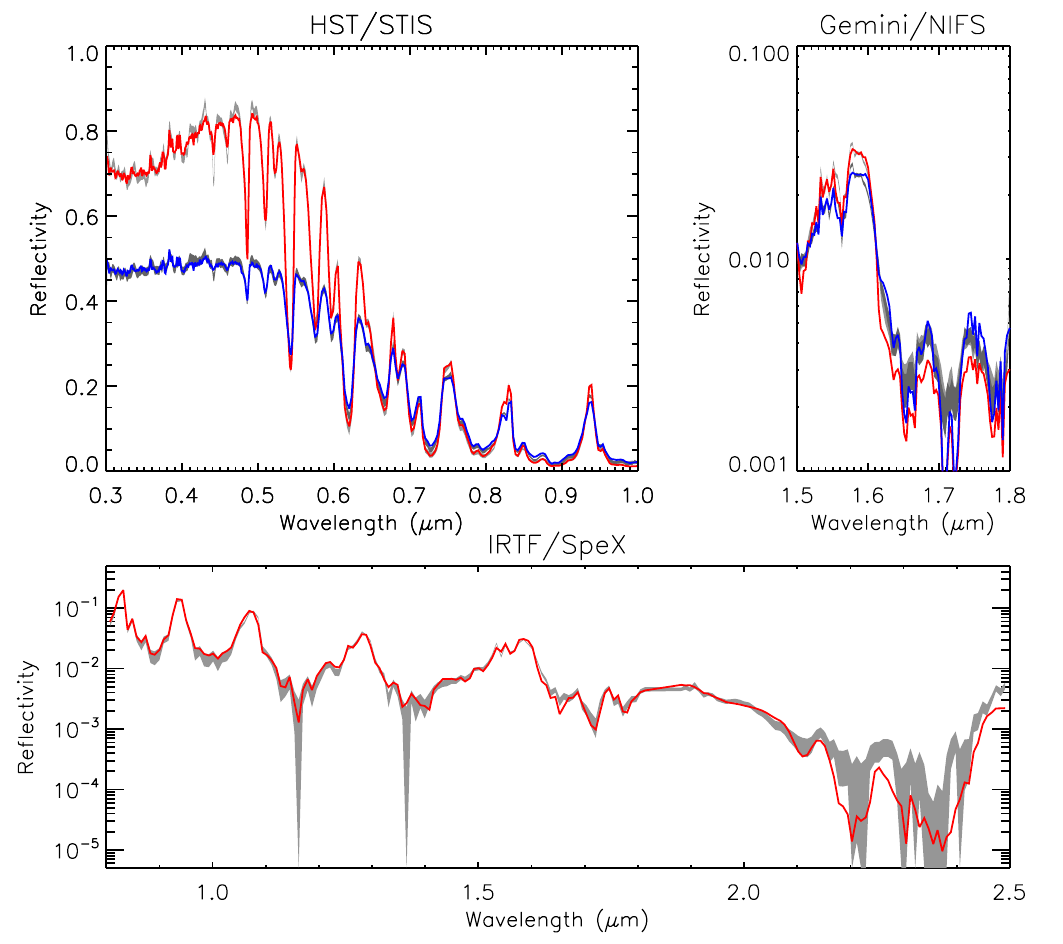}
\caption{Example of one of the best fitted spectra for Neptune, with $\chi^2/n =2.79$ for our assumed errors. Here, the measured spectra and assumed error range are shown in grey and the fits at 0$^\circ$ and 61.45$^\circ$ are shown in red and blue, respectively, for both the HST/STIS and Gemini/NIFS observations. For the IRTF/SpeX observation, the simulated central-meridian line average is shown in red. It can be seen that we achieve a very good fit to the data for all three data sets.} \label{fig:spectrum_neptune}
\end{figure*}

\begin{figure*} 
\centering
\includegraphics[width=1.0\textwidth]{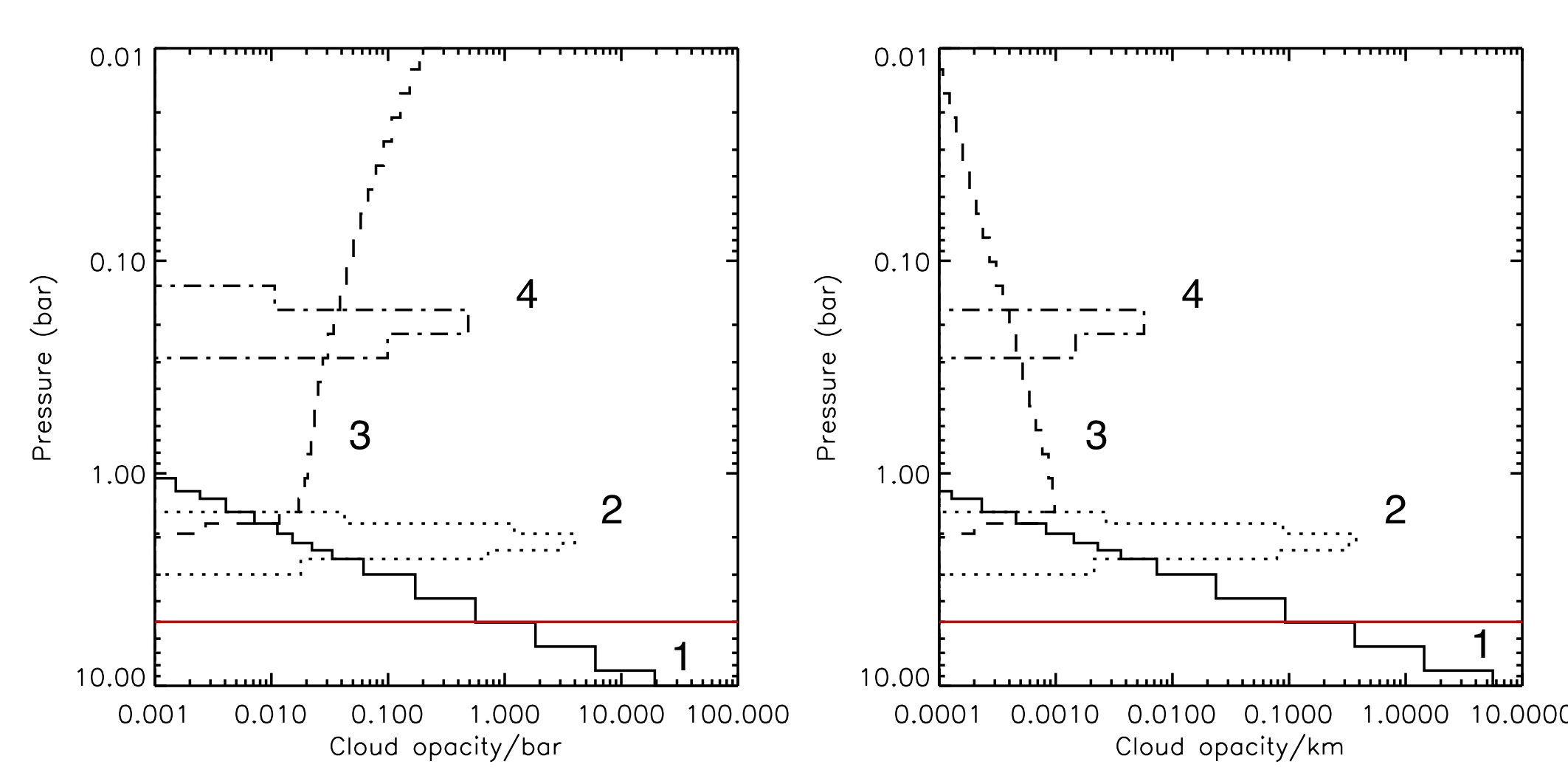}
\caption{Fitted vertical aerosol structure for Neptune's atmosphere from our example best-fit retrieval, presented in terms of opacity/bar (left) and opacity/km (right) at 0.8 $\mu$m. The deep Aerosol-1 layer is shown as a solid line, the middle Aerosol-2 layer as a dotted line, the tropospheric/stratospheric extended haze (Aerosol-3) as a dashed line, and the detached methane ice haze near the tropopause as the dashed-dotted line. The red line marks the 5-bar level, which is roughly the level where our sensitivity runs out.} \label{fig:cloud_neptune}
\end{figure*}

\subsection{Simultaneous retrieval of combined HST/STIS, IRTF/SpeX, and Gemini/NIFS Neptune observations}
\begin{figure*} 
\centering
\includegraphics[width=1.0\textwidth]{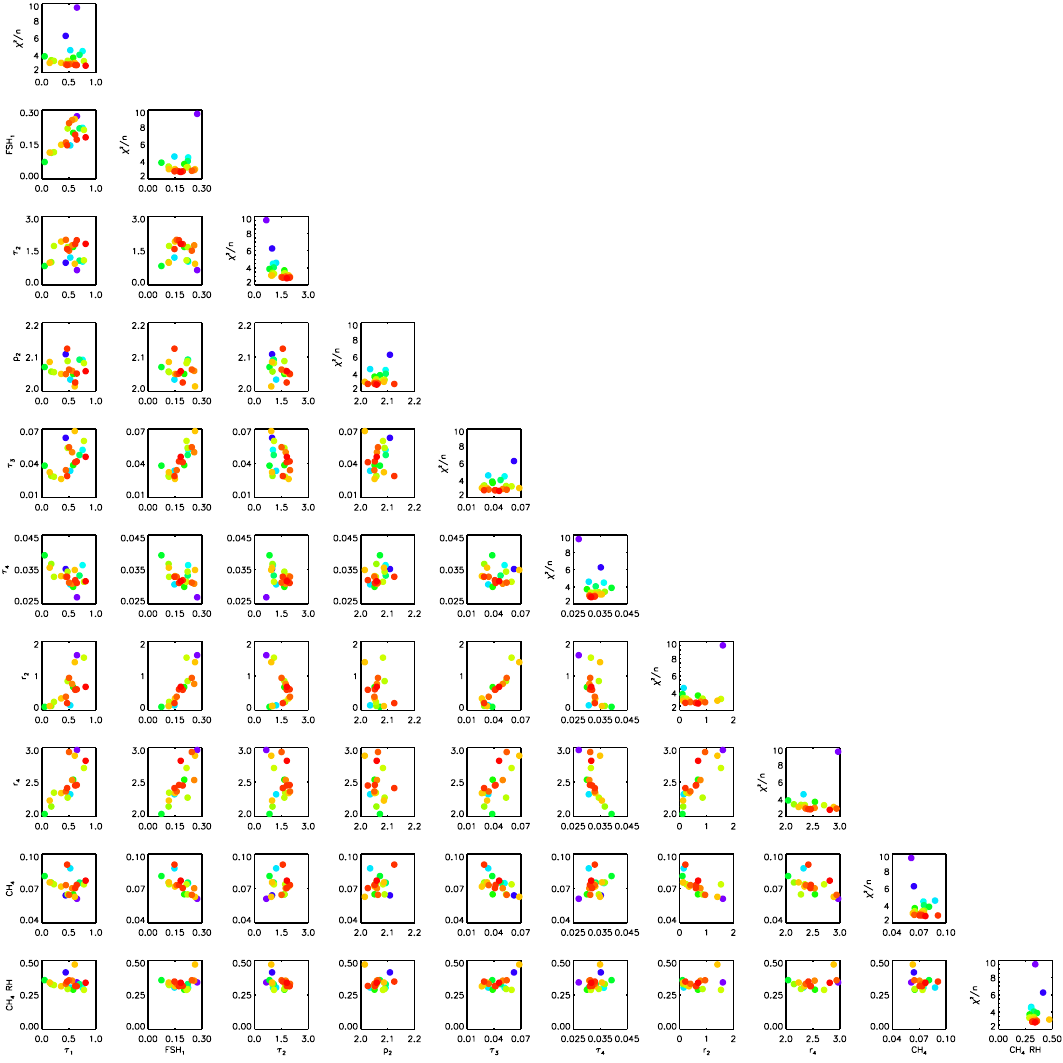}
\caption{`Corner' plot of 30 retrievals of the combined HST/STIS, IRTF/SpeX and Gemini/NIFS Neptune dataset with $\chi^2/n < 10$. The data points are colour-coded by the $\chi^2/n$ of the fit. Along the leading diagonal, instead of plotting the marginalised errors as would be usual for such plots, we plot $\chi^2/n$ as a function of fitted parameter. Although some parameters are better constrained than others, it can be seen that there is little cross-correlation between any of the fitted parameters.} \label{fig:corner_neptune}
\end{figure*}

Having found from the `snippet' analysis that our expected aerosol solution for Neptune was very similar to that for Uranus, we attempted to fit the combined STIS/SpeX/NIFS Neptune data set in the same way, except that we also added a thin layer of methane ice centred at 0.2 bar and with variable mean radius and fixed variance $\sigma = 0.3$.\footnote{ N.B., although we have modelled this layer as being homogeneous, it is also possible that this component is composed of sub-pixel-scale clouds.} 
We first fitted the 0.8--0.9 $\mu$m region separately with a range of randomly varied \textit{a priori} Aerosol-1 fractional scale heights and Aerosol-2 base pressures and radii and used those achieving $\chi^2/n < 1$ as the \textit{a priori} for the full retrievals, fixing the Aerosol-2 radii. Figure \ref{fig:spectrum_neptune} shows one of our best fits to the combined Neptune data set, where we can see that we again match the observed reflectivities and limb-darkening very well. The vertical aerosol structure retrieved for this case is shown in Fig. \ref{fig:cloud_neptune}, which is comprised of a deep Aerosol-1 layer at fixed base pressure 10 bar with fractional scale height 0.18 and integrated opacity to 5 bar (at 0.8 $\mu$m) 0.81 (coincidentally the same value as for our sampled Uranus retrieval shown earlier), a compact Aerosol-2 layer at 2.08 bar and opacity 1.8 composed of particles of mean radius 0.68 $\mu$m, an extended haze (Aerosol-3), based at 1.6 bar with fixed fractional scale height 2.0 and opacity 0.05, and a detached methane ice aerosol layer near the tropopause (0.2 bar), of opacity 0.03  comprised of particles of mean radius 2.8 $\mu$m. The methane deep mole fraction is 7.7\% and relative humidity at the tropopause is 34\% in this case. It should be stressed here that once again there is no single solution that fits better than all others; as we have performed multiple slightly perturbed retrievals there is a general class of solutions that fit best, of which this is one representative example. The complete set of retrieved parameters is shown in Fig. \ref{fig:corner_neptune} and the range and errors on our best fit solutions are listed in Table \ref{tab:retrieval_results}. It is worth noting, however, that the retrieved imaginary refractive index spectra of the main aerosol components, shown in Fig. \ref{fig:nimag_combined}, and listed in Table \ref{tab:refneptune} are similar to those retrieved for Uranus, suggesting a similar composition. It is also worth noting that the opacity of the middle Aerosol-2 layer at $\sim$ 2 bar is approximately half that determined for Uranus's atmosphere, while the opacity of the deeper Aerosol-1 layer is also greatly reduced. The lower Aerosol-2 opacity means that the Aerosol-1 layer has a greater contribution to the modelled spectra and so we have greater constraint on the Aerosol-1 imaginary refractive index spectrum as can be seen. For the methane ice layer at 0.2 bar, this is retrieved to be composed of moderately large particles ($r$ = 2--3 $\mu$m), which means they are most reflective at the longer wavelengths where the increased scattering at methane-absorbing wavelengths is seen. It also means that they are effectively conservatively forward-scattering at visible wavelengths, and thus they have very little effect here (as can be seen later in Fig.\ref{fig:colour}). Finally, since we discriminated against cloudy regions when we compiled our mean data sets, the opacity we retrieve for the Aerosol-4 layer will of course be significantly less than a true disc-average that includes such clouds. 

One curious feature of these retrievals is the significantly larger deep methane abundance we retrieve for Neptune of $7\pm1$\%, versus $3\pm1$\% for Uranus. An analysis of recent Neptune/MUSE observations \cite{irwin21} found values varying from $5\pm1$\% at the equator to $3\pm1$\% at the south pole, which compared reasonably well with previous analyses of these HST/STIS observations and others \cite<e.g.,>{kark09,kark11,sromovsky11,sromovsky14} who found values of $\sim$4\% at equatorial latitudes and $\sim$2\% at polar latitudes for both planets. In this study, we have analysed disc-averaged spectra so we might have expected to retrieve a methane abundance of $\sim$3--4\% for both Uranus and Neptune, and given that we would see more of the polar regions in the Neptune disc-averages than in the Uranus observations we might expect to retrieve lower methane abundances for Neptune than we do for Uranus.  In practice the determination of `deep' methane abundance (i.e., immediately below the methane condensation level) is almost inextricably tied up with the assumed cloud/haze parameterisation scheme, and also the assumed vertical distribution of methane. Our simple haze model, which is designed to match the observations at \textbf{all} wavelengths simultaneously, leads to fits that do not match the 800--860 nm range as well as other models that are able concentrate on this region alone \cite<e.g.,>{irwin21} and achieve closer fits, lower methane abundances, and slightly different aerosol profiles. Also, in this work we have assumed the same simple `step' model for the methane profile (constant mole fraction up to some fraction of the saturated vapour pressure curve set by the relative humidity) as \citeA{irwin21}, while other authors used a `descended' methane profile \cite<e.g.,>{sromovsky19}. Finally, although there is little cross-correlation seen in our `corner-plots' (Figs. \ref{fig:corner_uranus},\ref{fig:corner_neptune}) it is possible that the methane abundance is correlated with some of the cloud parameters, perhaps Aerosol-1, and hence for Neptune the methane abundance is being slightly overestimated and the Aerosol-1 parameters under- or over-estimated. Hence, in practice the methane retrieval problem is extremely degenerate and while here we find a difference in the retrieved deep methane abundances for Uranus and Neptune, there is not sufficient evidence to claim that this is a robust result. Perhaps with a revised haze parameterisation scheme, or different Aerosol-1 particle assumptions, we may determine values that are closer to each other. We hope to return to this question in future work.

\subsection{HST/WFC3  observations of dark spots in the atmospheres of Uranus and Neptune and their spectral characteristics.} \label{Sec:dark}

The analysis of available reflectance spectra of Uranus and Neptune reveals very similar aerosol structures for both planets, with a deep Aerosol-1 layer based at $p>$ 5--7 bar detectable at visible wavelengths, a middle Aerosol-2 layer at 1--2 bar seen at other wavelengths, a vertically extended stratospheric haze (Aerosol-3), detectable mostly at visible wavelengths, and finally, for Neptune, the presence of condensation layer of moderately large methane ice particles near the tropopause. Such a model can then be used simulate the general appearance at all wavelengths of either planet with good accuracy. However, there is one aspect in which Uranus and Neptune seem to differ substantially and that is the presence of `dark spots'

The first dark spots observed in an Ice Giant atmosphere were the Great Dark Spot (GDS) and Dark Spot 2 (DS2), seen by Voyager 2 in Neptune's atmosphere in 1989 \cite{smith89, sromovsky93}. The GDS was most visible at a wavelength of 0.48 $\mu$m, but was undetectable longward of 0.7 $\mu$m. Since the Voyager flyby several new dark spots have been seen in Neptune's atmosphere from Hubble Space Telescope observations in 1994, 1996, 2015 and 2018 \cite{hammel95,sromovsky01, wong18, hsu19}. These have all had similar spectral characteristics to those noted for the GDS, namely that they are visible near 0.5 $\mu$m, but not visible longward of 0.7 $\mu$m. For Uranus, there has to date only been one dark spot detected, which was observed in 2006 \cite{hammel09}. Unlike the Neptune dark spots the Uranus dark spot seems to have been detectable to longer wavelengths.

The dark spots seem to be some sort of vortex structure and the colouration must be due to either a darkening of the aerosol particles, or a change in opacity of the aerosol layer, or perhaps a combination of both effects. Putting aside the Uranus dark spot observation for one moment, the darkening effect needs to be confined to a rather small and particular range of wavelengths. It could very well be caused by a spectrally-limited darkening of the aerosol particles, but we would then need to have quite a particular perturbation of the cloud/haze scattering properties and simultaneously confine that change somehow to be within a vortex. On the other hand, if dark spots are caused by cloud/haze opacity changes, we need to explain how its signature is so spectrally confined. We believe that the retrievals described in previous sections provide the answer. The deep Aerosol-1 layer based at $p>$ 5--7 bar is only visible at wavelengths less than $\sim$0.7 $\mu$m and if that were to have lower opacity, or reduced reflectivity, then we might expect to see a dark spot at just those wavelengths.

To see if this phenomenon was apparent in our modelled Uranus observations we generated synthetic images of Uranus from our best-fitting cloud/haze model in the seven wavelength channels observed by HST/WFC3 with the OPAL program in 2014  \cite{simon15}. The location of the filters used (F467M, F547M, FQ619N, F658N, FQ727N, F845M, and FQ924N) are compared with the central-meridian line-averaged HST/STIS Uranus spectrum in Fig. \ref{fig:uranus_neptune_filter} and in Fig. \ref{fig:reconstruct_uranus} we compare the observations with synthetic images. The middle and bottom rows of Fig. \ref{fig:reconstruct_uranus} show synthetic images at each filter wavelength, which were constructed using the following procedure:

\begin{figure*}
\centering
\includegraphics[width=1.0\textwidth]{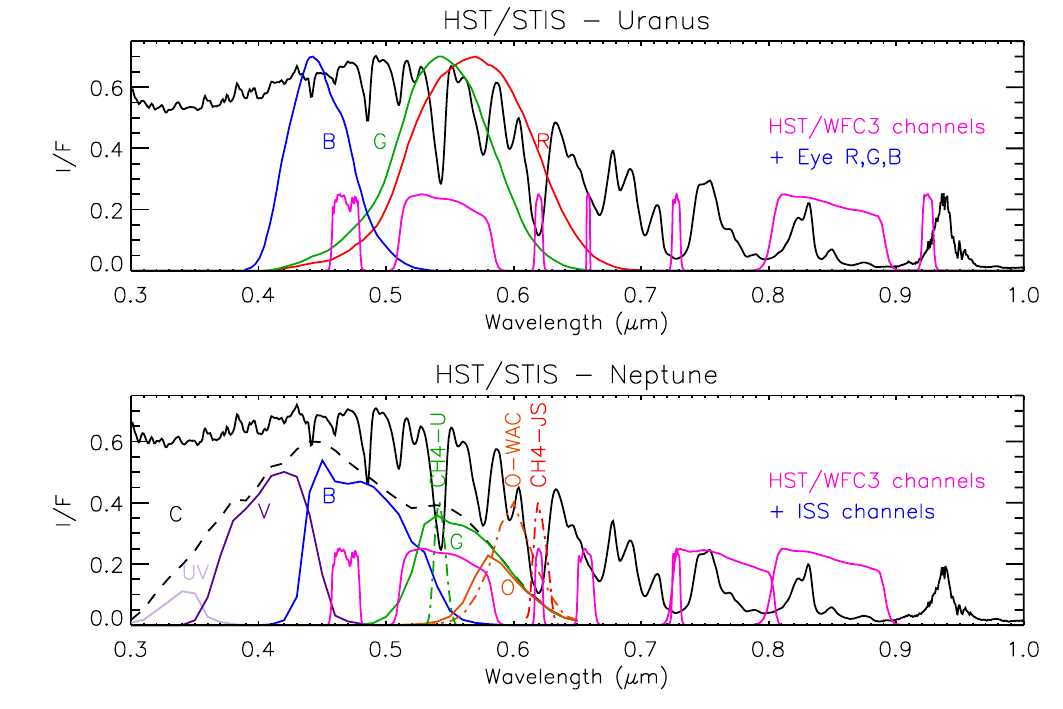}
\caption{HST/STIS central-meridian-averaged I/F spectra of Uranus (top panel) and Neptune (bottom panel). Overplotted in the top panel for Uranus are the red, green, blue sensitivities of the human eye \cite{stockmansharp00,stockman19}, together with the HST/WFC3 filters: F467M, F547M, FQ619N, F658N, FQ727N, F845M, and FQ924N \cite{dressel21}.  Overplotted in the bottom panel for Neptune are the HST/WFC3 filters: F467M, F547M, FQ619N, F657N, FQ727N, F763N and F845M, together with the `clear', `UV', `violet', `blue', `green' and `orange' Voyager-2/ISS NAC sensitivities \cite{smith77}, and the `CH4-U' (i.e., 547 nm), `CH4-JS' (i.e., 619 nm) and `orange' Voyager-2/ISS WAC sensitivities (dashed lines).}
\label{fig:uranus_neptune_filter}
\end{figure*}

\begin{figure*} 
\centering
\includegraphics[width=1.0\textwidth]{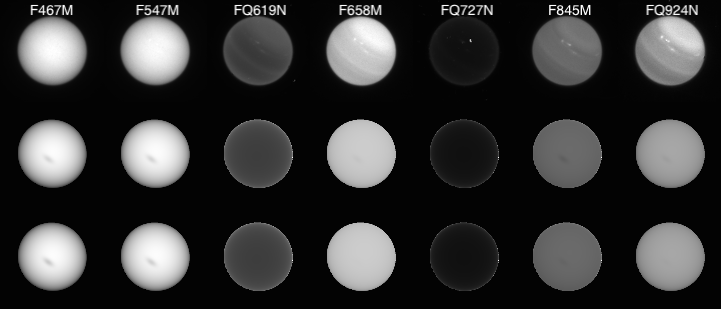}
\caption{Observed and reconstructed HST/WFC3 images of Uranus. Top row shows HST/WFC3 observations made in 2014 during the OPAL program, centred at the wavelengths: 467, 547, 619, 658, 727, 845, and 924 nm. Middle row shows images reconstructed from our fits to the HST/STIS data, which also includes a hole in the deep Aerosol-1 layer ($p>5-7$ bar) near the disc centre. Bottom row shows images reconstructed from our fits to the HST/STIS data, where the deep Aerosol-1 layer is darkened near the disc centre by setting $n_{imag}=0.001$ at all wavelengths. It can be seen that for the darkening case the modelled dark spot is visible at 467 and 547 nm, but not at longer wavelengths, whereas for the clearing simulation the modelled spot is visible at 467 and 547 nm, but also faintly at 658, 845 and 924 nm. In addition, the modelled spot is darker at 467 nm when the Aerosol-1 layer is darkened rather than removed. } \label{fig:reconstruct_uranus}
\end{figure*}

\begin{enumerate}
    \item In our baseline simulations we used our optimal fit to the reconstructed spectra at $0^\circ$ and $61.45^\circ$ to extract Minnaert parameters $I_0(\lambda)$ and $k(\lambda)$ and, using $I=I_0 \mu_0^k \mu^{k-1}$, constructed synthetic disc images of Uranus at each modelled HST/STIS wavelength, $I_{reference}(\lambda)$ for the HST/WFC3 observation zenith angles $\mu$ and $\mu_0$.
    \item Then, from our fitted cloud/haze model, we calculated modelled spectra (at 0$^\circ$ and 61.45$^\circ$) with either: A) the opacity of the Aerosol-1 layer based at $p>$ 5--7 bar set to zero; or B) the imaginary refractive indices of the particles in this layer set to 0.001 at all wavelengths (which darkens the particles at short wavelengths). We then extracted modified $R_0(\lambda)$ and $k(\lambda)$ Minnaert parameters for both cases, which we used to generate two additional sets of synthetic Uranus images, $I_{modified}(\lambda)$.
    \item We then constructed a weighting factor, $f_w$, to simulate a cloud/haze darkening/clearing near the disc centre, with latitude $\Phi_0=20^\circ$N and central meridian longitude $\Lambda_0=0^\circ$, setting $f_w = \exp(((\Phi-\Phi_0)/\Delta\Phi)^2+((\Lambda-\Lambda_0)/\Delta\Lambda)^2)$, where $\Phi$ is the latitude on the disc, $\Lambda$ is the longitude relative to the central meridian, and where  $\Delta\Phi=5^\circ$ and $\Delta\Lambda=10^\circ$.
    \item We then made weighted averages of our two sets of images at each HST/STIS wavelength: $I_{mean}(\lambda) = (1-f_w)I_{reference}(\lambda) + f_w I_{modified}(\lambda)$. 
    \item Finally, the combined images were convolved with the HST/WFC3 filter functions to create the synthetic images.
\end{enumerate}

From Fig. \ref{fig:reconstruct_uranus} it can be seen that aside from the latitudinal cloud/haze and methane abundance variations visible in the observed images (which we did not attempt to simulate here) we capture reasonably well the observed limb-darkening/limb-brightening at all wavelengths. Although the HST/WFC3 OPAL 2014 observations did not contain any dark spots, if there were a hole in the deep Aerosol-1 layer in Uranus's atmosphere, or a darkening, it can be seen in the middle and bottom rows of Fig. \ref{fig:reconstruct_uranus} that it would produce a dark spot at \textit{precisely} the same wavelengths as seen by HST/WFC3, i.e., at 467 and 547 nm. However, while for the Aerosol-1-darkening case the spot is invisible at longer wavelengths, for the clearing case it is still just visible at 658, 845 and 924 nm. In addition, the darkening simulation results in a spot that is darker at 467 nm than in the clearing simulation, which is again more consistent with the observations. 

\begin{figure*} 
\centering
\includegraphics[width=1.0\textwidth]{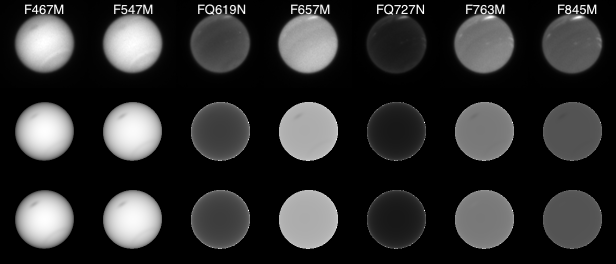}
\caption{Observed and reconstructed HST/WFC3 images of Neptune. Top row shows HST/WFC3 observations made in 2018 during the OPAL program, centred at the wavelengths: 467, 547, 619, 657, 727, 763, and 845 nm. Middle row shows images reconstructed from our fits to the HST/STIS data, which also includes a hole in the deep Aerosol-1 layer ($p>5-7$ bar) near the central meridian at 15$^\circ$N, and a clearing at $60^\circ$S. Bottom row shows images reconstructed from our fits to the HST/STIS data, where the Aerosol-1 layer is darkened near the central meridian at 15$^\circ$N and at all longitudes at $60^\circ$S by setting $n_{imag}=0.001$ at all wavelengths. As for the Uranus case shown in Fig. \ref{fig:reconstruct_uranus}, we can see that a dark spot is visible at 467 and 547 nm and that again, while for the Aerosol-1-darkening case the spot is invisible at longer wavelengths, for the clearing case it is still just visible at 657, 763 and 845 nm. Also, as for Uranus, the darkening simulation results in a spot that is darker at 467 nm than in the clearing simulation, which is more consistent with observations.} \label{fig:reconstruct_neptune}
\end{figure*}

Turning to Neptune, we generated synthetic images from our best-fitting model in the seven wavelength channels observed by HST/WFC3 with the OPAL program in 2018 \cite{simon19}, which reported the detection of a new dark spot, NDS-2018. The channel filter functions used (F467M, F547M, FQ619N, F657M, FQ727N, F763M and F845M) are shown in Fig. \ref{fig:uranus_neptune_filter} and the observed images can be seen in the top row of Fig. \ref{fig:reconstruct_neptune}. In the middle and  bottom rows, we present our simulations with a dark spot centred at $\Phi_0=15^\circ$N, $\Lambda_0=0^\circ$, and also with a clearing/darkening at latitude $\Phi_0=60^\circ$S.  Here, we can again see that we predict the appearance of the clearing of the deep Aerosol-1 layer to be most clearly detectable at the two shortest wavelengths, very similar to the observed spot. However, for the clearing simulation, the modelled spot is still just visible at 657, 763 and 845 nm, which is contrary to the observations and also contrary to the darkening simulation, where the modelled spot is completely invisible at the longer wavelengths. In addition, the spot modelled with the darkening hypothesis is darker at 467 nm than the clearing case, which is again more consistent with the observed properties.

Considering Uranus and Neptune together it can be seen that although a clearing of the Aerosol-1 layer based at $p>$ 5--7 bar  produces \textit{almost} the right response, it does not predict the spot to be darker at 467 nm than 547 nm, and the simulated spot can still just be seen at longer continuum wavelengths, contrary to the observations. In contrast, our simulated images where we darken the Aerosol-1 layer produces a feature that much more closely resembles the real NDS-2018 on Neptune. We find that removing the Aerosol-1 layer does not darken the spot sufficiently at short wavelengths since there is still significant Rayleigh scattering from the air itself at these depths. However, by darkening this aerosol layer, the contrast of the dark spot at short wavelengths is greatly increased. Add to that the disappearance of the feature at longer wavelengths by darkening the Aerosol-1 particles, rather than removing  them altogether, and on balance we conclude that darkening the particles provides a better match to the observations. Further evidence in support of this hypothesis comes from an analysis of earlier HST/WFC3 Neptune images \cite{karkoschka11_dark}, where the author suggested that the dark belt at 30--$60^\circ$S was best explained by a darkening of the haze at pressures $>$ 3 bar. \citeA{karkoschka11_dark} went on to speculate that since dark spots have very similar spectral characteristics to the 30--$60^\circ$S belt, they might be explained by vertical motions bringing dark material from depth up to altitudes where it can be observed. Finally, we also see that our predicted Neptune dark spot is slightly darker than the equivalent spot modelled for Uranus, which can be explained by the fact that the opacity of the overlying $\sim$2-bar Aerosol-2 layer on Neptune is lower than it is on Uranus. The higher opacity of the UV-absorbing Aerosol-2 layer on Uranus also explains why the UV reflectivity of Uranus is lower than Neptune. 

\begin{figure*} 
\centering
\includegraphics[width=1.0\textwidth]{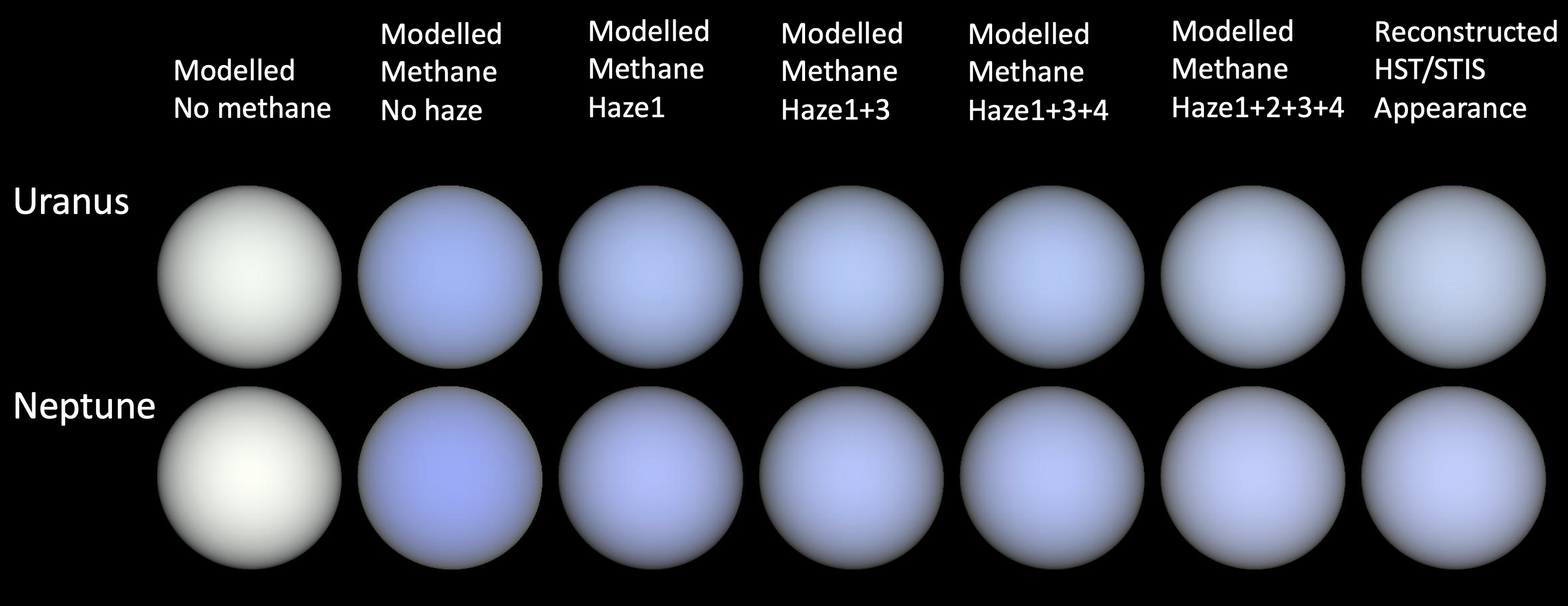}
\caption{Modelled colours of Uranus and Neptune. The right-hand column shows the simulated `observed' appearance of Uranus and Neptune, reconstructed using the fitted Minnaert limb-darkening coefficients extracted from the HST/STIS data and assuming the observing geometry of the 2014 Uranus HST/WFC3 observations (Fig. \ref{fig:reconstruct_uranus}) for both planets. The spectra across the discs were convolved with the Commission Internationale de l'\'Eclairage (CIE)-standard red, green, and blue human cone spectral sensitivities \cite{stockmansharp00,stockman19} to yield these apparent colours: it can be seen that Uranus is predicted to have a paler, more greenish colour than Neptune. The preceding columns show how our modelled appearance of these planets changes as we add different components to our radiative transfer model. Column 1 shows the modelled appearance of both planets if we remove all the haze and neglect methane absorption: in this case both planets would appear pearly-white since even though Rayleigh-scattering is more effective at blue wavelengths, we will eventually reach depths at even red wavelengths where Rayleigh scattering becomes effective. The effect of adding in absorption from the best-case retrieved methane profile for both planets is shown in Column 2, and underlines the fact that it is the presence of atmospheric methane that leads to the underlying blue colour of these planets. However, the difference in observed colour between Uranus and Neptune cannot be explained by just by the fact that we retrieve more methane in Neptune's atmosphere than Uranus's. Column 3 shows the effect of adding in the retrieved profiles for Aerosol-1, while Column 4 shows the effect of further adding Aerosol-3, which can be seen to have little effect. In Column 5 we also add in Aerosol-4. This is set to zero opacity for the Uranus retrievals and so there is no change, while the retrieved opacity for Neptune is very small, which when combined with the forward-scattering nature of the large (2--3 $\mu$m-sized) particles leads to negligible difference for Neptune also. Finally, in Column 6 we add in the opacity of Aerosol-2 to all the other components, which can be seen to lead to the greatest difference between the predicted colours of Uranus and Neptune, and also leads to final colours that are indistinguishable from those reconstructed from the initial HST/STIS limb-darkening coefficients in Column 7 (`Reconstructed HST/STIS Appearance'). (N.B., the overall brightness in each column as been separately scaled to make each column uniformly bright.)} \label{fig:colour}
\end{figure*}

Since the scattering cross-section of the approximately micron-sized Aerosol-2 particles is found to have a roughly white visible reflectivity spectrum, the higher opacity of the Aerosol-2 layer on Uranus also mostly explains why Uranus appears to have a paler blue colour to the human eye (to some more greenish) than Neptune, which we show in Fig. \ref{fig:colour}. In this figure we show the simulated colours of the background, masked regions of Uranus and Neptune reconstructed from the Minnaert limb-darkening approximations to the HST/STIS observations using Eq. \ref{eq:minnaert}. Here, as a representative case we assumed the solar and emission zenith angles varied across the disc of both Uranus and Neptune as observed for the Uranus 2014 HST/WFC3 observations shown in Fig. \ref{fig:reconstruct_uranus}. In these simulations the modelled radiance spectra across the planet were convolved with the CIE-standard (Commission Internationale de l'\'Eclairage) red, green, and blue human cone spectral sensitivities of \citeA{stockmansharp00,stockman19} and are, as near as we can determine the `natural' colours that the background atmospheres of these planets would have if we were to observe them from space. As can be seen Uranus appears paler and less blue than Neptune, although the colour differences are actually rather subtle. In Fig. \ref{fig:colour} we can see that if we modelled the planets' appearances for aerosol-free conditions and with no methane absorption then both planets would appear   pearly-white since even though Rayleigh-scattering is more effective at blue wavelengths, we eventually reach depths at even red wavelengths where Rayleigh scattering becomes effective. It is only when we add absorption by the red-absorbing methane that the characteristic blue colour of these planets develop. Then as we add the scattering effects of Aerosol-1, Aerosol-3 and Aerosol-4 components the modelled appearances become paler, but it is not until we add in scattering from Aerosol-2 that the main colour difference manifests itself. Why the Aerosol-2 layer on Uranus is thicker than that of Neptune is not clear. Potentially, we suggest that perhaps the more dynamically overturning atmosphere of Neptune is more efficient at clearing this haze layer through methane condensation at its base, as discussed later. Also, we note that during the Voyager-2 encounters with these planets, Uranus was close to southern summer solstice when the south polar region was covered in a `hood' of haze, which we presume to be Aerosol-2. Hence, the planet would likely have looked even paler than we have simulated here from HST/STIS observations in 2002 and 2003.

In summary we find that the aerosol structure that we have retrieved from the combined HST/STIS, IRTF/SpeX and Gemini/NIFS data for Uranus and Neptune are also consistent with HST/WFC3 observations and can also explain the observe colour difference between these planets. We also find that a darkening (or possibly opacity change) of the deepest aerosol layer (Aerosol-1) leads to a darkening that is consistent with the observed spectral characteristics of dark spots. Hence, we propose that the dark spots sometimes seen in Neptune's atmosphere (and occasionally Uranus's) are caused by a darkening of the deep Aerosol-1 layer. If dark spots in Uranus's atmosphere really are visible over a wider wavelength range, as reported by \citeA{hammel09}, then this suggests that Uranus's deep Aerosol-1 layer, presumably containing a significant fraction of H$_2$S ice, is brighter over a wider range of wavelengths, since retrieval tests conducted with a conservatively-scattering lower cloud/haze produced a detectable dark spot signature at all the HST/WFC3 wavelengths when the opacity of this layer was reduced, except in the strong methane bands at 619 nm and 727 nm.

\subsection{Voyager-2/ISS  observations of dark spots in Neptune's atmosphere.}

\begin{figure*} 
\centering
\includegraphics[width=0.9\textwidth]{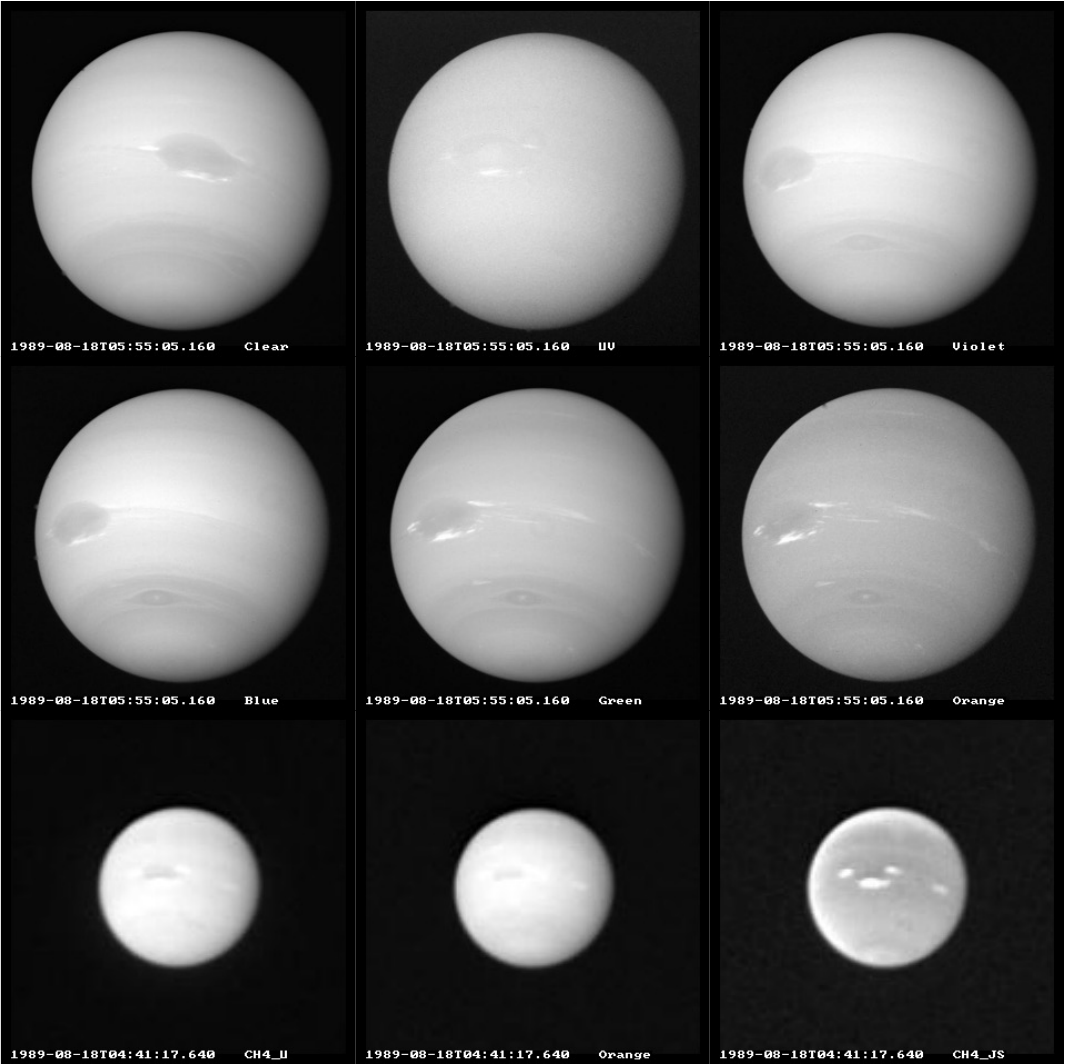}
\caption{Representative Voyager-2 ISS images of Neptune, observed in August 1989 in the Clear, UV, Violet, Blue, Green and Orange filters, respectively of the Narrow Angle Camera (NAC) and also CH4-U (541 nm), Orange and CH4-JS (619 nm) from the Wide Angle Camera (WAC). For the NAC images, the Great Dark Spot (GDS) and Dark Spot 2 (DS2) are clearly visible, except in the UV channel. Also visible (except again in the UV) is the generally darker belt at 45--55$^\circ$S. The WAC images are of poorer quality, but cover complementary wavelengths and the GDS and dark belt are still visible in the CH4-U and orange filters. For the CH4-JS filter, centred at 619 nm, the dark belt is less clear, and the white clouds around the GDS are relatively much brighter indicating that these are at a higher altitude than the main 1--2-bar Aerosol-2 layer.} \label{fig:voyager2_neptune}
\end{figure*}

Having found a good correspondence with the spectral properties of dark spots seen in Neptune's atmosphere with HST/WFC3 by darkening the Aerosol-1 layer, we wondered whether we might also be able to find a good correspondence with the original Voyager 2 ISS observations, which first detected such phenomena \cite{smith89}. During Voyager-2's approach to the Neptune system in August 1989, the Imaging Science Subsystem (ISS) took a particular sequence of observations, ``Full color set for atmospheric dynamics", consisting of 319 full-disc observations from 16th to 18th August in each of the main filters of its Narrow-Angle Camera: UV, Violet, Blue, Green, Orange and also with no spectral filter (Clear). Observations were also made in three filters of its Wide-Angle Camera: `CH4-U' (541 nm), `CH4-JS' (619 nm) and another Orange filter. The transmission functions of all these filters are shown in Fig. \ref{fig:uranus_neptune_filter}. In these images the Great Dark Spot (GDS) and also Dark Spot 2 (DS2) were observed at several phase angles. We examined the limb-darkening properties of the GDS and also the $50^\circ$S dark belt in these observations to see if they supported or contradicted our hypothesis that dark regions are caused by a clearing, or darkening, of the deep H$_2$S/haze Aerosol-1 layer. Representative Voyager-2 observations from this set are shown in Fig. \ref{fig:voyager2_neptune}. It can be seen that the GDS is particularly prominent in the Blue filter channel, but becomes less so as we move to both shorter and longer wavelengths. The bright companion clouds become more prominent at longer wavelengths, as the Rayleigh scattering becomes less significant, indicating these features to lie at higher altitudes, presumably in the upper troposphere. The contrast between the dark spots and the surrounding regions looks similar to that between the darker belt centred at 50$^\circ$S and surrounding latitudes and we assume, like \citeA{karkoschka11_dark}, that both are formed a similar way. Looking at the  50$^\circ$S dark belt it can be seen that this region is most visible near the central meridian and gets less clear towards the limbs. 


\begin{figure*} 
\centering
\includegraphics[width=0.9\textwidth]{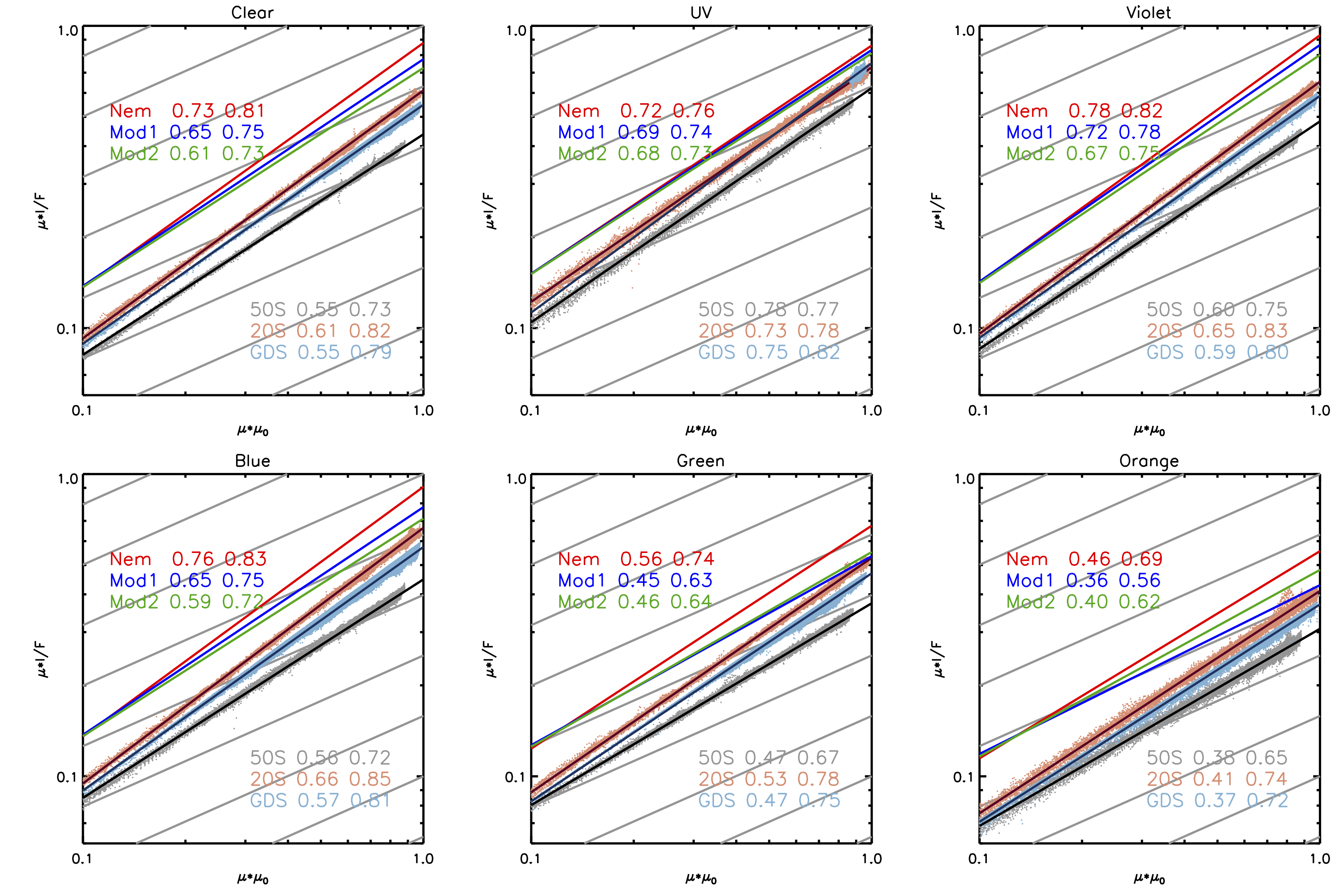}
\caption{Results of Minnaert analysis of the Voyager-2 ISS NAC observations of Neptune. In each panel (filter), we have first plotted on a log-log plot the observed reflectances multiplied by $\mu$ (the cosine of zenith angle) against $\mu \mu_0$ for the 15--25$^\circ$S and 45--55$^\circ$S bands, together with the fitted Minnaert lines and derived $(I/F)_0$ and $k$ values, showing the clear linear dependence expected from the Minnaert model, and less pronounced limb-darkening at 45--55$^\circ$S (smaller values of $k$) compared with 15--25$^\circ$S. Note the 45--55$^\circ$S reflectivities have been scaled by a factor of 0.8 for clarity. Also plotted are the observed limb-darkening dependencies of pixels within the GDS (and associated Minnaert fits), which have a smaller limb-darkening coefficient than the background at $20^\circ$S, but less so than at 45--55$^\circ$S. Finally, on each plot is shown the simulated limb-darkening behaviour of our best-fitting Neptune NEMESIS model (coloured lines, multiplied by 1.2 for clarity) including the deep $>$ 5--7-bar Aerosol-1 layer (`Nem'), removing the Aerosol-1 layer (`Mod1'), or darkening it (`Mod2'), which show similar differences indicating that dark regions are well fitted by our model with lower $>$ 5--7-bar aerosol layer reflectivity. The parallel grey lines on this plot are for reference and mark simulated $k=0.5$ lines, for which the disc has no limb-darkening or limb-brightening.} \label{fig:voyager2_neptune_minnaert}
\end{figure*}

To interpret these observations more quantitatively, we performed a limb-darkening analysis of the Voyager 2 ISS images, which we summarise in Fig. \ref{fig:voyager2_neptune_minnaert}. In this figure we have analysed the observations for each of the six NAC flters studied, first of all concentrating on images where the GDS and DS2 were not visible and analysing the limb-darkening of the background atmosphere in two latitude bands: 15 -- 25$^\circ$S (centred on the GDS) and 45 -- 55$^\circ$S (centred on the dark belt). Assuming the Minnaert approximation to hold, Eq. \ref{eq:minnaert} can be re-expressed as
\begin{linenomath*}
\begin{equation}
\log(\mu R) = \log(R_{0}) + k \log(\mu \mu_{0}),
\end{equation}
\end{linenomath*}
where $R$ is the measured reflectance (i.e., $I/F$), $\mu$ and $\mu_0$ are cosines of the viewing and solar zenith angles respectively, and $R_0$ and $k$ are the fitted Minnaert parameters. Hence, plotting $\mu R$ against $\mu \mu_0$ on a log-log plot should yield a straight line if the Minnaert approximation is good and in Fig. \ref{fig:voyager2_neptune_minnaert} we have plotted such curves for the two latitude belts for all filters. We find that the observed limb darkening is well described by the Minnaert model and also that the 45 -- 55$^\circ$S belt has a lower $(I/F)_0$ and is noticeably less limb-darkened than the 15 -- 25$^\circ$S belt, having a significantly shallower slope (i.e., lower $k$, but still greater than 0.5, which is the boundary between limb-darkening and limb-brightening). 

We then analysed individual observations in the data set containing the GDS at different zenith angles, selecting pixels within the GDS and over-plotting their reflectance in the same way in Fig. \ref{fig:voyager2_neptune_minnaert}. There was good sampling for the NAC images, but far fewer WAC images, which were also less spatially resolved making the limb-darkening less clear\footnote{In addition, the CH4-JS channel seems poorly flat-fielded and photometrically corrected}. Hence, the WAC images were discounted in the limb-darkening analysis. It can be seen in Fig. \ref{fig:voyager2_neptune_minnaert} that pixels within the GDS have a similar $(I/F)_0$ to those in the 50$^\circ$S belt and the limb-darkening is again less than the background 20$^\circ$S region, but less so than for 50$^\circ$S. However, the fact that the GDS and 50$^\circ$S belt have both lower $(I/F)_0$ \textbf{and} lower $k$ than the background at  20$^\circ$S suggests these dark features really do have a common origin. To determine if this common origin is due to lower reflectance from the deep Aerosol-1 layer we have also plotted in Fig. \ref{fig:voyager2_neptune_minnaert} the simulated limb-darkening from our best-fitting cloud/haze retrieval to the combined STIS/SpeX/NIFS Neptune data set, including the contribution from this layer, or modifying it by clearing or darkening. It can clearly be seen that reducing the contribution from the Aerosol-1 layer lowers both $(I/F)_0$ \textbf{and} the limb-darkening parameter $k$, just as is seen for the Voyager-2 dark regions, suggesting that this may indeed be the cause of the darkening. On balance, the difference in limb-darkening is again slightly better simulated by darkening the Aerosol-1 layer, rather than removing it.

In summary, the Voyager-2 ISS observations of Neptune support our suggestion that dark regions are caused by a darkening (or possibly clearing) of the deep Aerosol-1 H$_2$S/ photochemical haze layer. It has sometimes been suggested that the dark spots might be caused by a darkening of the overlying haze. We looked to see if darkening the 1--2-bar Aerosol-2 layer could explain the observations, but found that this is not supported by the limb-darkening evidence. Such a scenario would lead to stronger limb darkening than observed, since as the spot moved towards the limb we would be looking through more and more of this darker haze. Instead, we believe we are looking at the darker deep Aerosol-1 layer through the fairly uniform 1--2 bar Aerosol-2 layer, which is reasonably scattering at these wavelengths. Hence, as spots move towards the limb we see more reflectance from the Aerosol-2 layer at 1--2 bar and less from the deep Aerosol-1 layer, leading to the reduced limb darkening observed. 


\section{Discussion}\label{discussion}

\subsection{Vertical Aerosol Structure}

Excepting the additional upper tropospheric methane aerosol layer on Neptune, we find very similar vertical aerosol structures for both Uranus and Neptune composed of:
\begin{enumerate}
    \item Aerosol-1: A deep layer based at $p>$ 5--7 bar of moderately scattering particles (assumed, as previously discussed, to be sub-micron-sized) that we assume to be coincident with the main H$_2$S cloud/haze condensation layer, perhaps composed of a mixture of photochemically-produced haze particles and H$_2$S ice;
    \item Aerosol-2: A vertically-thin haze near the methane condensation level at 1--2 bar, composed of approximately micron-sized particles that are scattering at visible wavelengths, but more absorbing at UV and longer wavelengths; and
    \item Aerosol-3: A vertically-extended haze of sub-micron-sized particles with a fractional scale height of 2 and moderately similar refractive index spectra to the 1--2 bar Aerosol-2 layer.
    
\end{enumerate}

 The similarity of the structure (both vertical and spectral scattering) of the two planets should not be surprising given their similar tropospheric gaseous composition and temperatures, but how can we account for the fact that we apparently have a layer of photochemically-produced haze at the methane condensation level and not methane clouds? We hypothesise that to account for this structure we need to consider the vertical stability of Uranus's and Neptune's atmospheres.
 
 In \ref{App:A} we outline the theory behind buoyancy and static stability in planetary atmospheres. At pressures where the temperature profile falls more slowly with height than the lapse rate, the air is stable to vertical perturbations and will try, if moved vertically, to return to its original level, giving rise to gravity waves with a frequency defined by the Br\"unt-V\"ais\"al\"a frequency, $N$. For most atmospheres, this frequency just depends on the vertical gradient of the \textit{potential temperature} $\theta = T \big( \frac{p_0}{p}\big)^\gamma$, where $\gamma = R/C_p$, $T$ and $p$ are the local temperature and pressure, $p_0$ is a reference pressure level, $R$ is the gas constant and $C_p$ is the molar heat capacity at constant pressure. The Br\"unt-V\"ais\"al\"a frequency can be shown in \ref{App:A} to be:
 
 \begin{equation} \label{eq:brunt}
     N^2(z) = g
     \dv{}{z}\bigg(\ln \theta_0(z) \bigg)
 \end{equation}
where $\theta_0(z)$ is the potential temperature of the background air at altitude $z$. 

\begin{figure*} 
\centering
\includegraphics[width=\textwidth]{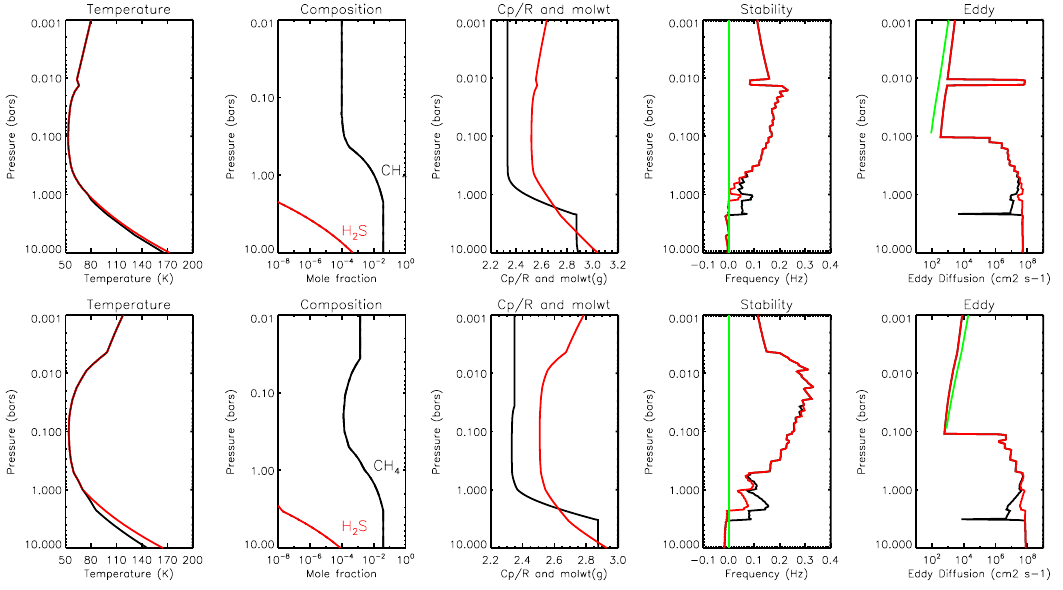}
\caption{Ice Giant stability plots, with Uranus on top row and Neptune on bottom. We have assumed He/H$_2$ = $1.06\times$ solar and CH$_4$/H$_2$ = $64\times$ solar for both planets (assuming protosolar composition of \cite{asplund09}), leading to a `deep' (i.e., at 10 bar)  CH$_4$ mole fraction of 4\%. For Uranus we have assumed H$_2$S/H$_2$ = $37\times$ solar and NH$_3$/H$_2$ = $1.4\times$ solar \cite{molter21}, and for Neptune we have assumed H$_2$S/H$_2$ = $54\times$ solar and NH$_3$/H$_2$ = $3.9\times$ solar \cite{tollefson21}, leading to mole fractions for H$_2$S at 20 -- 40 bar of $6.4\times 10^{-4}$ and $7.2\times 10^{-4}$, respectively. For each planet/row there are five plots: 1) temperature profiles - red line assumes a dry adiabatic lapse rate (DALR) at depth, while the black line assumes a saturated adiabatic lapse rate (SALR) and so includes latent heat released by condensation; 2) composition profiles showing the assumed methane and H$_2$S mole fraction profiles; 3) profiles of the mean molecular weight (black) and the molar heat capacity at constant pressure (red), $C_p$; 4) stability of atmosphere, represented in terms of the Br\"unt-V\"ais\"al\"a frequency, calculated from the temperature profile alone (red) and then also including the molecular weight gradient (black); and 5) Eddy-diffusion coefficient estimates calculated, including (black) or excluding (red) molecular weight changes, from the model of \citeA{ackerman01}, and compared with upper tropospheric determinations of $K_{zz}$ by \citeA{fouchet03} (green). In the Br\"unt-V\"ais\"al\"a frequency panels, the green line indicates the neutral static stability line.
} 
\label{fig:stability}
\end{figure*}

The latent heat released by the condensation of clouds lowers the rate at which temperature falls with height, making the air more stable for both planets, and this can be seen in our calculations 
of the Br\"unt-V\"ais\"al\"a frequency in Fig. \ref{fig:stability}, with a local region of higher Br\"unt-V\"ais\"al\"a frequency seen just above the methane condensation level. 
However, while Eq. \ref{eq:brunt} is perfectly acceptable for most atmospheres, where the molecular weight does not change greatly with height, this is not the case for Uranus and Neptune near the methane condensation layer. Assuming a deep mole fraction of 4\% for methane, the mean molecular weight reduces from $\sim$2.9 to $\sim$2.3 g/mol for both planets across the methane condensation region, a decrease of $\sim$20\%. This decrease in molecular weight at the condensation level increases the static stability even further since the lighter air is naturally more buoyant than the heavier air below. We show in \ref{App:A} that a more appropriate equation for the Br\"unt-V\"ais\"al\"a frequency for atmospheres where the molecular weight also changes significantly with height is:

\begin{equation}
    N^2(z) = g
     \dv{}{z}\bigg(\ln \bigg(\frac{\theta_0(z)}{M_0(z)}\bigg)\bigg)
\end{equation}

where $M_0(z)$ is the molecular weight of the background air at altitude $z$. It can be seen in Fig. \ref{fig:stability} that including this term greatly increases the static stability of the atmosphere at the methane condensation level. 

We hypothesise that this localised vertical region of static stability somehow leads to an enhanced opacity of larger haze particles that, at the lower boundary, act as seed particles for methane condensation. Given the large mole fraction of methane, and ready supply of cloud condensation nuclei (CCN), methane cloud formation will likely be extremely rapid and \citeA{carlson88} estimate that methane particles at 1--2 bar in Ice Giant atmospheres may grow to a size of $\sim$5 mm in as little as $\sim$100s. Hence, these large methane ice particles would almost immediately precipitate, or `snow', out, which would explain why pure methane ice clouds are almost never seen at these levels. The precipitating methane ice particles will drop to deeper levels before sublimating and releasing their payload of photochemical haze particles, which in turn may serve as CCNs for H$_2$S ice in the 4--10-bar region. The process is analogous to the ``smust'' concept that \citeA{hunten08} proposed to explain a vertical gap in the heavy hydrocarbon gas abundances of Jupiter. The difference is that in the case of Uranus and Neptune, methane ice transports haze particulates, while in the case of Jupiter, \citeA{hunten08}  proposed that haze particulates transported volatile hydrocarbon species. Since the mole fraction of H$_2$S is expected to be very much lower ($\sim$10$^{-4}$) than for methane, particle growth at 4--10-bar will not be so rapid \cite<e.g.,>{carlson88} and hence a stable layer of H$_2$S cloud could form. In addition, there are no vertical stability barriers since the latent heat release and change in mean molecular weight at this pressure level are very small, as can be seen in Fig. \ref{fig:stability}. 

Returning to the Aerosol-2 layer, methane condensation acts to create a clearing beneath it in two ways. Condensation onto haze particles at the base of the layer causes them to snow out directly, and the stable layer produced by latent heating and reduced molecular weight reduces $K_{zz}$ and slows the rate at which haze particles in the Aerosol-2 layer mix into the deeper levels. Within the layer, the mechanism for the local enhancement of haze opacity is currently unknown and understanding of this newly-discovered feature will require a future microphysical modelling study. Toward the top of the compact Aerosol-2 layer, the decrease in haze density with altitude (in the presence of eddy mixing) implies some sort of particle density source at the base of the layer, rather than above it, because eddy mixing alone can only modify the slope of a gradient from the source region to the sink region, not reverse it. The action of methane, near saturation, could be significant. Considering the likely hydrocarbon composition of haze particles, it is possible that methane condensation, even at subsaturated vapour pressures, could influence the surface chemistry of haze particles (in analogy to hydrophilic aerosols significant in terrestrial hazes). It is also possible that while methane condensation at the base is rapid, leading to the formation of large particles, there will be also be some component of smaller particles in the size distribution.  The fractional abundance of these smaller haze/ice particles is likely to get larger as we move up to lower pressures and lower temperatures in the Aerosol-2 layer, where the saturated vapour pressure of methane will be much lower. Hence, what we may be seeing in the Aerosol-2 layer is a mixture of photochemical haze and methane ice. Microphysics in the unique environment of the Aerosol-2 layer may thus be complex, and it is clear that the haze particles cannot be considered static tracers of mixing.

While this scenario provides an attractive model for the background aerosol structure of both planets, the atmosphere of Neptune is more energetic and contains frequent cloud features that appear to be methane condensation. For the most part these appear to form in the upwelling regions at 20--40$^\circ$N and 20--40$^\circ$S 
and at pressures of 0.6 to 0.1 bar \cite<e.g.,>{irwin16,molter19}. However, \citeA{irwin11} report the appearance of several clouds at 60--70$^\circ$S in 2009 that are deep in the atmosphere at 1--2 bar and may be rare examples of methane clouds forming near the methane condensation level itself.

As to the nature of the photochemical haze, there are several candidates. \citeA{khare93} report the spectra of `tholins' generated by irradiating H$_2$O/C$_2$H$_6$ mixtures, which have minimal absorption at 0.8 $\mu$m and increasing absorption as we move to UV, or  longer wavelengths. While the spectrum of these `tholins' is not identical to those derived here it is qualitatively similar. Another candidate is acetylene soot \cite{dalzell69}, which also has increasing absorption at long wavelengths. Indeed, such hazes share some characteristics with the phenomenon of `blue hazes' commonly observed in the Smoky Mountains of eastern Tennessee and the Blue Ridge Mountains of Virginia \cite{ferman81}. Examining the scattering cross-sections of the retrieved aerosol particles, we find that the small particles assumed to be present in the deep Aerosol-1 component, and found to be necessary in the extended Aerosol-3 component above $\sim 1.5$ bar, would appear to be blue at visible wavelengths. However, the larger particles in the 1--2 bar Aerosol-2 layer would appear white in the visible.

The stability of the 1--2-bar region in the Ice Giant atmospheres has also been concluded by \citeA{guillot95} and \citeA{leconte17}. Intriguingly, \citeA{teanby20} noted that a stable layer at this pressure level has important implications for the internal oxygen enrichment of Neptune \cite{teanby20,cavalie17}. Limited mixing at 1--2 bar could permit externally-sourced CO entering Neptune's atmosphere from comets to be trapped in the upper troposphere, which removes the need to have extreme internal oxygen enrichment in order to explain the wide wings in Neptune's sub-mm CO lines. 
Previous work had suggested CO to be present throughout the troposphere, which requires an extremely high internal oxygen enrichment compared with the solar composition by a factor of at least a few hundred \cite{lellouch05,Luszcz13,cavalie17,moses20} in order to act as a deep source for tropospheric CO. 
If externally sourced CO can be kept in the upper troposphere by limited mixing this opens up the possibility of a more rock-rich Uranus and Neptune \cite{teanby20}, in keeping with more modest oxygen enrichments of $\sim$50 inferred from the observed D/H ratio \cite{feuchtgruber97}. 

Finally, in \ref{App:D} we outline a review of methods for calculating the eddy diffusion coefficient, $K_{zz}$, in planetary atmospheres. With some modification of model parameters suggested by \citeA{ackerman01} we found we could derive a $K_{zz}$ profile that was consistent with upper stratospheric determinations of $K_{zz}$ by \citeA{fouchet03}, as can be seen in Fig. \ref{fig:stability}. With these parameters we find a local minimum of $K_{zz}$ at the methane condensation level, but also a smaller minimum near the tropopause; this second $K_{zz}$ minimum may help to explain the detached methane ice layer (Aerosol-4) seen near the tropopause of Neptune, where the atmosphere may be dynamically vigorous enough to mix methane up to such levels.

\subsection{Nature of dark regions in the atmospheres of the Ice Giants}
In this paper we have shown that dark regions seen occasionally in Neptune's atmosphere, and rarely in Uranus's atmosphere, can be explained by a darkening, or opacity change of the deep Aerosol-1 layer, which we have suggested may be composed of a mixture of H$_2$S ice and photochemically-produced haze. Such a condensation level is consistent with the estimate from ground-based microwave studies that the deep abundance of H$_2$S in Uranus's and Neptune's atmospheres are $53.8^{+18.9}_{-13.4}$ $\times$ solar \cite{tollefson21} and $37^{+13}_{-6}$ $\times$ solar \cite{molter21}, respectively.
We believe the reason that such regions are more commonly reported on Neptune than Uranus is that the overlying 1--2--bar Aerosol-2 layer on Neptune has lower opacity, making the deep Aerosol-1 layer more visible.

Interpreting the results of our radiative transfer analysis with respect to the structure of dark vortices is complicated, especially if the dark spots are to be explained by a darkening of particles in the deep haze layer. Particle colour (resulting from changes to $n_{imag}$) cannot be unambiguously attributed to composition and/or heterogeneous microphysical structure, and thus linking these quantities to vortex dynamics and structure would be speculative given the limited information currently available. Even for Jupiter, where more detailed information is available, links between vortex structure and aerosol properties remain tentative \cite<e.g.,>{wong11}. It is possible that dark spots are generated by  secondary-circulation uplift within vortices dredging up and concentrating a `chromophore' material from warmer depths below, which is dark at visible wavelengths. However, it is not clear what this material might be, and photolysis, which might be thought necessary to generate dark particles, is unlikely to be significant at pressures of $\sim$ 5--7 bar since the UV flux will be very low. Rather than interpreting such features as being caused by the addition of a new `chromophore', an alternative explanation might be that such regions are anomalously warm at the H$_2$S condensation level and so H$_2$S ice sublimates into the vapour state, revealing the darker CCN photochemical haze core, similar to the mantling process proposed in \cite{west86} to explain belt-zone color differences in Jupiter's atmosphere. Such a scenario may be consistent with cooling above the anticyclone mid-plane and warming below, as seen in numerical
simulations \cite{stratman01,hadland20,lemasquerier20} and theoretical models \cite{marcus13}. However, more observations are needed to fully constrain why the particles in the Aerosol-1 layer appear to be darker in the centre of dark spots and at dark latitudes.

Our results place interesting constraints on the vertical structure of dark spots. Geostrophic anticyclones in stratified fluids---whether salt lenses in the Earth's oceans or the Great Red Spot in Jupiter's atmosphere---are characterized by high density anomalies in their upper regions, and low density anomalies in their lower regions (for example, dark contours in Fig. 2 of \citeA{barcelo-llull17}, Fig. 3 of \citeA{marcus13}, Fig. S8 of \citeA{lemasquerier20}). Thus, how giant planet atmospheric vortices affect the observable aerosol structure depends on the location of aerosol layers with respect to the vortex mid-plane. Anticylones are also more efficiently mixed, compared to the stratified background atmosphere \cite<e.g.,>{hassanzadeh12}.

\begin{figure*} 
\centering
\includegraphics[width=\textwidth]{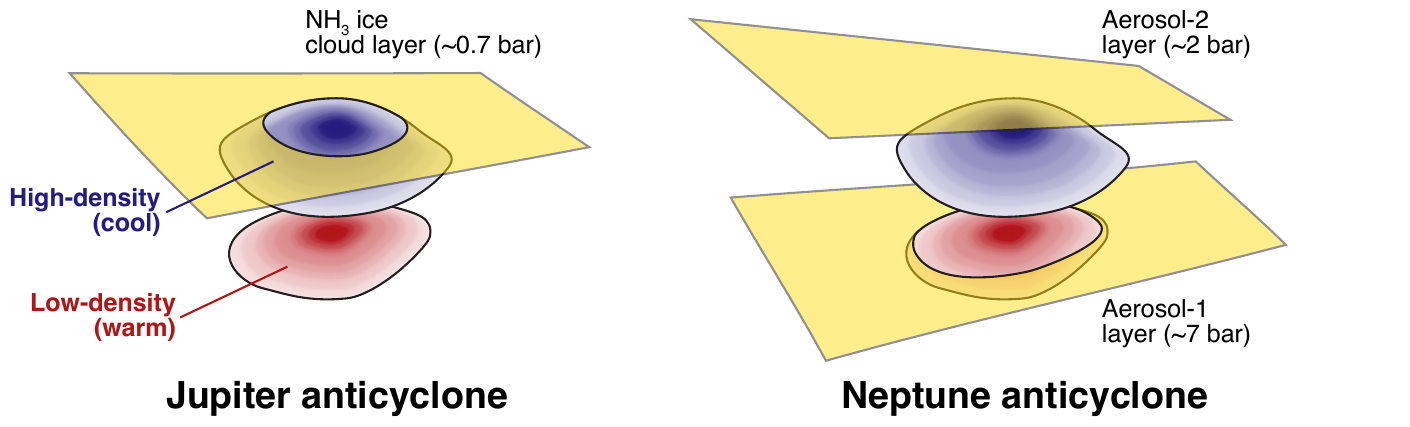}
\caption{Possible explanations for vertical structure of dark spots in Uranus and Neptune's atmospheres, compared with vortex model for Jupiter's atmosphere. Cloud structure may be affected differently, depending on whether cloud layers intersect with the high-density (cool) or low-density (warm) anomalies associated with anticyclonic vortices. Left hand panel shows a model of vortices in Jupiter's atmosphere, where the upper cooler part of the vortex intersects with the NH$_3$ condensation layer, leading to enhanced ice formation there, and enhanced haze formation above it (extending close to the tropopause). In contrast, the right hand panel shows a similar model for Neptune's atmosphere, where the lower warm region overlaps with the H$_2$S condensation level, causing a darkening, or clearing, of the Aerosol-1 layer. To explain the lack of any changes in the Aerosol-2 layer at the location of dark spots, the high-density (cool) anomaly may need to reside deeper than the Aerosol-2 layer. However, a deep vortex top is challenging to reconcile with observations of orographic companion clouds near dark spots. Alternatively, the mid-plane may need to coincide with the Aerosol-2 layer.}
\label{fig:vortexslice}
\end{figure*}

Figure \ref{fig:vortexslice} compares differences in model Jupiter and Neptune vortex structures that might be consistent with our deep aerosol results. In Jupiter's case, observable aerosol layers (the NH$_3$ ice cloud layer plus the tropospheric haze above it) intersect the upper part of the vortex, where the high density anomaly is associated with cooler temperatures \cite<e.g.,>{cheng08}. The cooler temperatures enhance haze (e.g., N$_2$H$_4$)  condensation \cite{west86}, and efficient mixing delivers cloud and haze precursors to high altitudes, producing the increased aerosols that are seen within the upper portion of Jovian anticyclones \cite{banfield98}.

Depleted, or darkened, aerosols in Neptune's dark spots, as possibly suggested by our analysis, may indicate that the vortex midplane is located at higher altitudes than the deep Aerosol-1 layer. Higher temperatures in the low-density anomaly would inhibit H$_2$S condensation. Depending on the vertical distribution of the dark haze CCNs populating the deep layer, mixing within the anticyclone may also play a role by altering the concentration of CCNs at the H$_2$S condensation layer. This scenario may also constrain the top of the vortex in this model. Efficient mixing within the anticyclone would disrupt the middle Aerosol-2 layer near 1--2 bar, which is not observed. We have demonstrated that we can model the dark spot appearance at all HST/WFC3 wavelengths (Fig. 10) without any changes to the middle Aerosol-2 layer, which would mean that the high-density anomaly of the dark vortex may reside entirely below this level. Models where both the Aerosol-1 and Aerosol-2 layers were depleted rendered vortices visible at longer HST/WFC3 wavelengths, in contradiction with the observations. If vortices are indeed bounded by approximately the 2-bar and 10-bar surfaces, their thickness is less than two atmospheric pressure scale heights. This thickness is much smaller than that for Jupiter, where remote sensing of the visible clouds constrains vortex tops to about the 0.5-bar level \cite<e.g.,>{banfield98,west04,cheng08} and vortex bases to levels near the 10-bar level, or perhaps as deep as 800 bar in the special case of the Great Red Spot \cite{parisi21,bolton21}. Thermal infrared observations, however, \cite<e.g.,>{fletcher10} find that the thermal effects of vortices reach even higher to the tropopause. Jovian vortices thus span ranges of 3--7 scale heights. The potential difference in vortex thicknesses between Jupiter and Neptune may perhaps be related to the deep stratification of these outer planet atmospheres, which is still not known to high precision. Direct measurements at relevant pressure levels are only available for a single time and place: the Galileo Probe entry site at Jupiter \cite{seiff98,magalhaes02}. Given the dependence of vortex vertical/horizontal aspect ratio on the difference between internal and environmental stratification \cite{hassanzadeh12}, and the roughly comparable horizontal dimensions of vortices on Jupiter and Neptune \cite{li04,wong18}, the thicker vertical dimensions of Jupiter's vortices may indicate that the environmental stratification is weaker on Jupiter compared to Neptune.
However, this tightly constrained thickness scenario runs counter to the observation that such dark spot disturbances are often accompanied by what appear to be orographic clouds near the tropopause,  e.g., the `scooter' cloud that accompanied Voyager-2's GDS. Hence, it would seem that the effect of the vortices must actually extend much higher. A possible scenario that might be consistent with all the evidence may be that the mid-plane of the vortex coincides with the statically stable 1--2-bar Aerosol-2 layer, i.e., the methane condensation level, which could leave this layer unaffected, but would affect both upper and lower clouds.  

\subsection{Dependence of solutions on methane absorption spectra}
The main retrievals presented here have been conducted using k-tables generated from the methane band model coefficients of \citeA{kark10}. We used these coefficients as they cover the entire spectral range of our observations leading, we hope, to self-consistent results. The line data sets that are available, e.g., WKLMC@80K \cite{campargue13} have been shown to provide better fits to the H-band observations \cite<e.g.,>{irwin12lbl}, but are limited in their wavelength coverage and so cannot cover the entire 0.3 -- 2.5 $\mu$m region. Later databases, such as HITRAN16 \cite{hitran16}, and TheoRETS \cite{Rey18} extend the lower wavelength limit to 1.0 and 0.75 $\mu$m, respectively, but these still fail to cover the spectral region where dark spots are most visible. We repeated some of our retrievals for 0.76 -- 2.5 $\mu$m region alone, using k-tables generated from the  methane band parameters \cite{kark10} and also from the TheoRETS database and found better fits to the observations with TheoRETS, and broadly similar retrieved aerosol structures and scattering parameters for the two different sources of methane absorption data. However, significant differences were found between the aerosol structures retrieved from the 0.76 -- 2.5 $\mu$m using using the methane band/k-data and those retrieved from the full 0.3 -- 2.5 $\mu$m range, with significantly more opacity of the deep Aerosol-1 layer estimated when using the restricted wavelength range. This may go some way to explaining the difference between the vertical profiles of aerosol retrieved here and those from previous studies, concentrating on the 820 nm or H-band regions alone, that did not include a separate deep Aerosol-1 layer and generally put the main Aerosol-2 layer deeper at 2--3 bar, and for the 820 nm retrievals deduced a lower methane abundance for Neptune \cite{irwin21}. It is only the methane band data of \citeA{kark10} that allows us to analyse the whole range simultaneously and deduce the presence of the deeper layer Aerosol-1 layer, whose spatial variations are, we suggest, responsible for dark spot features.

We did attempt trying to combine the band and line data together, but there are significant differences in the wavelength regions of overlap and we found, generally, that the fits were worse when using a combination of methane data from different sources. In order to extend this analysis we really need for the line data sources to extend to 0.3 $\mu$m so that we are able to analyse the whole spectral range with a self-consistent set of absorption data. We look forward to the time when this might be possible.

\subsection{Spatial Deconvolution}
While the HST/STIS observations presented here have been spatially deconvolved with a Lucy-Richardson deconvolution scheme, this has not yet been possible with the Gemini/NIFS  data we have analysed. For Uranus we do not consider this to be too significant a problem since the disc of Uranus seen in 2009 was moderately featureless. However, for the Neptune observations, there were significant levels of high methane ice clouds. Although we masked these regions when extracting the Minnaert coefficients of the background regions, given the likely shape of the Point-Spread-Function (PSF) it is possible that the unmasked regions were still contaminated by light from the bright, cloudy regions, which may help to explain why the NIFS Neptune data show very little limb-brightening at methane-absorbing wavelengths, in contrast to the expectations from the fits to the combined data and also to the behaviour seen in the Uranus NIFS observations. We hope, when possible, in future work to attempt to spatially deconvolve these Gemini/NIFS observations to explore this. 

\begin{figure*} 
\centering
\includegraphics[width=\textwidth]{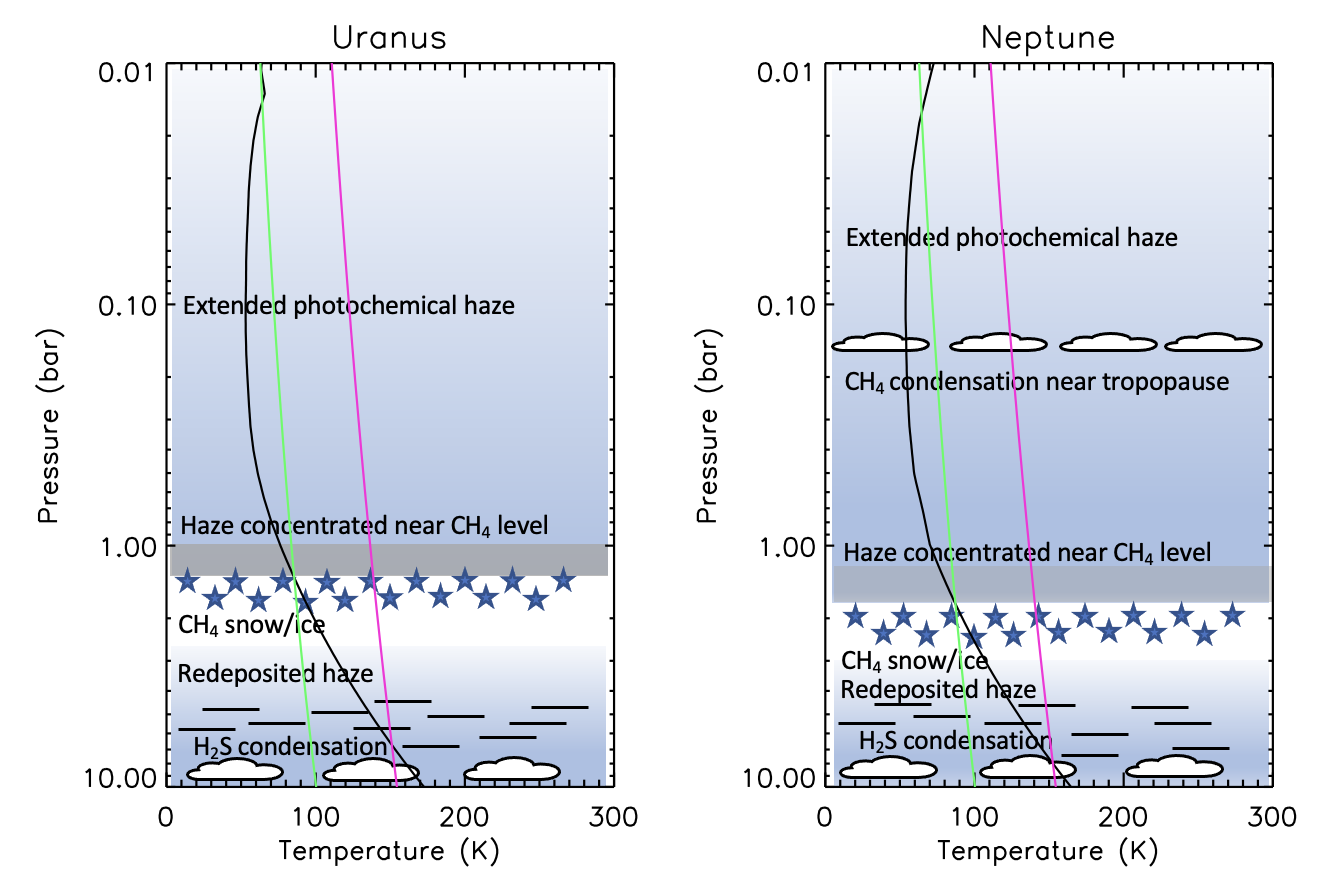}
\caption{Summary of retrieved aerosol distributions for Uranus (left) and Neptune (right), compared with the assumed temperature/pressure profiles. On each plot is also shown the condensation lines for CH$_4$ (green) and H$_2$S (pink), assuming mole fractions at 10 bar of 4\% for CH$_4$ and $1\times 10^{-3}$ for H$_2$S, to move the condensation levels to the approximate levels of the Aerosol-1 and Aerosol-2 layers. The default solution is composed of: A) an extended layer of haze, photochemically-produced in the stratosphere and mixed by eddy diffusion to lower levels (Aerosol-3); B) a thicker haze layer of larger particles ($r \sim 1$ $\mu$m) near the CH$_4$ condensation level (Aerosol-2); C) rapid formation of large methane ice/snow particles at the base of this layer, which rapidly fall and redeposit the haze cores at lower altitudes; and D) an H$_2$S cloud based at $p>5-7$ bar, which forms on the haze particles (Aerosol-1). For Neptune, we find we also need a component of moderate-sized ($\sim$2 $\mu$m) methane ice particles near the tropopause.  Note that the Aerosol-2 layer on Neptune has noticeably less opacity than that on Uranus, by a factor of $\sim$ 2.} 
\label{fig:cartoon}
\end{figure*}

\section{Conclusions}\label{conclude}
Modelling the visible/near-infrared reflectivity spectra of Uranus and Neptune over a wide wavelength range of 0.3--2.5 $\mu$m represents a considerable challenge. In effect we are trying to find a cloud/haze structure of unknown constituents, of unknown size and unknown complex refractive index spectra, using as our main probe of vertical level the absorption spectrum of gaseous methane, which also has unknown systematic errors. Add to that the uncertain measurement error of the observed data themselves and it can be seen that this is a considerably degenerate problem! Hence, there are multiple solutions that fit equally well and all may need to be revised once the available line data of methane have been extended to the visible. However, by performing our `snippet analyses', we have shown how we can differentiate between spectral and vertical variations of the Ice Giant aerosols. Using the snippet analyses to constrain our traditional retrieval approach (modified to use multiple starting points to avoid the solutions becoming trapped in local $\chi^2/n$ minima) we have determined an aerosol model for Uranus and Neptune that matches the observations well and is simple, elegant, has some basis in terms of haze production and likely pressure levels of static stability, and serendipitously can also explain the observed characteristics of Neptune's (and Uranus's) dark spots.

In summary, in this work we have found that we can model the observed reflectivity spectra of both Uranus and Neptune from 0.3 to 2.5 $\mu$m with a single, simple aerosol model, summarised in Fig. \ref{fig:cartoon}, comprised of:
\begin{enumerate}
    \item A deep layer based at $p>$ 5--7 bar (Aerosol-1) of what we assume to be sub-micron-sized particles, which we suggest to be coincident with the main H$_2$S cloud/haze condensation layer and composed of a mixture of photochemically-produced haze particles and H$_2$S ice. These particles are highly scattering at 500 nm, but become more absorbing at both shorter and longer wavelengths;
    \item A vertically thin aerosol layer of micron-sized particles just above the methane condensation level at 1--2 bar (Aerosol-2), possibly composed of a mixture of photochemically-produced haze and methane ice, that are less reflective than the deeper Aerosol-1 haze particles at visible wavelengths and are also more absorbing at both shorter and longer wavelengths;
    \item A vertically extended haze of small particles based at 1--2 bar (Aerosol-3) with a fractional scale height of $\sim$2 and similar refractive index spectrum to the main 1--2-bar Aerosol-2 layer;
    \item For Neptune, an additional vertically thin component of moderate-sized methane ice particles ($\sim$2 $\mu$m), located at $\sim$ 0.2 bar just below the tropopause .
\end{enumerate}

We find this model to be consistent with HST/STIS, HST/WFC3, IRTF/SpeX, Gemini/NIFS and Voyager-2/ISS observations. Our main conclusions are:
\begin{enumerate}
    \item The UV- and long-wavelength-absorbing nature of the retrieved imaginary refractive index spectrum of the 1--2-bar Aerosol-2 layer is not consistent with our expectation for CH$_4$ or H$_2$S ice. The spectral dependence is more consistent with the imaginary refractive index spectrum derived for the tropospheric/stratospheric haze. Hence, we suggest that the 1--2-bar Aerosol-2 layer contains a considerable fraction of photochemical haze, produced at higher altitudes, which has somehow become concentrated and modified at this level in a region of static stability created by vertical gradient in molecular weight and also latent heat release from methane condensation.
    \item The haze particles in the 1--2-bar Aerosol-2 layer act as cloud-condensation nuclei (CCN) for methane condensation at the lower boundary, which condenses so quickly that methane ice immediately `snows out' to re-evaporate at deeper levels, redepositing the haze particles there. The larger retrieved size of the particles in the Aerosol-2 layer compared with the higher, vertically extended Aerosol-3 layer may in part reflect this cloud-seeding process and may also just arise due to coagulation and coalescence in this vertically stable layer.
    \item The haze particles at deeper levels act as CCNs for H$_2$S condensation at $\sim$ 4--10 bar (Aerosol-1). At the low partial pressures of H$_2$S seen at these levels and the higher pressure, this condensation is slower, forming a cloud of what we have assumed to be sub-micron-sized particles, which have spectral properties consistent with a mix of dark haze and ice. 
    \item In Neptune's atmosphere, increased reflectance in methane absorption bands at wavelengths longer than  $\sim$1 $\mu$m can be accounted for by the addition of an optically and vertically thin component of micron-sized methane ice particles, based near 0.2 bar, just below the tropopause.
    \item Darkening of the particles in the deep haze/H$_2$S-ice Aerosol-1 layer (or to a lesser extent a clearing of this layer) leads to spectral perturbations that match well the observed characteristics of dark spots and the dark latitude bands seen in Voyager-2/ISS and HST/WFC3 observations of Neptune  (i.e., visibility only at $\lambda <$ $\sim$700 nm). Such perturbations also provide a good fit to observed limb-darkening properties of these features. At the same time, the Aerosol-2 layer (1--2 bar) appears to be unperturbed by dark regions. This potentially limits the vertical thickness of dark spots (and dark latitudes) to disturbances at pressures greater than $\sim$3-bar, and perhaps spanning less than 2 scale heights, a significant difference from thicker anticyclones seen in Jupiter's atmosphere. However, such an interpretation does not rest easily with the observation that such features are frequently accompanied by orographic clouds, based near the tropopause. Hence, it may be that the vortex mid-plane coincides with the 1--2-bar Aerosol-2 layer.
    \item The opacity of the 1--2-bar Aerosol-2 layer in Uranus's atmosphere is found to be significantly thicker than that of Neptune by a factor of $\sim 2$; since these particles are found to be UV-absorbing, this explains Uranus's lower observed UV reflectivity and also explains why Uranus appears to have a paler blue colour to the human eye than Neptune since these particles are found to have a roughly white visible reflectivity spectrum. The lower opacity of Neptune's Aerosol-2 layer also explains why dark spots, caused, we suggest, by perturbations of the deep Aerosol-1 H$_2$S/haze layer are easier to observe in Neptune's atmosphere than in Uranus's. We suggest that the Aerosol-2 layer on Neptune may be thinner than that of Uranus due to Neptune's dynamically overturning atmosphere being more efficient at clearing this haze layer through methane condensation.
    \item Future observations of Uranus and Neptune with instruments such as HST/STIS or VLT/MUSE, able to return high spectral resolution hyperspectral cubes at visible wavelengths, may help to resolve the question of whether dark spots and dark regions are caused by a darkening or a clearing of the Aerosol-1 layer. This will, we hope, be the focus of future work.

\end{enumerate}

\acknowledgments

We are grateful to the United Kingdom Science and Technology Facilities Council for funding this research (Irwin: ST/S000461/1, Teanby: ST/R000980/1). Glenn Orton was supported by funding to the Jet Propulsion Laboratory, California Institute of Technology, under a contract with the National Aeronautics and Space Administration (80NM0018D0004). Leigh Fletcher and Mike Roman were supported by a European Research Council Consolidator Grant (under the European Union's Horizon 2020 research and innovation programme, grant agreement No 723890) at the University of Leicester.  Santiago P\'{e}rez-Hoyos is supported by grant PID2019-109467GB-I00Z, funded by \url{MCIN/AEI/10.13039/501100011033}. We are also grateful for the assistance of Larry Sromovsky and Pat Fry in locating and reading the HST/STIS data. We are also grateful to the PDS Ring-Moon Systems Node's OPUS search service for providing access to the Voyager-2 ISS images. Finally, the Gemini/NIFS observations used were obtained at the international Gemini Observatory, a program of NSF’s NOIRLab, which is managed by the Association of Universities for Research in Astronomy (AURA) under a cooperative agreement with the National Science Foundation. on behalf of the Gemini Observatory partnership: the National Science Foundation (United States), National Research Council (Canada), Agencia Nacional de Investigaci\'{o}n y Desarrollo (Chile), Ministerio de Ciencia, Tecnolog\'{i}a e Innovaci\'{o}n (Argentina), Minist\'{e}rio da Ci\^{e}ncia, Tecnologia, Inova\c{c}\~{o}es e Comunica\c{c}\~{o}es (Brazil), and Korea Astronomy and Space Science Institute (Republic of Korea). 

\section{Data Availability}

The Uranus HST/STIS datasets used in this study are available from \citeA{frykarkoschka18}, 
while the Neptune/STIS dataset is available from \citeA{irwin22d}.  
The IRTF/SpeX observations are available from \url{http://irtfweb.ifa.hawaii.edu/~spex/IRTF_Spectral_Library/References_files/Planets.html}. The Uranus Gemini/NIFS data are available at \url{https://archive.gemini.edu/searchform/GN-2009B-Q-52/}, while the Neptune Gemini/NIFS data are available at \url{https://archive.gemini.edu/searchform/GN-2009B-Q-85/}. The Voyager ISS images were downloaded from 
\url{https://bit.ly/3qjuqk6}.
The spectral fitting and retrievals were performed using the NEMESIS radiative transfer and retrieval algorithm \citeA{irwin08} and can be downloaded from \citeA{irwin22a}, with supporting website information at \citeA{irwin22b}. The data products generated in this study (processed spectra and retrieved atmospheric parameters) are available from \citeA{irwin22c}.

\appendix
\section{Bouyancy Forces} \label{App:A}

Consider the force acting in the vertical direction $z$ on a parcel of air at altitude $z_0$ of cross-sectional area $A$, height $\dd z$, and density $\rho$:
\begin{equation}
\rho A \dd z \dv[2]{z}{t} = -\rho g A \dd z + \rho_0 g A \dd z
\end{equation}
where $\rho_0$ is the density of the surrounding air and $g$ is the gravitational acceleration. Dividing by $A \rho \dd z$ we have:
\begin{equation}
\dv[2]{z}{t} = -g \frac{(\rho - \rho_0)}{\rho}.
\end{equation}
The density can be written as $\rho = \frac{Mp}{RT}$, where $M$ is the molecular weight of the air, $p$ is the pressure, $T$ is the temperature and $R$ is the gas constant. Hence, we can rewrite this equation as:
\begin{equation}
\dv[2]{z}{t} = -g \frac{(\frac{Mp}{RT}-\frac{M_0 p}{RT_0})}{\frac{Mp}{RT}}
\end{equation}
where we have assumed the parcel and the surroundings have the same pressure. Rearranging this equation we find:
\begin{equation}
 \dv[2]{z}{t} = -g \bigg( 1-\frac{M_0}{M} \frac{T}{T_0} \bigg) = -g \bigg( 1-\frac{M_0}{M} \frac{\theta}{\theta_0} \bigg)
\end{equation}
 where $\theta$ is the potential temperature of the parcel (i.e., the temperature it would have if compressed or expanded adiabatically to a reference pressure $p_0$), defined for the parcel to be $\theta = T \big( \frac{p_0}{p}\big)^\gamma$, where $\gamma = R/C_p$ and $C_p$ is the molar heat capacity at constant pressure. The potential temperature of the surrounding air is similarly $\theta_0 = T_0 \big(\frac{p_0}{p}\big)^\gamma$ and hence $\frac{T}{T_0} = \frac{\theta}{\theta_0}$. 
 
 If the parcel moves adiabatically then its potential temperature and molecular weight remain constant (if no condensation) so any buoyancy must be due to changes in $M_0$ and $T_0$. We know that the acceleration $a = \dv[2]{z}{t} = 0$ at $z=z_0$ by definition and so 
\begin{equation}
 \dv[2]{z}{t} \sim \bigg(\dv{a}{z}\bigg)_{z_0} (z-z_0)
\end{equation}

where
\begin{equation}
\bigg(\dv{a}{z}\bigg)_{z_0} = g \frac{\theta}{M} \dv{}{z} \bigg(\frac{M_0}{\theta_0}\bigg) = g \frac{\theta}{M} \frac{\theta_0 \dv{M_0}{z} - M_0 \dv{\theta_0}{z}}{\theta_0^2}.
\end{equation}

At $z=z_0$, $\theta=\theta_0$ and $M=M_0$, and hence:
\begin{equation}
\bigg(\dv{a}{z}\bigg)_{z_0} = g \bigg( -\frac{1}{\theta_0} \dv{\theta_0}{z} + \frac{1}{M_0} \dv{M_0}{z} \bigg)
\end{equation}

or
\begin{equation}
    \bigg(\dv{a}{z}\bigg)_{z_0} = g \dv{}{z}\bigg(\ln \bigg(\frac{M_0}{\theta_0}\bigg)\bigg).
\end{equation}

Hence
\begin{equation}
    \dv[2]{z}{t} = -g
     \dv{}{z}\bigg(\ln \bigg(\frac{\theta_0}{M_0}\bigg)\bigg)\big( z-z_0 \big).
\end{equation}

This is simple harmonic motion, with an angular frequency, $N$, known as the Br\"unt-V\"ais\"al\"a frequency, given by

\begin{equation}
    N^2 = g
     \dv{}{z}\bigg(\ln \bigg(\frac{\theta_0}{M_0}\bigg)\bigg).
\end{equation}

This is different from the more familiar definition of the Br\"unt-V\"ais\"al\"a frequency that ignores molecular weight changes:
\begin{equation}
    N^2 = g
     \dv{}{z}\bigg(\ln \theta_0 \bigg).
\end{equation}

\section{Disc-averaging} \label{App:B}
The disc-averaged radiance $\overline{I}$ of a planet of apparent radius $R$ is defined as:
\begin{equation}
    \overline{I} = \frac{1}{\pi R^2}\int_{r=0}^{r=R} \int_{\phi=0}^{\phi=2 \pi} I r \dd r \dd \phi
\end{equation}

where $r$ is the radial position on the disc and $\phi$ is the azimuth angle. Substituting $r=R \sin \theta$, where $\theta$ is the local zenith angle, this becomes
\begin{equation}
    \overline{I} = \frac{1}{\pi}\int_{\theta=0}^{\theta = \pi /2} \int_{\phi=0}^{\phi=2 \pi} I \sin \theta \cos \theta \dd \theta \dd \phi.
\end{equation}

Assuming the radiance $I$ is azimuthally symmetric, i.e., does not depend on $\phi$, this simplifies to
\begin{equation}
    \overline{I} = 2 \int_0^{\pi /2} I(\theta) \sin \theta \cos \theta \dd \theta = \int_0^{\pi /2} I(\theta) \sin  2\theta  \dd \theta
\end{equation}

or
\begin{equation}
    \overline{I} = 2 \int_0^1 I(\mu) \mu \dd \mu 
\end{equation}

where $\mu = \cos\theta$. If we assume that $I(\mu)$ is well approximated by Minnaert limb darkening and that $\mu = \mu_0$, then $I(\mu)=I_0 \mu^{2k-1}$. Hence, the disc-averaged radiance $\overline{I}$ will be
\begin{equation}
    \overline{I} = 2 I_0 \int_0^1 \mu^{2k} \dd \mu = \frac{2I_0}{2k+1}. 
\end{equation}

The `disc-averaged' IRTF/SpeX spectra are actually line-integrals along the central meridian, for which:
\begin{equation}
    \overline{I} = \frac{\int I \dd y \dd x}{\int \dd y \dd x} \sim \frac{\Delta x \int I \dd y}{2R\Delta x} = \frac{\int I \dd y}{2R}
\end{equation}

where $y$ is the distance along the slit, $x$ is the position across it, and $\Delta x$ is the slit width (assumed to be small). Substituting $y = R \sin \theta$ this becomes
\begin{equation}
    \overline{I} = \frac{1}{2}\int_{-\pi /2}^{\pi/2} I(\theta) \cos \theta \dd \theta = \int_0^{\pi/2} I(\theta) \cos \theta \dd \theta.
\end{equation}

Assuming $I(\theta)$ is Minnaert-dependent (i.e., $I(\theta)=I_0(\cos\theta)^{2k-1}$) this becomes
\begin{equation}
    \overline{I} = I_0 \int_0^{\pi/2}  (\cos\theta)^{2k} \dd \theta.
\end{equation}

Sadly, this is not simply integrable for a general, non-integer $k$, but it may be evaluated numerically. This can be done either by pre-tabulating $\overline{I}/I_0$ as a function of $k$, or alternatively calculating $I(\theta)$ at all the quadrature zenith angles of the radiative transfer model and integrating $I(\theta)\cos(\theta)$ using the quadrature scheme weights. Both approaches were tested and found to give similar results. However, with the 5-point Gaussian-Lobatto scheme we generally use, the latter approach was found to be much slower  since calculations at the higher zenith angles need many more Fourier components to fully resolve the azimuthal part of the radiance calculation. Instead, calculating at just two zenith angles (0, $42.47^\circ$), extracting the minnaert-$k$ coefficients $I_0$ and $k$, and using the pre-tabulated dependence of  $\overline{I}/I_0$ on $k$ to calculate $\overline{I}$ was found to be more than twice as fast and of equivalent accuracy. 

\section{Cloud Opacity Units} \label{App:C}

The default units for particle cross-sections, $\chi(\lambda)$, in NEMESIS are cm$^2$/particle, while the default units of aerosol density, $D(p)$, are particles/gram. Hence, $\chi(\lambda) D(p)$ gives aerosol cross-section in units of cm$^2$/gram (i.e., cm$^2$ per gram of atmosphere).

To get opacity we need to multiply this cross-section by the path amount, $u$, which has units of gram/cm$^2$, and can be calculated as $u = \rho z$, where $z$ is the path length in cm and $\rho$ is the atmospheric density in units of gram/cm$^3$. Hence, the opacity of a path at wavelength, $\lambda$, is $\tau(\lambda) = \chi(\lambda) D(p) u = \chi(\lambda) D(p) \rho z$, which is unit-less. The opacity per km in a vertical path of atmosphere is then simply $\dv{\tau}{z}=10^5 \chi(\lambda) D(p) \rho$. 

Other authors commonly express cloud profiles in units of opacity/bar, $\dv{\tau(\lambda)}{p}$, which by the chain rule can be calculated as $\dv{\tau(\lambda)}{z}.\dv{z}{p}$. Assuming hydrostatic equilibrium, $p=p_0 e^{z/H}$, where $H$ is the Scale Height and $z$ is measured downwards, and thus $\dv{p}{z} =  \frac{p}{H}$ and $\dv{\tau(\lambda)}{p} = 10^5 \chi(\lambda) D(p) \rho \frac{H}{p}$. Substituting for the Scale Height, $H=RT/Mg$ and for atmospheric density, $\rho=pM/RT$, we conclude that the cloud opacity per bar is 

\begin{equation}
\dv{\tau(\lambda)}{p} = 10^5 \frac{\chi(\lambda) D(p)}{g(p)}
\end{equation}

where $g(p)$ is the gravitational acceleration at pressure level, $p$.

In NEMESIS we usually normalise the aerosol density profile, $D(p)$ and cross-section spectra, $\chi(\lambda)$, such that the cross section is 1.0 at a reference wavelength, $\lambda_0$, and the normalised dust opacity profile $D'(p)=D(p)\chi(\lambda_0)$. In this case, the opacity/bar at $\lambda_0$ is related to the scaled aerosol density, $D'(p)$, as 

\begin{equation}
\dv{\tau(\lambda_0)}{p} = 10^5 \frac{D'(p)}{g(p)}.
\end{equation}

\section{Eddy diffusion coefficients} \label{App:D}

Following the method of \citeA{ackerman01} we assume the eddy diffusion coefficient $K_{zz}$ is the same as that for free convection, which \citeA{gierasch85} derive to be:

\begin{equation}
K_{zz} = \frac{H}{3} \biggl( \frac{L}{H} \biggr)^{4/3} \biggl( \frac{RF}{\mu \rho_a c_p}\biggr)^{1/3}
\end{equation}

where $H$ is the atmospheric scale height, $\mu$ is the atmospheric molecular weight, $c_p$ is the specific heat capacity at constant pressure (i.e., J K$^{-1}$ kg$^{-1}$), $R$ is the gas constant (8.31 J K$^{-1}$ mol$^{-1}$), $\rho_a$ is the atmospheric density (kg m$^{-3}$), and $F$ is the convective heat flux (W m$^{-2}$). This equation can also be slightly more simply re-expressed as:

\begin{equation}
K_{zz} = \frac{H}{3} \biggl( \frac{L}{H} \biggr)^{4/3} \biggl( \frac{RF}{\rho_a C_p}\biggr)^{1/3}
\end{equation}

where $C_p$ is the molar heat capacity at constant pressure (i.e., J K$^{-1}$ mol$^{-1}$). 

In this equation, $L$ is the \textit{turbulent mixing length}. For a freely convecting atmosphere, this is typically assumed to be the atmospheric scale height, but in regions of atmospheric stability, \citeA{ackerman01} apply a scaling factor:

\begin{equation}
L = H \max (\Lambda, \Gamma/\Gamma_d)
\end{equation}

where $\Gamma$ is the local lapse rate (i.e., $-\dv{T}{z}$) and $\Gamma_d$ is the dry adiabatic lapse rate. The parameter $\Lambda$ is present to prevent $L$ becoming negative and was assumed by \citeA{ackerman01} to have a value of 0.1. In this study, where a large part of the atmospheric stability is assumed to come from the vertical variation of molecular weight, we modified this scaling to:

\begin{equation}
L = H \max \biggl(\Lambda, \frac{\Gamma}{\Gamma_d}\biggl(1-\alpha \dv{\mu}{z}\biggr ) \biggr )
\end{equation}

where the factor $\alpha$ was set to be:

\begin{equation}
\alpha  = \frac{1-0.0001}{ \max \biggl ( \abs{\dv{\mu}{z}} \biggr )}.
\end{equation}

Also, to force the profile of $K_{zz}$ to be consistent with stratospheric determinations for \citeA{fouchet03}, we reduced the factor $\Lambda$ from 0.1 to 0.0001.

\bibliography{NeptuneUranusHaze}

%
%
%
%
%

\end{document}